\documentclass[acmlarge]{acmart}

\AtBeginDocument{%
  \providecommand\BibTeX{{%
    \normalfont B\kern-0.5em{\scshape i\kern-0.25em b}\kern-0.8em\TeX}}}

\setcopyright{none}

\usepackage[normalem]{ulem} 
\usepackage{mathtools}
\usepackage{array}
\usepackage[export]{adjustbox}
\usepackage{graphicx}
\usepackage{caption}
\usepackage{makecell}

\usepackage{outlines}
\usepackage{dirtytalk}
\usepackage{tikz}
\usetikzlibrary{shapes, arrows, positioning, matrix, fit}
\usepackage{booktabs}
\usepackage{multirow}
\usepackage{xcolor}
\usepackage{subfigure}
\usepackage{longtable}
\usepackage{comment}

\newcolumntype{C}[1]{>{\centering\arraybackslash}m{#1}}
\newcolumntype{P}[1]{>{\arraybackslash}m{#1}}

\begin{document}

\title{A survey on the impacts of recommender systems on users, items, and human-AI ecosystems}

\author{Luca Pappalardo}
\email{luca.pappalardo@isti.cnr.it}
\orcid{0000-0002-1547-6007}
\affiliation{%
  \institution{ISTI-CNR}
  \streetaddress{Via G. Moruzzi 1}
  \city{Pisa}
  \state{Tuscany}
  \country{Italy}
  \postcode{56124}
}
\affiliation{%
  \institution{Scuola Normale Superiore of Pisa}
  \streetaddress{Piazza dei Cavalieri, 7}
  \city{Pisa}
  \state{Tuscany}
  \country{Italy}
  \postcode{56126}
}

\author{Salvatore Citraro, Giuliano Cornacchia, Mirco Nanni, Valentina Pansanella, Giulio Rossetti}
\affiliation{%
  \institution{ISTI-CNR}
  \streetaddress{Via G. Moruzzi 1}
  \city{Pisa}
  \state{Tuscany}
  \country{Italy}}

\author{Gizem Gezici, Fosca Giannotti, Margherita Lalli, Giovanni Mauro}
\affiliation{%
\institution{Scuola Normale Superiore of Pisa}
  \streetaddress{Piazza dei Cavalieri, 7}
  \city{Pisa}
  \state{Tuscany}
  \country{Italy}
  \postcode{56126}
  }

\author{Gabriele Barlacchi, Daniele Gambetta, Virginia
Morini, Dino Pedreschi}
\affiliation{%
  \institution{University of Pisa}
  \streetaddress{Largo Bruno Pontecorvo 3}
  \city{Pisa}
  \state{Tuscany}
  \country{Italy}
  \postcode{56127}}

\author{Emanuele Ferragina}
\orcid{0000-0002-6308-1413}
\affiliation{%
  \institution{Sciences Po}
  \streetaddress{27, Rue Saint Guillaume}
  \city{Paris}
  \state{Ile de France}
  \country{France},
  \postcode{75337 Cedex 07}
  }
\email{emanuele.ferragina@sciencespo.fr}
  
\renewcommand{\shortauthors}{Pappalardo et al.}

\begin{abstract}
Recommendation systems and assistants (in short, recommenders) influence through online platforms most actions of our daily lives, suggesting items or providing solutions based on users' preferences or requests.
This survey systematically reviews, categories, and discusses the impact of recommenders in four human-AI ecosystems -- social media, online retail, urban mapping and generative AI ecosystems.
Its scope is to systematise a fast-growing field in which terminologies employed to classify methodologies and outcomes are fragmented and unsystematic.
This is a crucial contribution to the literature because terminologies vary substantially across disciplines and ecosystems, hindering comparison and accumulation of knowledge in the field.
We follow the customary steps of qualitative systematic review, gathering 154 articles from different disciplines to develop a parsimonious taxonomy of methodologies employed (empirical, simulation, observational, controlled), outcomes observed (concentration, content degradation, discrimination, diversity, echo chamber, filter bubble, homogenisation, polarisation, radicalisation, volume), and their level of analysis (individual, item, and ecosystem).
We systematically discuss substantive and methodological commonalities across ecosystems, and highlight  potential avenues for future research.
The survey is addressed to scholars and practitioners interested in different human-AI ecosystems, policymakers and institutional stakeholders who want to understand better the measurable outcomes of recommenders, and tech companies who wish to obtain a systematic view of the impact of their recommenders.
\end{abstract}




\maketitle

\section{Introduction}
\label{sec:intro}
Recommendation systems and assistants (from now on, \emph{recommenders}) -- algorithms suggesting items or providing solutions based on users' preferences or requests \cite{ricci2015recommender, li2023recent, zhang2019deep} -- 
influence through online platforms most actions of our daily life.  
For example, recommendations on social media platforms suggest new posts and social connections, those on online retail platforms guide users' product choices, urban mapping platforms recommend locations to visit and offer routes to reach them, and generative AI platforms create content based on users' prompts.
Unlike other AI tools, such as medical diagnostic support systems, robotic vision systems, or autonomous driving, which assist in specific tasks, recommenders shape instantly many of our decisions.
The interactions between users and recommenders may generate long-lasting and often unintended impacts on human-AI ecosystems \cite{pedreschi2023human}, such as amplifying political radicalisation processes \cite{huszar2022algorithmic}, increasing CO2 emissions \cite{cornacchia2022routing} and amplifying discrimination \cite{mehrabi2021survey}.

A substantial body of surveys maps methodologically and conceptually the evolution of recommenders. 
Some surveys focus on algorithmic paradigms -- such as knowledge-based approaches \cite{tarus2018knowledge} and deep learning architectures \cite{zhang2019deep} -- and review their main applications. 
Transparency, interpretability and explainability are also central themes \cite{explainableRecommendersSurvey2021,zhang2020explainable}, with research examining how intelligibility, user trust, and accountability can be enhanced. 
Other contributions take a domain-specific perspective or systematise evaluation practices, discussing accuracy, robustness, and reproducibility in experimental workflows \cite{silveira2019good, EvaluateRecommenderSurvey2022}. 
Beyond accuracy, researchers have surveyed the techniques to achieve diversity, serendipity, and user satisfaction in recommender systems \cite{kunaver2017diversity, farinessAndDiversitySurvey2025, diversitySerendipityNoveltySurvey2016}. 
Other works focused on ethical and societal concerns \cite{trustworthyRecomSystemSurvey, fairnessSurvey2023} as well as sources of biases and associated debiasing strategies \cite{chen2023bias}. 
Temporal and behavioural dynamics constitute another frontier, with work examining sequence modelling and time-aware mechanisms \cite{sequentialRecommenderSurvey2024, timeAwareRecommenderEvalution2014}. 
Finally, surveys on graph neural networks and knowledge-graph-enhanced recommenders elucidate how graph structures enrich representation learning, reasoning, and explainability in recommender pipelines \cite{graphNeuralNetworksRecommendersSurvey2022, knowledgeGraphBasedSurvey2022}.
 
\subsection*{Gap in the literature}
In summary, existing surveys focus only on recommenders' characteristics (e.g., bias, efficiency, and transparency), but do not consider the impact of recommenders on users, items, and ecosystems. 
Moreover, articles have examined and measured this impact employing different nomenclatures, mathematical notations, methods, and datasets.

We utilise the term ``impact'' to encompass a broad range of studies that look causally or observationally at the outcomes generated in the ``relational process'' between users and recommenders.
This allows us to consider studies that look at the impact of recommenders on users, items and ecosystems, but also emerging research that considers this relational process not unidirectionally.
We refer in particular to the human-AI feedback loop, a process defined as follows \cite{pedreschi2023human}: users' choices determine the datasets on which recommenders are trained; the trained recommenders then exert an influence on users' subsequent choices, which in turn affect the next round of training, initiating a potentially never-ending cycle.

On these bases, our survey is motivated by the following questions:
What is the impact of recommenders on users, items, and human-AI ecosystems?
And what are the outcomes that emerge?
What insights emerge when constructing a holistic interpretation of the outcomes observed and the methodologies employed in the literature?
What are the key commonalities and differences of the outcomes of human-recommender interactions across ecosystems? 

\subsection*{Novel contribution} 
Our survey systematically reviews, categorises, and discusses the impact of recommenders on four human-AI ecosystems -- social media, online retail, urban mapping and generative AI -- and the methodologies employed to study these outcomes.
Online platforms in these four ecosystems offer prototypical examples of user–recommender interactions, making them a vantage point for understanding similar dynamics in other emerging human–AI ecosystems.
The terminologies employed to define the outcomes of user-recommender interactions and the methodologies employed to measure them are highly fragmented and unsystematic \cite{pedreschi2023human}.
Our survey addresses this issue, providing four novel contributions.

First, we gather and standardise in a parsimonious taxonomy (illustrated in Table \ref{tab:outcomes}) the outcomes observed in the literature -- concentration, content degradation, discrimination, diversity,  echo chamber, filter bubble, homogenisation, polarisation, radicalisation, volume.
This step is crucial because terminologies vary substantially across disciplines and ecosystems, hindering comparison and accumulation of knowledge in the field.
To provide an example, both concentration and popularity bias describe the amplification of attention toward a subset of items or individuals: the former is common in the social media literature, the latter in research on location-based recommender systems.  
By harmonising terminologies, our taxonomy eases comparisons at a wider scale. 
We define all outcomes on the basis of four key components of human–recommender interactions: the \emph{user}, the \emph{item}, the \emph{interaction}, and the \emph{recommendation}. 
This notation abstracts away ecosystem-specific details while remaining sufficiently flexible to describe a wide range of domains. 
By relying on these components, we provide a consistent and extensible foundation for classifying outcomes also in emerging ecosystems that future research may explore.

Second, we categorise the methodologies employed to assess the outcomes of user-recommender interactions  -- empirical, simulation, observational, and controlled studies (see Figure \ref{fig:methodologies}).
This helps investigating how each outcome has been derived.
Empirical studies offer insights into real-world behaviour, whereas simulation studies examine simplified artificial environments constructed through mathematical models.
Controlled studies divide users into treatment and control groups exposed to different recommendation strategies, while observational studies assume a single recommendation mechanism applied uniformly to all users.
Our categorisation also provides an overview of which methodological approaches prevail in specific ecosystems. 
For example, controlled empirical studies are common in online retail, but rare in the urban mapping ecosystem.
In a further step, we harmonise the mathematical notations of measures employed in the literature (Table \ref{tab:measures}) and provide visualisations that enable readers to quickly grasp what has been accomplished and where gaps remain (Figure \ref{fig:heatmaps}).

Third, we distinguish the levels at which outcomes are measured -- individual, item, and ecosystem -- and highlight what remains understudied across ecosystems.
This perspective offers a more comprehensive approach than is customary in the literature and enables the study of interactions across levels of analysis. 
For example, we show that an increase in individual diversity can coexist with a decrease in ecosystem diversity.
Fourth, we address data-access and ethical challenges that emerge from user–recommender interactions, offer recommendations for developers and policy makers, and outline methodological and substantive directions for future research.

Our survey can be helpful to several public and private stakeholders: \emph{(i)} scholars and practitioners may obtain guidance on recent advancements in different human-AI ecosystems; \emph{(ii)} policy makers and institutional stakeholders may better understand measurable outcomes of actual or potential recommenders and their societal consequences, such as polarisation, concentration, and discrimination; and \emph{(iii)} tech companies developing and employing recommenders may obtain a systematic view of the impact of their services to increase revenues and contribute to societal development.

The remainder of the paper is organised as follows.
Section \ref{sec:survey_building} details how we collected and classified the articles and built our taxonomy of outcomes and methodologies. 
Sections \ref{sec:social_media}-\ref{sec:GenAI} discuss the evidence gathered in each human-AI ecosystem.
Section \ref{sec:commonalities_differences} examines commonalities and differences across ecosystems.
Section \ref{sec:open_challenges}, approaches open challenges, provide recommendations and outline future research avenues.

\section{Construction of the Survey}
\label{sec:survey_building}

\begin{figure}[htb!]
\centering
\includegraphics[width=1\linewidth]{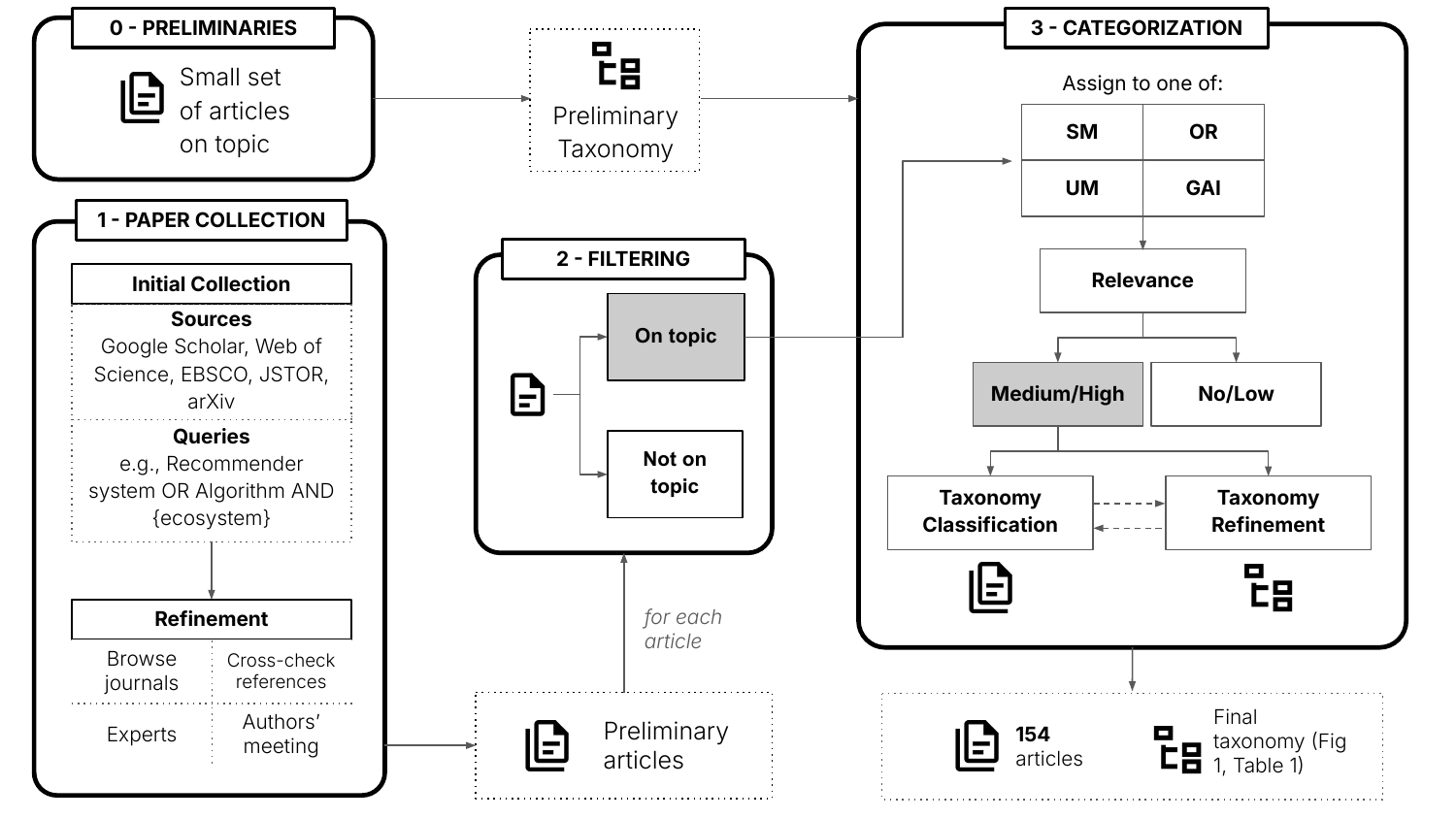}
\hfill
\caption{Flowchart summarising the steps employed to construct the survey: preliminaries, where a small initial set of relevant articles was used to define a first taxonomy (box 0); paper collection, where we combined database searches, journal browsing, reference cross-checking, and expert input (box 1); filtering, where articles were screened for relevance (box 2); and categorisation, where relevant articles were assigned to one of the four ecosystems (SM, OR, UM, GAI), evaluated for relevance, and integrated into the final taxonomy (box 3).}
\label{fig:flowchart}
\end{figure}

We gathered 154 articles from different disciplines (e.g., computer science, network science and complex systems, computational social science, marketing and management, human mobility, urban studies) in leading international journals, conferences and preprint databases on the basis of the customary steps of qualitative systematic reviews \cite{green2001writing}. 
Studies were collected from Google Scholar, Web of Science, EBSCO, JSTOR, and ArXiv by scanning titles and abstracts for keywords related to the impact of recommenders in social media, online retail, urban mapping and generative AI ecosystems (see Figure \ref{fig:flowchart}, box 0).
In some studies, the recommender under examination is explicitly described or implemented by the authors, whereas in others it remains implicit and the analysis focuses on the platform’s overall impact.
We include the latter when the observed effects can reasonably be attributed to the platform’s recommender.

We refined our initial results with a series of additional steps (Figure \ref{fig:flowchart}, boxes 1 and 2). 
We browsed all issues of journals and proceedings of conferences where the original articles were gathered in the initial search and cross-checked the bibliography of each selected article. 
Then, we called upon the expertise of two senior scholars and presented the article selection in a group meeting with all authors. 
Moreover, we eliminated the articles that did not fit our search definition after the initial classification process.\footnote{We only consider articles that measure the impact of recommenders on users, items and ecosystems; therefore, we exclude those only aimed at improving recommenders' performance, for which surveys already exist \cite{tarus2018knowledge, zhang2020explainable, silveira2019good, EvaluateRecommenderSurvey2022, chen2023bias, sequentialRecommenderSurvey2024, timeAwareRecommenderEvalution2014, graphNeuralNetworksRecommendersSurvey2022, knowledgeGraphBasedSurvey2022}.} 

We categorise the outcomes and methodologies of each article through the following process (see Figure \ref{fig:flowchart}, box 3).
We divided the authors into four teams, including in each group experts for the specific ecosystem analysed.
All articles were assigned to at least two coders, who independently read the paper, evaluated its relevance for the survey, and manually classified it based on a preliminary taxonomy.
Ecosystem teams discussed each article, solving disagreements on the coders' classification.
This step allowed each team to present a preliminary classification of the articles to the entire research group.
During this presentation, the ecosystem teams illustrated doubts concerning the keywords employed and the articles that were difficult to classify under the preliminary taxonomy.
These doubts were progressively solved to build the final taxonomy.
The outcomes and their definitions are summarised in Table \ref{tab:outcomes}.

\subsection{Taxonomy}
\label{sec:taxonomy}
We designed a taxonomy that classifies studies based on the outcomes measured (Table \ref{tab:outcomes}) and the methodologies employed (Figure \ref{fig:methodologies}).
We built the taxonomy through a consensus exercise among the authors. 
Initially, the taxonomy was constructed through a deductive process based on the characteristics of a sample of articles already known by the authors. 
All these articles have been then reclassified by the authors to validate or question each category in the taxonomy.
The iterative nature of the process allowed us to progressively improve the initial categorisation, proposing, in the end, a robust and comprehensive taxonomy.

\subsubsection*{Human-AI Ecosystems} We gather articles from four human-AI ecosystems: social media, online retail, urban mapping, and generative AI. 
Articles in the social media ecosystem analyse recommenders that filter content or suggest users to follow, across platforms such as Facebook, Google News, Apple News, Instagram, X, Reddit, Gab, YouTube, and TikTok.
Research in the online retail ecosystem focuses on recommenders that suggest products and services (such as goods, songs, books, and movies) across platforms like Amazon, Alibaba, eBay, Netflix, and Spotify.
Studies in the urban mapping ecosystem examine recommenders that suggest routes, rides, accommodations, or points of interest, across platforms such as Google Maps, Waze, Uber, Lyft, Airbnb, Booking.com, Tripadvisor, and Yelp.
The generative AI ecosystem covers research on tools that generate content (e.g., text, images, audio, or video) from user prompts, including ChatGPT, Llama, Mistral, Gemini, DeepSeek, and DALL·E.
The interdisciplinary team of authors -- including computer scientists, complexity scientists, mathematicians, and sociologists -- has been constructed to cover expertise on these four human-AI ecosystems and the different methodologies employed. 
This classification is mirrored in the paper's organisation: Sections \ref{sec:social_media}-\ref{sec:GenAI} correspond to specific ecosystems, enhancing readability and accessibility for readers interested in specific domains. 
Some recommenders can span multiple ecosystems. 
For example, accommodation platforms could be placed under online retail, but we classify them within urban mapping because they pertain to the spatial dimension of human behaviour. 
We group recommenders of routes, rides, accommodations and points of interest in the urban mapping ecosystem to provide a coherent perspective for scholars in human mobility and urban studies while remaining accessible to researchers interested in online retail.

We focus on these four human-AI ecosystems because they underpin large-scale platforms that influence the behaviour of vast populations. Moreover, some of these platforms fall within the scope of Very Large Online Platforms and Search Engines (VLOPs/VLOSEs) as defined by the EU's Digital Services Act (DSA).
These online platforms steer content consumption, purchasing choices, mobility decisions, and content generation, posing potential systemic risks to citizens and society. 
Other digital platforms based on recommenders -- such as medical diagnosis systems, autonomous vehicles, or educational tutors -- currently operate in more individualised settings but may scale up into mainstream ecosystems. 
Our taxonomy could be extended and refined in the future to encompass these and other emerging domains.

\begin{figure}[H]
\centering
\begin{tikzpicture}[node distance=1.5cm, auto]
    \tikzstyle{level1} = [rectangle, draw, text centered, rounded corners, minimum height=1cm, minimum width=4cm, fill=white!80!gray, font=\Large\bfseries];
    \tikzstyle{level2} = [rectangle, draw, text centered, font=\bfseries, minimum height=1cm, minimum width=3cm];
    \tikzstyle{level3} = [rectangle, draw, text centered, 
    minimum height=1cm, minimum width=3cm];

    \node [level1] (methodology) {Methodologies};

    \matrix (level2matrix) [matrix of nodes, nodes={level2}, below=0.5cm of methodology, column sep=5cm, row sep=0.5cm] {
        Empirical          & Simulation\\
    };

    \matrix (level3matrix) [matrix of nodes, nodes={level3}, below=0.5cm of level2matrix, column sep=1cm, row sep=0.5cm] {
        Controlled          & Observational & Controlled & Observational \\
    };

    \draw (methodology.south) -- (level2matrix-1-1.north);
    \draw (methodology.south) -- (level2matrix-1-2.north);

    \draw (level2matrix-1-1.south) -- (level3matrix-1-1.north);

    \draw (level2matrix-1-1.south) -- (level3matrix-1-2.north);

    \draw (level2matrix-1-2.south) -- (level3matrix-1-3.north);

    \draw (level2matrix-1-2.south) -- (level3matrix-1-4.north);
    
\end{tikzpicture}
\caption{Categorisation of methodologies of the surveyed articles.
At a first level, we categorise articles into empirical or simulation; at the second level, we classify them as controlled or observational.}
\label{fig:methodologies}
\end{figure}

\subsubsection*{Methodologies} 
We systematically classify articles within each human-AI ecosystem into empirical and simulation studies and within each category, we distinguish between controlled and observational studies.

\emph{Empirical studies} derive insights from data produced by user-recommender interactions. 
These data describe users' choices and actions on platforms and interactions with recommenders (e.g., posts and likes, routes or rides requested, products searched or purchased, prompts submitted and response received).
When datasets are large and diverse, these studies allow for deriving useful insights.
However, the ability to draw universal conclusions is constrained by specific geographic, temporal and contextual circumstances. 
Moreover, reproducing these studies is challenging because data are often owned by online platforms that are reluctant to share them.

\emph{Simulation studies} are anchored to synthetic data, whether generated by mechanistic models, AI-driven models, digital twins, or simulation frameworks. 
They offer an alternative methodological pathway to deal with large-scale ecosystems when empirical data is not readily available, and allow reproducibility of experiments under the same initial conditions. 
By manipulating parameters, scholars can scrutinise how the impact of recommenders can vary under different conditions (e.g., kind of recommender, user profile). 
However, simulations are based on heavy assumptions and therefore do not necessarily reflect real-world dynamics and are limited in unveiling unexpected outcomes. 

Both empirical and simulation methodologies can employ controlled or observational approaches. 
\emph{Controlled studies} comprehend quasi-experiments, randomised controlled trials, and A/B tests \cite{deaton2018understanding, hariton2018randomised}.
These studies divide user samples into control and treatment groups exposed to different recommendations.
Sample randomisation may reduce selection biases, ensuring that participants in both groups have an equal chance of receiving the recommendation.
Controlled studies enable researchers to control for various factors and conditions, allowing the isolation of the effect produced by a specific intervening variable. 
Their main advantage is inferring causal relationships and attributing observed effects to the recommendation. 
However, in complex social systems, individuals within the control group can be influenced by those in the treatment group \cite{aronowet2017estimating}.
For example, users on social media platforms interact by viewing each other's posts, while drivers using location-based recommenders influence one another at road intersections.
Therefore, controlled experiments may not satisfy the Stable Unit Treatment Value Assumption (SUTVA) for causal inference -- which states that potential outcomes for each individual should be unrelated to the treatment status of other individuals \cite{cox1958planning, rubin1986whichifs} -- and might not provide unbiased estimates of causal quantities of interest.
Moreover, the inclusion and exclusion criteria of the controlled settings might limit the generalisability of findings; and there is limited flexibility in adapting to changes intercurring during the experiments. 
Controlled studies are hard to design because they require direct access to platforms' users and recommenders \cite{knott2022transparency}.

\emph{Observational studies}, whether grounded in empirical or synthetic data, examine a single recommendation principle for the entire population. 
For example, they can include the analysis of users' choices on social media platforms, the routes requested on location-based platforms, the products purchased on online retail platforms, the prompts submitted on generative AI platforms and the data gleaned from browser loggers and platform APIs \cite{knott2021responsible}. 
While offering broad insights when data is large and representative, they are ill-equipped to establish causal relationships, necessitating supplementary evidence. 
Observational studies are also susceptible to confounding variables, which may compromise the capacity to causally evaluate the impact of recommenders. 

Some examples can help clarify how we categorise the methodologies.
If a study analyses data reflecting users' behaviour on an online platform, it is categorised as an observational empirical study. 
However, if only a subset of platform users is exposed to a recommender, and the study compares the behaviours of those exposed to those who are not, it is categorised as a controlled empirical study.
Differently, when a study generates synthetic data, we categorise it as a simulation study, which can be controlled or observational.
Quasi-experiments, in which an exogenous element (e.g., the introduction of a new recommender) splits the population into two or more groups \cite{gangl2015making}, are considered as controlled studies.
In contrast, when an exogenous element does not segment the population into different groups, studies are categorised in the observational category. 
An article may be categorised under different methodologies if it employs two or more of them.
A note of caution is warranted regarding the categorisation of methodologies in the generative AI ecosystem. 
User-recommender interactions occur through prompting, and each prompt generates synthetic content. 
The outputs, therefore, share characteristics of both empirical and simulated data: a real user submits the request, but the response is fully model-generated. 
In this survey, we classify such studies as simulations. 
When researchers issue multiple prompts using different criteria or techniques and treat prompts as simulated agents, we interpret these variations as treatment groups in a controlled study. 
However, in the absence of treatment and control groups, we categorise these studies as observational rather than controlled.

\subsubsection*{Outcomes}
We defined the taxonomy of outcomes employing the following analytical procedure. 
First, we extracted a comprehensive list of keywords from the reviewed literature, capturing the outcomes related the impact of recommenders.
Next, we examined semantic overlaps among these keywords to identify those that referred to similar underlying concepts.
When overlaps were found, we harmonised them under a more general label.
Finally, we defined a set of outcomes balancing parsimony and conceptual coverage.

Because individuals interact with different types of items across ecosystems, we introduced a unified notation to describe these interactions. 
We use the term \emph{user} to denote the human agent involved in the interaction with the recommender: this may correspond to a social media user, an online retail customer, a user, driver or passenger in urban mapping, or a user of a generative AI platform. 
Likewise, the notion of \emph{item} is ecosystem-dependent. 
In social media, it includes content such as posts or videos, as well as users themselves when new friendships are recommended. 
In online retail, it refers to commodities such as products, songs, podcasts, or movies. 
In urban mapping, it encompasses locations or routes suggested by location-based platforms, or rides and houses suggested by ride-hailing and accommodation platforms. 
In the generative AI ecosystem, it corresponds to model-generated content or to the semantic entities contained within it (e.g., recommended movies).
We define \emph{interactions} as any action linking a user to an item, such as views, clicks, likes, or purchases. 
We distinguish interactions from \emph{recommendations}: an item may be displayed to the user, who elicits no subsequent action. 
The notation used throughout the paper follows this unified conceptual framework, ensuring consistency in the description of outcomes across ecosystems (see Table \ref{tab:outcomes})
and their measures (see Table \ref{tab:measures}).

We categorise the outcomes at the \emph{individual}, \emph{item}, and \emph{ecosystem} levels.
Individual outcomes are defined as the effects of user-recommender interactions on users, whereas item outcomes refer to the effects on specific items.
For example, when the interactions with the recommender lead a user to engage with a more varied set of items over time, this represents an individual outcome. 
Instead, when these interactions cause certain items to become increasingly popular over time, this reflects an item outcome.
Ecosystem outcomes operate at an aggregate scale and capture collective dynamics that emerge across a population of users or items.
For example, if user–recommender interactions increase the overall volume of purchases or intensify the concentration of product popularity on an online platform, we consider them as ecosystem outcomes.

Table \ref{tab:outcomes} summarises the definitions and analytical levels of the outcomes examined in the literature and indicates the ecosystems where they emerge.
In our taxonomy, each outcome is associated with a single analytical level, except for \emph{diversity} and \emph{volume}, which can be at the individual, item, and ecosystem levels.
To illustrate this point, we showcase examples from the online retail ecosystem. 
Studies examine changes in revenue and purchased products, measuring whether users purchase more products (individual outcome), specific items are purchased more (item), and aggregate consumption increases or decreases (ecosystem). 
Similarly, they explore whether users engage with a more diverse range of products (individual), items are purchased by a more diverse set of users (item), and how items' popularity is distributed (ecosystem).
The other outcomes are: \emph{radicalisation} (individual) and \emph{polarisation} (ecosystem) in social media, \emph{filter bubbles} in social media and online retail (individual); \emph{content degradation} in generative AI (item); \emph{concentration}, \emph{discrimination}, \emph{echo chamber}, and \emph{homogenisation} in all ecosystems (ecosystem level).
Definitions of all these outcomes are provided in Table \ref{tab:outcomes}.

The distinction between different analytical levels allows us to disambiguate outcomes that are often confused in the literature.
For example, the terms filter bubble and echo chambers are often used interchangeably in social media research, while we clearly distinguish them in our taxonomy.
Filter bubbles describe an individual outcome in which a user is exposed primarily to items that align with their past online behaviour. 
Echo chambers are instead an ecosystem outcome, referring to a situation in which users predominantly interact with like-minded others or with items preferred by similar users.

Some outcomes appear under different names in the literature, both within and across ecosystems.
A clear example is the distinction between polarisation and fragmentation. 
While polarisation traditionally refers to the emergence of two opposing groups -- reflecting its origins in studies of the U.S. two‐party system -- fragmentation denotes the presence of multiple opinion clusters. 
Over time, however, polarisation has been broadened to describe a wide variety of configurations (with at least twelve definitions identified in the literature \cite{bramson2016disambiguation}). 
For analytical parsimony, we unify these terms under a single outcome label: polarisation.
A similar issue arises with concentration and popularity bias. 
Although popularity bias is commonly used in the social media literature, both terms ultimately describe the same underlying tendency: recommenders amplifying attention toward a subset of users or items. 
Because concentration is more general and transferable across ecosystems, we adopt it as our preferred outcome label.
In the urban mapping ecosystem, we introduce one further simplification by treating congestion as a specific form of concentration that manifests in the spatial domain through the build-up of vehicular traffic.
It is important to acknowledge that some outcomes in the classification may partially overlap.
For example, the concentration of user purchases can also be associated with a reduction of ecosystem diversity.
However, the opposite is not always true: diversity reduction does not imply an increase in concentration.

\begin{table}[H]
\caption{Taxonomy of outcomes, with the level of analysis, description, and where the outcome is encountered.
We use the following acronyms: social media (SM), online retail (OR), urban mapping (UM), and generative AI (GAI).}
\vspace{-0.4cm}
\centering
{\scriptsize
\renewcommand{\arraystretch}{1.4}
\begin{tabular}{C{1.3cm} C{2.3cm} P{8.8cm} C{2.5cm}}
\toprule
\textbf{Level of analysis} & \textbf{Outcome} & \textbf{Description of the outcome (in parentheses, we provide examples about each ecosystem)} & \textbf{Where the outcome is encountered} \\
\midrule

\multirow{8}{*}{%
  \makebox[1.3cm][c]{%
    \rotatebox[origin=c]{90}{%
      {\large Individual}
    }%
  }%
}

& {\footnotesize Diversity} & 
    Extent to which a user distributes interactions evenly across items (e.g., content consumed in SM, products purchased in OR, locations visited in UM, content generated in GAI).
    & SM, OR, UM, GAI\\
\cmidrule{2-4}
  & {\footnotesize Filter Bubble} & 
    Situation in which a user is predominantly exposed to items that reinforce their past preferences and behaviours (e.g., same-viewpoint content in SM, similar products repeatedly shown in OR). 
    & SM, OR\\
\cmidrule{2-4}
  & {\footnotesize Radicalisation} & 
    Process through which a user progressively adopts more extreme ideological positions over time (e.g., increasingly extreme content consumed in SM).
    & SM\\
\cmidrule{2-4}
  & {\footnotesize Volume} & 
    Amount of user’s engagement or exposure and user's attribute values (e.g., posts interacted in SM, items purchased or amount spent in OR, routes taken in UM).
    & SM, OR, UM\\

\midrule
\midrule
\multirow{5.5}{*}{%
  \makebox[1.3cm][c]{%
    \rotatebox[origin=c]{90}{%
      {\large Item}
    }%
  }%
}
& {\footnotesize Diversity} & 
   Extent to which engagement with an item is distributed across different users (e.g., users interacting with a post in SM, purchasing a product in OR, visiting a location in UM).
    & SM, OR, UM\\ 
\cmidrule{2-4}
  & {\footnotesize Content Degradation} & Decline in the quality or realism of AI-generated outputs (e.g., repetitive or incoherent text, reduced image fidelity in GAI).
    & GAI\\
\cmidrule{2-4}
  & {\footnotesize Volume} & 
    Amount of engagement or exposure an item receives, or changes in its attributes (e.g., number of users interacting with a piece of content in SM, purchase counts in OR, visits to a location in UM).
    & SM, OR, UM\\
\midrule
\midrule
\multirow{14}{*}{%
  \makebox[1.3cm][c]{%
    \rotatebox[origin=c]{90}{%
      {\large Ecosystem}
    }%
  }%
}
& {\footnotesize Concentration} & 
    Extent to which overall engagement is dominated by a small subset of items (e.g., most views going to a few influencers in SM, most purchases to a few products in OR, most visits to a few locations in UM, repeated generation of certain terms in GAI). 
    & SM, OR, UM, GAI
    \\
\cmidrule{2-4}
  & {\footnotesize Discrimination} & 
Extent to which interactions systematically favour users or items with certain characteristics (e.g., visibility skewed toward specific demographic groups in SM and OR, preferential booking of hosts in UM).
& SM, OR, UM\\
\cmidrule{2-4}
  & {\footnotesize Diversity} &
   Extent to which engagement or exposure is distributed evenly across items in the entire ecosystem (e.g., distribution of followers in SM, product popularity distribution in OR, location popularity distribution in UM, variety of patterns in GAI).
   & SM, OR, UM, GAI\\
\cmidrule{2-4}
  & {\footnotesize Echo Chamber} & 
    Situation in which users predominantly interact with like-minded others or with items preferred by similar users (e.g., ideologically homogeneous content in SM, similar products co-purchased in OR, socio-demographically matched host–guest interactions in UM).
    & SM, OR, UM\\
    \cmidrule{2-4}
    & {\footnotesize Homogenisation} & Extent to which individuals become similar in their exposure or interaction patterns over time (e.g., users consuming the same content in SM, purchasing similar products in OR, or visiting the same locations in UM, similar writing styles in GAI).
    & SM, OR, UM, GAI\\
\cmidrule{2-4}
  & {\footnotesize Polarisation} & 
    Separation of users into distinct and opposing groups with limited overlap and divergent attributes or preferences (e.g., ideological families in SM). 
    & SM\\
\cmidrule{2-4}
  & {\footnotesize Volume} & 
    Overall amount of engagement, exposure, or attribute values across the ecosystem (e.g., total posts produced in SM, total purchases in OR, total trips in UM, writing ideas in GAI). 
    & SM, OR, UM, GAI\\
\bottomrule
\end{tabular}}
\label{tab:outcomes}
\end{table}

\paragraph{Measurements of outcomes.} The fragmentation and lack of consistency in the definition of outcomes are reflected by the panoply of measures used to quantify them. 
To address this gap, we collected and harmonised the main measures employed across the literature. 
This required standardising the notation for users, items, interactions, and recommendations so that the formulas describing each measure are consistent and comparable.
Table \ref{tab:measures} reports this unified notation alongside the definition of each measure, its mathematical formulation, the outcomes it captures, and the articles in which it is used.

{\scriptsize
\begin{longtable}[H]{P{5.25cm}  P{6.25cm} P{1.5cm} C{1cm}}
\caption{Main measures to quantify outcomes of user-recommender interactions.}
\vspace{-0.4cm}
\label{tab:measures} \\
\toprule
\textbf{Measure and Definition} &
\textbf{Formula} &
\textbf{Outcomes} &
\textbf{Papers} \\
\midrule
\endfirsthead

\toprule
\textbf{measure and definition} &
\textbf{formula} &
\textbf{outcomes} &
\textbf{papers} \\

\midrule
\endhead

\endfoot

\bottomrule
\endlastfoot

\multicolumn{4}{p{15cm}}{
\textbf{Notation of the table:} $\mathcal{U}$ denotes the set of all users, and $u \in \mathcal{U}$ represents a single user. Similarly, $\mathcal{I}$ denotes the set of all items, and $i \in \mathcal{I}$ represents a single item. $\hat{\mathcal{I}}$ denotes the set of items recommended at least once to any user.
The set of items recommended to user $u$ is denoted by $\hat{\mathcal{I}}_u$ and the set of items $u$ has actually interacted with is denoted by $\mathcal{I}_u$.
We denote with $\mathcal{T}$ the set of all user-item interactions. $\mathcal{T}_i$ is the set of user-item interactions involving item $i$. 
We denote with $\hat{\mathcal{T}}$ the set of all user-item recommendations, and with $\hat{\mathcal{T}}_i$ the set of all user-item recommendations involving item $i$.
}\\
\midrule

\textbf{Amplification}: it measures how much more likely is that users in the treatment group encounter certain items compared to those in the control group. 
Higher values indicate that treatment users are exposed to a larger volume of those items. 
& 
$A_d(T){=} \left(
\frac{|\mathcal{U}_{T,d} \cap \mathcal{U}_{\text{treat}}| + 1}{|\mathcal{U}_{T,d} \cap \mathcal{U}_{\text{control}}| + 1}-1
\right) \cdot 100\%$,
where $T {\subseteq} \mathcal{I}$ is a set of specific items (e.g., tweets from a specific political party), 
$\mathcal{U}_{\text{treat}}$ and $\mathcal{U}_{\text{control}}$ 
are the treatment and control user groups, 
and $\mathcal{U}_{T,d}$ is the set of users who were exposed to items in $T$ on day $d$.
& item volume & \cite{huszar2022algorithmic, bouchaud2023algorithmic, ye2025auditing} \\
\midrule

\textbf{Gini index}: it measures how evenly interactions are distributed across items. 
Low values indicate that interactions are evenly distributed among items, while high values indicate that a small number of items receive most of the interactions.
&  
$G = 1 - 2 \int_0^1 L(p) \, dp$, where $L(p)$ is the Lorenz curve, representing the cumulative share of the interactions held by the bottom $p$ fraction of items, sorted in ascending order of share.
For a finite distribution $x_1, \dots, x_n$ with mean $\mu$:
$G = \frac{\sum_{i=1}^n \sum_{j=1}^n |x_i - x_j|}{2n^2\mu}$.
It ranges from $0$ (absolute equality) to $1$ (absolute inequality). 
& 
ecosystem diversity, individual diversity, concentration, discrimination, content degradation
&
\cite{hazrati2022recommender,lee2020, lee2019How, chaney2018algorithmic, lee2014impact,flederHosanagar2009blockbuster,matt2013,barlacchi2025simulation,fleder2007recommender,matt2019factual,wu2011,bokanyi2020understanding,mauro2025urban,sanchez2023bias,bartley2021auditing,bartley2024impacts,zhou2010impact,fabbri2022exposure,ferrara2022link,ye2025auditing,gambetta2025} \\
\midrule

\textbf{Intragroup Diversity}: the degree to which individuals differ from one another in the composition of the categories they consume. 
Higher values indicate greater variation in habits across individuals. 
& $ID = \frac{1}{|\mathcal{U}|} \sum_{u \in \mathcal{U}} \left[ 1-cos(\mathbf{r}_u, \overline{\mathbf{r}} ) \right]$, where $\mathbf{r}_u$ is a vector describing the
fraction of user $u$'s interactions belonging to each item category, and $\overline{\mathbf{r}}$ is the average of all such vectors.
& homogenisation & \cite{holtz2020engagement} 
\\ 
\midrule

\textbf{Coverage}: the percentage of items recommended or with which the users interacted at least once. 
& 
$Cov = |\hat{\mathcal{I}}|/|\mathcal{I}|$
& ecosystem diversity, individual diversity & \cite{hazrati2022recommender, yi2022recommendation, matt2019factual, mansouryAbdollahpouri2020, li2022recommender}
\\
\midrule

\textbf{Shannon Entropy}: it measures how a user's interactions are distributed across items or categories of items.
High values indicate balanced and diverse consumption; low values indicate concentration on fewer items or categories. 
& $S_u = - \sum_{i \in \mathcal{I}} p_{u,i} \log p_{u,i}$, where 
$\mathcal{I}$ 
$p_{u,i}$ is the proportion of interactions (or occurrences, or session time) that $u$ allocates to $i$.
The measure is typically normalized by dividing it by $\log |\mathcal{I}|$.
& individual diversity, content degradation, filter bubble & \cite{holtz2020engagement, chen2022more, gambetta2025}  \\ 
\midrule

\textbf{Pairwise Item Distance}: the average dissimilarity between the items a user interacts with. Higher values indicate that the user engages with items that are distinct from each other, according to a domain-specific distance function. 
& $D_u = \frac{1}{|\mathcal{I}|} \frac{1}{|\mathcal{I}_u|(|\mathcal{I}_u|-1)} \sum_{j,l \in \mathcal{I}_u, j\neq l } d(j,l)$, where $d(j, l)$ is a domain-specific distance function between two items $j, l \in \mathcal{I}$. 
&  individual diversity & \cite{aridor2020deconstructing, noordeh2020echo, ge2020understanding, nguyen2014exploring, anderson2020algorithmic} \\
\midrule

\textbf{Average  Popularity}: the mean frequency with which items appear in recommendations. 
Higher values means that recommendations are biased towards more popular items.
& 
$AIP_u = \frac{1}{|\mathcal{I}_u|} \sum_{i \in \mathcal{I}_u} \text{pop}_i$, where $\text{pop}_i$ is the popularity of item $i$, typically computed as:
$\text{pop}_i = n_i/(\sum_{k \in \mathcal{I}} n_k)$.
Here, $n_i$ is the total number of interactions received by $i$. 
& 
 concentration & \cite{hazrati2022recommender, mansouryAbdollahpouri2020} \\
\midrule

\textbf{Items share}: the fraction of total interactions held by the most (least) popular items.
Higher values indicate concentration of users' interactions on the most (least) popular items. 
& $IS_n = \sum_{i \in \mathcal{I}^{(n)}} |\mathcal{T}_i|/|\mathcal{T}|$ where $\mathcal{I}^{(n)}$ is the set of the $n$ most (least) popular items.
& concentration & \cite{lee2019How, matt2013, matt2019factual, yi2022recommendation}  \\
\midrule
\textbf{Ideological Share}: the proportion of recommended items that belong to different ideological categories (e.g., Democrats, Republicans) relative to the user’s own political positioning. 
More even proportions indicate higher ideological diversity; skewed proportions indicate lower diversity. 
& 
$IS_u = \left\{\frac{|\hat{\mathcal{I}}_u(j)|}{|\hat{\mathcal{I}}_u|} \right\}_{j {\in} \mathcal{J}}$, where $u {\in} \mathcal{U}$ and $\hat{\mathcal{I}}_u(j) {\in} \hat{\mathcal{I}}_u$ is the subset of items recommender to $u$ with ideology $j$, and $\mathcal{J}$ is a set of ideological categories. & individual diversity, filter bubble, radicalisation & \cite{haroon2022youtube, ribeiro2023amplification} \\
\midrule
\textbf{Cross-Cutting Content Ratio}: the percentage of content a user encounters or interacts with that originates from users with different political or ideological views. Low values indicate limited exposure to such content, suggesting a potential filter bubble effect.
& 
The cross-cutting content ratio for user $u {\in} \mathcal{U}$ is defined as
$CCR_u
{=} 
\frac{P(\text{click} \mid \text{exposed}, \text{cross})}
{P(\text{click} \mid \text{exposed}, \text{not cross})}.
$
To compute these probabilities, let $\hat{\mathcal{I}}_u^{\text{cross}}$ and $\hat{\mathcal{I}}_u^{\text{noncross}}$ 
denote the sets of cross-cutting and non-cross-cutting items exposed to user $u$, and let 
$\mathcal{T}_u^{\text{cross}}$ and $\mathcal{T}_u^{\text{noncross}}$ be the corresponding sets of user–item interactions. 
Then:
$P(\text{click}|\text{exposed}, \text{cross})
{=} |\mathcal{T}_u^{\text{cross}}|/|\hat{\mathcal{I}}_u^{\text{cross}}|$ and $P(\text{click}| \text{exposed}, \text{not cross})
{=} |\mathcal{T}_u^{\text{noncross}}|/|\hat{\mathcal{I}}_u^{\text{noncross}}|$.
 & filter bubble & \cite{bakshy2015exposure, nyhan2023like, chen2021neutral} \\
\midrule
\textbf{Attribute Variance}: the dispersion of users' attributes (e.g., opinions) around the mean. Higher values indicate that attributes are more spread out across users. 
&
$Var_x = \sum_{u \in \mathcal{U}} \bigl(x_u - \mu_x\bigr)^2
$, where $x_u$ is the attribute value for user $u$, and $\mu_x$ is the average over the population. 
& polarisation & \cite{chitra2020analyzing} \\
\midrule
\textbf{Random Walk Controversy}: the probability that a random walker starting from a user in a social network encounters users with the same attribute. Higher values indicate that users are more likely to remain within attribute-homogeneous regions of the network, reflecting polarisation. 
&  
$RWC = P_{AA} P_{BB} - P_{AB} P_{BA}$,
where $P_{ab}$ denotes the probability that a random walker transitions from a user with attribute $a \in \{A, B\}$ to a user with attribute $b \in \{A, B\}$.
& polarisation & \cite{cinus2022effect} \\
\midrule
\textbf{DER Polarisation}: the average distance between individuals’ attributes (e.g., opinions) in a multidimensional space, assigning greater weight to distances involving attributes held by many people. It rewards configurations in which large, dense groups are far apart. Higher DER values indicate stronger polarisation. 
& $
DER_{\gamma}(\theta) \;=\; 
\int_{d} \int_{d} 
\theta(x_u)^{\,1+\gamma} \, \theta(x_v) \, \lvert x_u - x_v \rvert \, dx_u \, dx_v,
\ \gamma \in \left[\tfrac{1}{4}, 1\right]$, where $x_u$ is the attribute (e.g. opinion) of user $u$, $d$ denotes an ideological dimension (a coordinate axis of the multidimensional ideological space), $\theta$ is the attribute distribution on $d$, and $\gamma$ is a sensitivity parameter. The attribute is assumed to be continuous. 
& polarisation & \cite{ramaciotti2021auditing} \\
\midrule
\textbf{Mean and Std of Ideological Slant}: the mean and standard deviation of the ideological orientation of congenial items recommended to an user over time. A shift in the user-level mean toward more extreme values indicates a potential trajectory toward radicalisation.
& $\mu_y = \tfrac{1}{|\hat{\mathcal{I}}_u|}\sum_{i=1}^{|\hat{\mathcal{I}}_u|} y_i,\;\; 
\sigma = \sqrt{\tfrac{1}{{|\hat{\mathcal{I}}_u|}}\sum_{i=1}^{|\hat{\mathcal{I}}_u|} (y_i -\mu_y)^2}$, 
where $y_j$ is the slant of item $i$ (ideologically congenial).   
& radicalisation & \cite{haroon2023auditing} \\
\midrule
\textbf{Degeneracy}: the amount of divergence from its initial state of an user’s current interest vector. The interest vector assigns to each item a value reflecting the strength of the user’s interest. Higher degeneracy indicates a larger shift in interest (often toward more extreme content) and signals a greater degree of radicalisation.
& $D = \lVert x(t) - x(0) \rVert, x(t) : \mathcal{I} \rightarrow \mathbb{R}$, where $x(0)$ is the user's attribute (e.g. opinion) vector at $t=0$, and $x(t)$ is the user's attribute vector at time $t$. $x(0)$ is sampled from a uniform distribution. & radicalisation & \cite{ jiang2019degenerate} \\
\midrule
\textbf{Segregation Index}: the extent to which different groups consume or interact with dissimilar sets of items. Higher values indicate stronger separation between groups, signalling a more pronounced echo chamber effect in the system. 
& $S{=}\sum_{i \in \mathcal{I}}
\left( \frac{A_i}{A} \cdot \frac{A_i}{v_i} \right)
{-}
\sum_{i \in \mathcal{I}}
\left( \frac{B_i}{B} \cdot \frac{A_i}{v_i} \right)$,
where $A_i$ and $B_i$ denote the number of interactions with item $i$ from groups $A$ and $B$, 
$A {=} \sum_{i \in \mathcal{I}} A_i$ and $B {=} \sum_{i \in \mathcal{I}} B_i$ are the total interactions from each group, 
and $v_i {=} |\mathcal{T}_i|$ is the total number of interactions with item $i$.
The first term represents the interaction-weighted exposure of group $A$ to group $A$, 
and the second term represents the exposure of group $B$ to group $A$.
& echo chamber & \cite{gonzalez2023asymmetric} \\

\midrule
\textbf{Attribute Alignment KDE}: 
the kernel density estimate of the joint probability distribution between a user’s attribute and the weighted average attribute of their neighbors. Bimodality in the resulting density indicates the presence of two distinct echo chambers. 
& 
$P(x, \bar{x}) 
{=} \frac{1}{|\mathcal{U}|\, h^2} 
\sum_{u \in \mathcal{U}}
K\!\left(
\frac{x - x_u}{h},\;
\frac{\bar{x} - \bar{x}_{\Gamma_u}}{h}
\right)$,
where $x_u$ is the attribute value (e.g., opinion) of user $u \in \mathcal{U}$, $\Gamma_u$ is the set of neighbors of $u$, and $\bar{x}_{\Gamma_u}$ is the average attribute of those neighbors.
Here, $K(\cdot,\cdot)$ is a bivariate kernel function and $h$ is the bandwidth parameter.
& echo chamber 
& \cite{valensise2023drivers} \\
\midrule
\textbf{Favorability Score}: the normalized difference in group exposure for each item.
It captures the ideological composition of the audience consuming an item: a value of $+1$ indicates that the audience is exclusively from group $A$, $-1$ indicates an audience exclusively from group $B$, both cases implying zero diversity.
& $ FAV_i {=} (A_i - B_i)/(A_i + B_i)$ , where $A_i$ and $B_i$ denote the number of exposures of groups $A$ and $B$ to item $i \in \mathcal{I}$. 
& item diversity & \cite{gonzalez2023asymmetric} \\
\midrule

\textbf{Click-Through Rate}: the proportion of exposures to an item that result in an interaction. It captures how effectively an item converts exposure into engagement. Higher values indicate higher engagement volume. 
& $ CTR_i = |\mathcal{T}_i|/|\hat{\mathcal{T}}_i|$, where $i \in \mathcal{I}$. 
& item volume, individual volume, ecosystem volume & \cite{zhou2010impact, long2022choice, donnelly2021longtail} \\
\midrule

\textbf{Minority Exposure}: the share of recommendations a minority group receives relative to the total number of recommendations provided to users. 
A lower-than-expected exposure indicates inequality in visibility. 
& 
$\varepsilon_m {=} \frac{|\hat{\mathcal{T}}_m|}{|\hat{\mathcal{T}}|}$, where 
$|\hat{\mathcal{T}}| = k \, |\mathcal{U}^{\alpha}|$, and $\hat{\mathcal{T}}_m$ is the set of recommendations delivered to a minority group $m$, 
$k$ is the number of link recommendations each user received, which is constant and equal for all users, and $\mathcal{U}^{\alpha}$ is the set of active users receiving recommendations.
&  Discrimination & \cite{fabbri2022exposure} \\
\midrule
\textbf{Daily New Edges}: the overall rate of new connections formed in a social network per day. It captures ecosystem-level growth in network connections. Higher values indicate a greater volume of tie formation.  
& $\{ E_t \}_{t \in T}$,

$E_t {=} |\mathcal{E}_t|, \;
\mathcal{E}_t {=} \{(u,v)|u,v {\in} \mathcal{U}, \text{ edge } u {\to} v \text{ created on day } t \}$, where
$T$ is the set of days. 
& ecosystem volume & \cite{su2016effect} \\
\midrule
\textbf{User Distances}: these measures quantify how dissimilar users are in terms of what they see or interact with. Different formulations use overlap measures, rank correlations, semantic similarity, entropy-based indices, or average pairwise distances. 
& Examples include: Jaccard $J(A,B)=\tfrac{|A \cap B|}{|A \cup B|}$; Edit distance $\min\{\text{ins.}+\text{del.}+\text{subst.}\}$; Cosine $\tfrac{\vec{A}\cdot\vec{B}}{\|\vec{A}\|\|\vec{B}\|}$; Kendall $\tau=1-\tfrac{2\,\#\text{discordant}}{n(n-1)/2}$; Rao’s entropy $H=\sum_i\sum_j p_ip_j d_{ij}$; Opinion clustering $C=\tfrac{(\sum_i c_i)^2}{\sum_i c_i^2}$; Avg. pairwise distance $\tfrac{1}{N(N-1)}\sum_i\sum_{j\neq i}|x_i-x_j|$; Kullback–Leibler divergence $D_{\mathrm{KL}}(G_1 \,\|\, G_2)
= \sum_{x} G_1(x) \, \log \frac{G_1(x)}{G_2(x)}$, where $G_1$ and $G_2$ are the distribution of items' ratings relative two two different groups o users. 
& homogenisation, polarisation & \cite{le2019measuring, le2023modeling, yang2023bubbles, pansanella2022mean, pansanella2022modeling, sirbu2019algorithmic, barlacchi2025simulation, chaney2018algorithmic, mansouryAbdollahpouri2020} \\
\midrule

\textbf{Delay}: the time lost per unit distance, defined as the travel time required to cover one kilometer, expressed as the reverse of speed. 
This measure is typically applied to items such as road segments or entire routes.
& $Delay = \tfrac{T_{actual}}{d}$, where $T_{actual}$ is the observed travel time (minutes) and $d$ is distance (km). 

& item volume & \cite{hanna2017citywide} \\

\midrule
\textbf{Volume over Capacity ratio}: it measures traffic concentration (i.e., congestion) by comparing observed traffic volume on a road to the road's maximum capacity. High values indicate severe congestion, with values above 1 implying that demand exceeds capacity. 
& $VOC = V/C$, where $V$ is the observed traffic volume on a road (vehicles/hour), and $C$ is the road capacity (vehicles/hour) & concentration & \cite{colak2016understanding} \\

\midrule
\textbf{Demand-to-Supply Ratio}: the ratio between the total distance generated by all user-item interactions (vehicles travelling on road segments) and the maximum distance that the system (the road network) can support per hour. Higher values indicate that demand is approaching or exceeding the system’s capacity.
& $DSR = \frac{\sum_{e \in E} l_e x_e}{\sum_{e \in E, x_e > 0} l_e C_e}$, where $l_e$ is the length of road segment $e$, $x_e$ is the vehicle flow on $e$, and $C_e$ is the capacity of $e$.
& ecosystem volume& \cite{colak2016understanding} \\

\midrule
\textbf{Perplexity}: it measures how well a model predicts a sample of data, such as a sequence of words. A lower perplexity indicates the model is less "surprised" by the data and is therefore better at predicting the next item (e.g., word); a higher perplexity means the model is less accurate. 
& $PPL=exp\left(-\frac{1}{|\mathcal{I}|}\sum_{i\in \mathcal{I}}\log(P(i))\right)$, where $i$ is an item (e.g., a product or a token in a text document). 
& content degradation & \cite{shumailov2023curse}, \cite{zhu2024synthesize} \\

\midrule

\textbf{n-gram diversity/TTR}: it quantifies how many unique $n$-grams a document contains relative to the total number of $n$-grams. It captures how varied the vocabulary or phrasing is. 
For $n=1$, it reduces to the classic type–token ratio (i.e., the number of unique words divided by the number of words in a text document).
& $D_n(d) = \frac{|\mathcal{V}_n(d)|}{|\mathcal{G}_n(d)|}$, where $\mathcal{V}_n(d)$ is the set of unique $n$-grams in document $d$ and $\mathcal{G}_n(d)$ is the multiset of all $n$-grams in $d$.
& content degradation & \cite{guo2024curious}, \cite{zhu2024synthesize}\\

\midrule
\textbf{Normalized Linguistic Entropy}: it measures how evenly tokens are distributed in the document. 
Higher values indicate a broader and more varied vocabulary, while lower values signal repetitive or limited language use.  & $H(d) = -\frac{\sum_{w \in W(d)} q_w \log(q_w)}{\log |W(d)|}
$, where $q_w$ is the empirical probability of token $w$ in $d$, computed as the frequency of $w$ divided by the total number of terms in $d$ & content degradation & \cite{gambetta2025} \\
\bottomrule
\end{longtable}}

\begin{figure}[htb!]
\centering
\subfigure[]{\includegraphics[width=0.4155\linewidth]{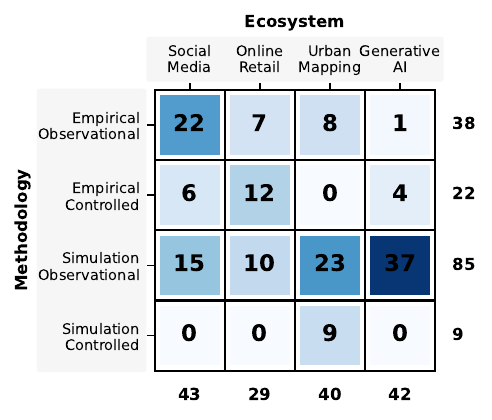}}
\hfill
\subfigure[]{
\includegraphics[width=0.56\linewidth]{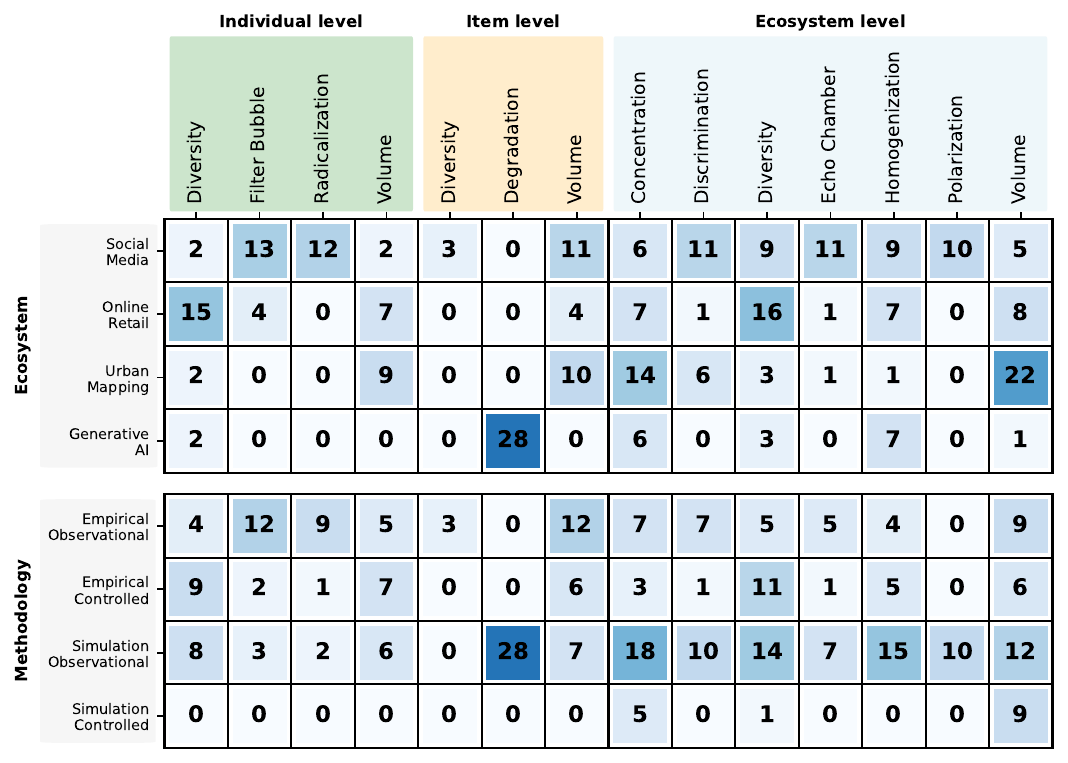}}
\caption{(a) Distribution of methodologies employed across ecosystems. Each cell reports the number of studies employing a
methodology within an ecosystem. The bottom row reports total counts per methodology and the rightmost column reports total counts per ecosystem (170 studies in total).
(b) Distribution of individual, item and ecosystem outcomes across ecosystems and methodologies. Each cell reports the number of outcomes found per ecosystem/methodology.}
\label{fig:heatmaps}
\end{figure}

\begin{table}[h]
\caption{Social Media Ecosystem. Classification of selected papers based on methodology, outcomes and level of analysis.} 
\vspace{-0.4cm}
\centering
{
\renewcommand{\arraystretch}{1.2}
\scriptsize
\begin{tabular}{@{}
l
l
P{2.425cm}
P{2.425cm}
P{2.425cm}
P{2.425cm}
@{}}

\toprule

\multicolumn{2}{c}{\multirow{2}{*}{\centering {\LARGE \textbf{Social Media}}}} &
\multicolumn{2}{c}{\large \textbf{Empirical}} & 
\multicolumn{2}{c}{\large \textbf{Simulation}} \\

\cmidrule(lr){3-6}

\multicolumn{2}{c}{} & 
\multicolumn{1}{c}{\textbf{Observational}} & \multicolumn{1}{c}{\textbf{Controlled}} &
\multicolumn{1}{c}{\textbf{Observational}} & \multicolumn{1}{c}{\textbf{Controlled}} \\

\midrule

\multirow{2.5}{*}{\footnotesize \textbf{Individual}}

 & \textbf{Filter Bubble}  &\cite{le2023modeling,bouchaud2023algorithmic,haroon2023auditing,hosseinmardi2024causally,bakshy2015exposure,ibrahim2023youtube,ledwich2019algorithmic,chen2021neutral,ye2025auditing,le2019measuring,gonzalez2023asymmetric}       
&\cite{nyhan2023like} 
&\cite{gausen2022using}    
&\\
 & \textbf{Radicalisation} &\cite{haroon2023auditing,hosseinmardi2021examining,ledwich2019algorithmic,hosseinmardi2024causally,ibrahim2023youtube,chen2021neutral,le2023modeling,ribeiro2020auditing,whittaker2021recommender}
&\cite{markmann2021youtube}                    
&\cite{rossi2021closed, ribeiro2023amplification}
&\\

\midrule
\midrule

{\footnotesize \textbf{Item}}
 & \textbf{Content Degradation} & & & & \\

\midrule
\midrule

\multirow{6}{*}{\footnotesize \textbf{Ecosystem}}
 & \textbf{Concentration}  &\cite{ledwich2019algorithmic,su2016effect,santini2023recommending,cakmak2024unveiling} 
&
&\cite{fabbri2022exposure, ferrara2022link}
&\\
 & \textbf{Discrimination} &\cite{santini2023recommending,haroon2023auditing,ledwich2019algorithmic,ye2025auditing,bouchaud2023algorithmic}
&\cite{huszar2022algorithmic}                
&\cite{fabbri2022exposure, ferrara2022link,perra2019modelling, peralta2021effect, peralta2021opinion}
&\\
 & \textbf{Echo Chamber}  &\cite{gonzalez2023asymmetric,chen2021neutral,bakshy2015exposure}
& \cite{nyhan2023like}
&\cite{chitra2020analyzing, perra2019modelling, cinus2022effect, pansanella2022modeling,  valensise2023drivers, peralta2021effect, gausen2022using}    
&\\
 & \textbf{Homogenisation}  & \cite{le2023modeling,yang2023bubbles}
& \cite{markmann2021youtube}     
&  \cite{sirbu2019algorithmic, pansanella2022mean, pansanella2022modeling,perra2019modelling,peralta2021effect,peralta2021opinion}     
&\\
 & \textbf{Polarisation} &
&
&\cite{chitra2020analyzing, cinus2022effect, pansanella2022mean, pansanella2022modeling, peralta2021effect, peralta2021opinion, ramaciotti2021auditing,sirbu2019algorithmic, valensise2023drivers,perra2019modelling}
&\\

\midrule
\midrule

\multirow{4}{*}{\begin{tabular}[c]{@{}l@{}}{\footnotesize \textbf{Individual}}\\ {\footnotesize \textbf{Item}}\\ {\footnotesize \textbf{Ecosystem}}\end{tabular}}

& \textbf{Diversity} & individual: \cite{haroon2023auditing},

item: \cite{gonzalez2023asymmetric,bouchaud2023algorithmic,boeker2022empirical}, 

ecosystem: \cite{santini2023recommending,zhou2010impact,kirdemir2021assessing,bartley2024impacts}          & 
ecosystem: \cite{bartley2021auditing,nyhan2023like}
&
individual: \cite{gausen2022using},

ecosystem: \cite{fabbri2022exposure, ferrara2022link, peralta2021opinion}
&\\
& \textbf{Volume} & item: \cite{bouchaud2023algorithmic,santini2023recommending,zhou2010impact,su2016effect,boeker2022empirical,ibrahim2023youtube,ye2025auditing,whittaker2021recommender},

ecosystem: \cite{su2016effect,haroon2023auditing,ledwich2019algorithmic}
&
individual: \cite{guess2023social,guess2023reshares},

item: \cite{huszar2022algorithmic,guess2023social,guess2023reshares},

ecosystem: \cite{nyhan2023like} 
&  
ecosystem: \cite{gausen2022using}
&
\\
  
\bottomrule
\end{tabular}}
\label{tab:social}
\end{table}

\section{Social Media Ecosystem}
\label{sec:social_media}
The social media ecosystem comprises online platforms designed for connecting users and for content creation, sharing, and interaction. 
It includes social networking services such as Facebook, Instagram, TikTok, and X; community-driven platforms like Reddit, Gab, and YouTube that organize individuals around shared interests or content consumption; and content aggregators such as Google News and Apple News that curate and distribute information. 
Figure~\ref{fig:heatmaps}a summarises the distribution of methodologies employed across studies in this ecosystem, while Table \ref{tab:social} presents an overview of the outcomes examined. 
Appendix \ref{app:social_media} provides details about each study.

The analysis of this ecosystem includes 43 papers, comprising 28 empirical studies \cite{bakshy2015exposure,bartley2021auditing,bartley2024impacts,boeker2022empirical,bouchaud2023algorithmic,cakmak2024unveiling,chen2021neutral,gonzalez2023asymmetric,guess2023reshares,guess2023social,haroon2023auditing,hosseinmardi2021examining,hosseinmardi2024causally,huszar2022algorithmic,ibrahim2023youtube,kirdemir2021assessing,le2019measuring,le2023modeling,ledwich2019algorithmic,markmann2021youtube,nyhan2023like,ribeiro2020auditing,santini2023recommending,su2016effect,whittaker2021recommender,yang2023bubbles,ye2025auditing,zhou2010impact} and 15 simulation studies \cite{chitra2020analyzing,cinus2022effect,fabbri2022exposure,ferrara2022link,gausen2022using,pansanella2022mean,pansanella2022modeling,peralta2021effect,peralta2021opinion,perra2019modelling,ramaciotti2021auditing,ribeiro2023amplification,rossi2021closed,sirbu2019algorithmic,valensise2023drivers}.
We observe a prevalence of observational studies, comprising as many as 37 papers \cite{bakshy2015exposure,bartley2024impacts,boeker2022empirical,bouchaud2023algorithmic,cakmak2024unveiling,chen2021neutral,gonzalez2023asymmetric,haroon2023auditing,hosseinmardi2021examining,hosseinmardi2024causally,ibrahim2023youtube,kirdemir2021assessing,le2019measuring,le2023modeling,ledwich2019algorithmic,ribeiro2020auditing,santini2023recommending,su2016effect,whittaker2021recommender,yang2023bubbles,ye2025auditing,zhou2010impact,chitra2020analyzing,cinus2022effect,fabbri2022exposure,ferrara2022link,gausen2022using,pansanella2022mean,pansanella2022modeling,peralta2021effect,peralta2021opinion,perra2019modelling,ramaciotti2021auditing,ribeiro2023amplification,rossi2021closed,sirbu2019algorithmic,valensise2023drivers}.
Observational empirical studies often rely on artificial bots or sock-puppets\footnote{Bots are automated accounts programmed to perform predefined actions on an online platform (e.g., viewing, liking, or following content) to systematically collect data. 
Sock-puppets are human-controlled fake accounts used to interact with an online platform in a realistic manner while keeping the operator’s identity undisclosed.} to mimic user behaviour and collect data from online platforms. Whereas observational simulation studies typically employ agent-based models to analyse how recommendations and interactions evolve under different recommendation strategies.
Only six empirical studies are controlled \cite{bartley2021auditing,guess2023reshares,guess2023social,huszar2022algorithmic,markmann2021youtube,nyhan2023like}, reflecting the practical constraints of developing controlled experiments in this ecosystem. 
These experiments require privileged access to platform data, the ability to adjust recommender parameters, and the capacity to assign users to treatment and control groups -- all conditions rarely available to researchers external to online platforms.
To the best of our knowledge, there are no controlled simulation studies published in the social media ecosystem. 

Several outcomes are controversial and highly debated in the literature, leading to diverging interpretations of the same phenomena -- e.g., individual outcomes like filter bubble and ecosystem level outcomes such as radicalisation and echo chamber. 
These controversies stem from differences in data access and methodological choices (such as the selection of measures, baselines, and user samples).

Research on YouTube's recommender -- the most extensively studied platform with a total of 13 studies -- offers a clear illustration of these controversies.
Except for a handful of observational studies \cite{haroon2023auditing, ribeiro2020auditing, whittaker2021recommender}, there is no evidence that YouTube recommender creates or exacerbates radicalisation \cite{hosseinmardi2021examining, hosseinmardi2024causally, ibrahim2023youtube, le2023modeling, ledwich2019algorithmic, markmann2021youtube}. 
Notably, a study shows that radicalisation arises only when recommendations are followed indiscriminately, and disappears when users select content according to their own preferences \cite{ribeiro2023amplification}.
A comparable pattern is observed for filter bubbles: while some studies find that YouTube's recommender reinforces them \cite{ibrahim2023youtube,le2023modeling,haroon2023auditing}, another study shows that user preferences rather than algorithmic personalisation are the primary driver of filter bubble effects \cite{hosseinmardi2024causally}.
This aligns with observational studies on other social media platforms.  
For example, individual choices have a more significant impact than Facebook's News Feed in limiting exposure to diverse political news \cite{bakshy2015exposure}.
Similarly, there is little evidence of Twitter's recommender generating partisan bias, with users primarily receiving content aligned with their friendship networks \cite{chen2021neutral}.  

All these studies are based on an observational approach and cannot establish causal relationships on the impact of recommenders. 
Therefore, controlled experiments are essential for quantifying this impact at different scales.
\citet{huszar2022algorithmic}, using proprietary Twitter data in a controlled study involving nearly two million users, led the way in this direction. 
Users were split into a control group receiving reverse-chronological feeds and a treatment group visualising algorithmically ranked tweets. 
Comparing behaviours across groups allows the authors to isolate the recommender’s causal effect on discrimination, showing that mainstream right-leaning content is amplified more than left-leaning content.
This finding is consistent across empirical studies: some research provides evidence of content amplification driven by recommendations \cite{boeker2022empirical,bouchaud2023algorithmic,guess2023reshares,guess2023social,huszar2022algorithmic,ibrahim2023youtube,santini2023recommending,su2016effect,whittaker2021recommender,ye2025auditing,zhou2010impact}, while other work shows that this amplification is selective, disproportionately boosting certain content while suppressing others \cite{bouchaud2023algorithmic,huszar2022algorithmic,santini2023recommending,ye2025auditing}. 

Simulation studies yield more convergent findings than observational empirical analyses: for example, polarisation reliably emerges across diverse modelling frameworks \cite{sirbu2019algorithmic,valensise2023drivers,peralta2021effect,pansanella2022modeling,cinus2022effect,perra2019modelling,peralta2021opinion,chitra2020analyzing,ramaciotti2021auditing}.
However, a few models identify conditions under which recommenders may reduce fragmentation or promote homogenisation \cite{ramaciotti2021auditing,peralta2021opinion}.
To the best of our knowledge, no empirical study has yet examined polarisation.

Studies on echo chambers also offer controversial findings, both when employing empirical and simulation approaches. 
Some research shows that biased "who-to-follow" recommendations\footnote{“Who-to-follow” recommendations are algorithms that suggest accounts a user may want to follow on a social media platform, typically based on their past interactions, network connections, or the similarity of interests inferred from their activity.} can foster echo-chamber formation \cite{chitra2020analyzing,peralta2021opinion,perra2019modelling, cinus2022effect}, while other work suggests that these recommendations may actually slow down or weaken this process \cite{pansanella2022modeling,valensise2023drivers}. 
Some simulation studies examine how social networks evolve under these recommenders, finding that visibility concentrates around popular accounts and disadvantages minority groups \cite{ferrara2022link,fabbri2022exposure,peralta2021effect,peralta2021opinion,perra2019modelling}. 
Combined with users' preferences in tie formation, who-to-follow recommenders can also accelerate the emergence of echo chambers and polarisation \cite{cinus2022effect,ramaciotti2021auditing}.

\begin{table}[H]
\caption{Online Retail Ecosystem. Classification of selected papers based on methodology, outcomes and level of analysis.}
\vspace{-0.4cm}
\centering
{
\renewcommand{\arraystretch}{1.2}
\scriptsize
\begin{tabular}{@{}
l
l
P{2.425cm}
P{2.425cm}
P{2.425cm}
P{2.425cm}
@{}}

\toprule

\multicolumn{2}{c}{\multirow{2}{*}{\centering {\LARGE \textbf{Online Retail}}}} &
\multicolumn{2}{c}{\large \textbf{Empirical}} & 
\multicolumn{2}{c}{\large \textbf{Simulation}} \\

\cmidrule(lr){3-6}

\multicolumn{2}{c}{} & 
\multicolumn{1}{c}{\textbf{Observational}} & \multicolumn{1}{c}{\textbf{Controlled}} &
\multicolumn{1}{c}{\textbf{Observational}} & \multicolumn{1}{c}{\textbf{Controlled}} \\

\midrule

\multirow{2.5}{*}{\footnotesize \textbf{Individual}}

 & \textbf{Filter Bubble}  &  \cite{nguyen2014exploring}                 &         \cite{chen2022more}     &
{\cite{noordeh2020echo,aridor2020deconstructing}}
&\\
 & \textbf{Radicalisation} & & & & \\

\midrule
\midrule

{\footnotesize \textbf{Item}}
 & \textbf{Content Degradation} & & & & \\

\midrule
\midrule

\multirow{6}{*}{\footnotesize \textbf{Ecosystem}}
 & \textbf{Concentration}  & 
\cite{pathak2010empirical}
&     
\cite{yi2022recommendation,matt2013,lee2019How}
&
\cite{mansouryAbdollahpouri2020, hazrati2022recommender, barlacchi2025simulation}
&                     
\\
 & \textbf{Discrimination} &                         
&                     
& \cite{mansouryAbdollahpouri2020}        
&                     
\\
 & \textbf{Echo Chamber} &              
\cite{ge2020understanding}          
&                     
&      
&                     
\\
 & \textbf{Homogenisation}  &  
\cite{hosanagar2014will} 
&  \cite{holtz2020engagement}                   
&\cite{mansouryAbdollahpouri2020, barlacchi2025simulation, aridor2020deconstructing, chaney2018algorithmic, de2023recommender} 
&\\
 & \textbf{Polarisation} & & & & \\

\midrule
\midrule

\multirow{5}{*}{\begin{tabular}[c]{@{}l@{}}{\footnotesize \textbf{Individual}}\\ {\footnotesize \textbf{Item}}\\ {\footnotesize \textbf{Ecosystem}}\end{tabular}}

& \textbf{Diversity}  &   individual: \cite{nguyen2014exploring, ge2020understanding, anderson2020algorithmic},

ecosystem: \cite{ pathak2010empirical}
 &    individual: \cite{yi2022recommendation, matt2019factual, liang2022exploring, holtz2020engagement, li2022recommender, lee2014impact, lee2019How},
 
ecosystem:~\cite{yi2022recommendation, matt2013, matt2019factual, donnelly2021longtail, lee2014impact, lee2019How, lee2020, chen2022more}      

&  individual: \cite{barlacchi2025simulation, noordeh2020echo, aridor2020deconstructing, de2023recommender, flederHosanagar2009blockbuster},

ecosystem: \cite{mansouryAbdollahpouri2020, hazrati2022recommender, barlacchi2025simulation, flederHosanagar2009blockbuster, chaney2018algorithmic, fleder2007recommender, wu2011},
  &                \\
& \textbf{Volume}    &    individual: \cite{FENG2024103905, hosanagar2014will},

item: \cite{chen2004impact},

ecosystem: \cite{anderson2020algorithmic, pathak2010empirical, chen2004impact, hosanagar2014will}              &  individual: \cite{holtz2020engagement, li2022recommender, long2022choice, donnelly2021longtail, lee2014impact},

item: \cite{lee2014impact, lee2019How, lee2020},

ecosystem:~\cite{long2022choice, lee2014impact, lee2019How, lee2020}        &  
 &                    \\
  
\bottomrule
\end{tabular}}
\label{tab:market}
\end{table}

\section{Online Retail Ecosystem}
\label{sec:markets}

The online retail ecosystem comprises online platforms that allow users to purchase products or services. It includes e-commerce marketplaces such as Amazon, eBay, and Alibaba; and digital media streaming services such as Netflix and Spotify for movies, series, and music.  
Figure \ref{fig:heatmaps}a
summarises the distribution of methodologies employed across studies in this ecosystem, while Table \ref{tab:market} presents
an overview of the outcomes examined. Appendix \ref{app:online_retail} provides details about each study.

The analysis of this ecosystem comprises 29 studies, with a prevalence of empirical (19 out of 29 \cite{nguyen2014exploring, ge2020understanding, anderson2020algorithmic, pathak2010empirical, chen2004impact, FENG2024103905, hosanagar2014will, yi2022recommendation, matt2013, matt2019factual, liang2022exploring, holtz2020engagement, li2022recommender, long2022choice, donnelly2021longtail, lee2014impact, lee2019How, lee2020, chen2022more}) over simulation studies (which are 10 in total \cite{mansouryAbdollahpouri2020, hazrati2022recommender, chaney2018algorithmic, fleder2007recommender, wu2011, barlacchi2025simulation, noordeh2020echo, flederHosanagar2009blockbuster, aridor2020deconstructing, de2023recommender}).
The predominance of empirical studies reflects the economic motivation of e-commerce platforms in understanding how recommenders influence engagement, sales, and revenues. 
Empirical studies tend to analyse user activity directly on commercial online platforms or design laboratory experiments, while simulation studies are based on models of user preferences that rely on ad-hoc assumptions or platform-derived data.
Among empirical contributions, controlled studies (12 out of 29 \cite{yi2022recommendation, matt2013, matt2019factual, liang2022exploring, holtz2020engagement, li2022recommender, long2022choice, donnelly2021longtail, lee2014impact, lee2019How, lee2020, chen2022more}) prevail over observational ones (seven in total \cite{nguyen2014exploring, ge2020understanding,
anderson2020algorithmic, pathak2010empirical, chen2004impact, FENG2024103905, hosanagar2014will}). 
This pattern contrasts with other ecosystems and relates to the relative ease of running controlled experiments on online retail platforms. 
Indeed, users can be randomly assigned to treatment and control groups with minimal risk of violating the SUTVA, as interactions across groups are limited. 
To the best of our knowledge, no studies employ a controlled simulation methodology.

There is a broad agreement across the reviewed studies that the introduction of recommenders increases overall platform activity, leading to higher volumes of interactions, sales, and revenues \cite{feng2024beyond,hosanagar2014will,lee2020,chen2004impact,lee2019How,donnelly2021longtail,long2022choice,li2022recommender,holtz2020engagement,pathak2010empirical,anderson2020algorithmic}.
Findings on other outcomes are more heterogeneous and context-dependent, with controversies arising even across studies run on the same online platform. 
For example, some studies on Spotify report that recommendations increase individual diversity in listening \cite{liang2022exploring}, whereas others find that they reduce the exploration of novel music content \cite{anderson2020algorithmic,holtz2020engagement}. 
These conclusions reflect differences in recommenders' technology, the heterogeneity of the metrics used to quantify diversity and the variability of the benchmarks against which these metrics are evaluated.

A recurring pattern across both empirical and simulation studies is that recommendations may induce at the same time gains in individual diversity and losses in ecosystem diversity \cite{yi2022recommendation,barlacchi2025simulation,matt2019factual,lee2014impact,lee2019How}.
This is because, although users progressively tend to reduce their consumption over time, recommenders can broaden their exploration \cite{nguyen2014exploring,aridor2020deconstructing}.
At the same time, recommenders may increase the homogenisation of users’ interactions or purchasing patterns \cite{aridor2020deconstructing} -- an effect frequently accompanied by declines in ecosystem diversity \cite{barlacchi2025simulation,chaney2018algorithmic,mansouryAbdollahpouri2020}.

The study of \citet{lee2014impact, lee2019How} is a key contribution that causally examines the impact of recommenders on diversity. 
Users of a major Canadian retailer were split into a control group with no recommendations, two treatment groups exposed to collaborative filtering algorithms, and a treatment group exposed to a recommender displaying items the user had recently browsed. 
The collaborative filtering algorithms increase the individual diversity of sales but reduce ecosystem diversity and concentrate purchases on a few popular item categories. 
Instead, the “recently viewed’’ recommender lowers ecosystem diversity while leaving individual diversity unaffected.

Other studies employ different methodologies to investigate how observed outcomes vary according to the conditions under which experiments are conducted \cite{hazrati2022recommender, holtz2020engagement}.
A simulation study uses log data from three Amazon datasets (Kindle, Games and Apps) to train recommenders based on collaborative filtering, item popularity, and item average rating \cite{hazrati2022recommender}. 
They find that the impact on item and ecosystem diversity depends both on the specific recommendation technology and the dataset considered. 
A controlled study on Spotify finds that when a popularity-based recommender -- which tends to homogenise user behaviour -- serves as the benchmark, the personalised recommender induces less homogenisation and a smaller decline in individual diversity \cite{holtz2020engagement}. 

\begin{table}[H]
\caption{Urban Mapping Ecosystem. Classification of selected papers based on methodology, outcomes and level of analysis.}
\vspace{-0.4cm}
\centering
{
\renewcommand{\arraystretch}{1.2}
\scriptsize
\begin{tabular}{@{}
l
l
P{2.425cm}
P{2.425cm}
P{2.425cm}
P{2.425cm}
@{}}

\toprule

\multicolumn{2}{c}{\multirow{2}{*}{\centering {\LARGE \textbf{Urban Mapping}}}} &
\multicolumn{2}{c}{\large \textbf{Empirical}} & 
\multicolumn{2}{c}{\large \textbf{Simulation}} \\

\cmidrule(lr){3-6}

\multicolumn{2}{c}{} & 
\multicolumn{1}{c}{\textbf{Observational}} & \multicolumn{1}{c}{\textbf{Controlled}} &
\multicolumn{1}{c}{\textbf{Observational}} & \multicolumn{1}{c}{\textbf{Controlled}} \\

\midrule

\multirow{2.5}{*}{\footnotesize \textbf{Individual}}

 & \textbf{Filter Bubble}  & & & & \\
 & \textbf{Radicalisation} & & & & \\

\midrule
\midrule

{\footnotesize \textbf{Item}}
 & \textbf{Content Degradation} & & & & \\

\midrule
\midrule

\multirow{5}{*}{\footnotesize \textbf{Ecosystem}}
 & \textbf{Concentration}  &  \cite{hanna2017citywide}                      &                   &      \cite{johnson2017beautiful, sanchez2023bias, mauro2025urban, sanchez2021effects, colak2016understanding, erhardt2019transportation, martinez2017assessing, garcia2020short}                  &   \cite{perezprada2017managing, cornacchia2022routing, cornacchia2024navigation, thai2016negative, sabet2025exploring}\\
 & \textbf{Discrimination} &   \cite{edelman2014digital, zhang2021frontiers}                     &                 &    \cite{afeche2023ridehailing, bokanyi2020understanding, sanchez2021effects, johnson2017beautiful} & \\
 & \textbf{Echo Chamber}  &  \cite{koh2019offline}  & & & \\
 & \textbf{Homogenisation}  & & &    \cite{mauro2025urban} & \\
 & \textbf{Polarisation} & & & & \\

\midrule
\midrule

\multirow{6}{*}{\begin{tabular}[c]{@{}l@{}}{\footnotesize \textbf{Individual}}\\ {\footnotesize \textbf{Item}}\\ {\footnotesize \textbf{Ecosystem}}\end{tabular}}
& \textbf{Diversity} &                        &                     &     individual: \cite{sanchez2023bias, mauro2025urban},

ecosystem: \cite{mauro2025urban, cornacchia2023oneshot}

&    ecosystem: \cite{cornacchia2024navigation}
\\
& \textbf{Volume}    &   individual:~\cite{schwieterman2019uber, santi2014quantifying, zhang2021frontiers},

item: \cite{santi2014quantifying, hanna2017citywide, zhang2021frontiers},

ecosystem: \cite{jalali2017emission, jiang2025carbon}             &       & individual: \cite{fagnant2018dynamic, colak2016understanding, maciejewski2016assignment, maciejewski2016largescale, Maciejewski2013SimulationAD, mori2022developing},

item: \cite{maciejewski2016assignment, maciejewski2016largescale, zhu2017reducing, alonsomora2017ondemand, martinez2017assessing, mori2022developing, garcia2020short},

ecosystem:  \cite{mehrvarz2020optimal, barth2007environmental, colak2016understanding, cornacchia2023oneshot, erhardt2019transportation, zhu2017reducing, fagnant2014travel, agarwal2022impact, martinez2017assessing, alonsomora2017ondemand, storch2021incentive}                  & ecosystem: \cite{arora2021quantifying, valdes2016eco, perezprada2017managing, cornacchia2022routing, cornacchia2024navigation, cornacchia2023effects, thai2016negative, ahn2013eco, sabet2025exploring}                  \\
  
\bottomrule
\end{tabular}}
\label{tab:urban}
\end{table}

\section{Urban Mapping Ecosystem}
The urban mapping ecosystem comprises a broad range of recommenders designed to support and guide city dwellers. It includes location-based platforms such as TripAdvisor, Yelp, and Google Places; accommodation services including Airbnb, Booking.com, and Zillow; navigation services such as Google Maps, Waze, and Baidu Maps; and ride-hailing and micromobility platforms like Uber, Lyft, Bolt, Free Now, Bird, and Lime. 
Figure \ref{fig:heatmaps}a summarises the distribution of methodologies
employed across studies in this ecosystem, while Table \ref{tab:urban} presents an overview of the outcomes examined.
Appendix \ref{app:urban_mapping} provides details about each study.

The analysis of this ecosystem comprises 40 studies, with a prevalence of simulation studies (32), which are widespread for at least two reasons. First, since urban mapping platforms rarely disclose data or algorithms, it is hard to conduct and reproduce empirical assessments.
Second, running controlled empirical experiments in real urban environments is challenging. 
For example, since drivers share the road network, any intervention applied to one group (e.g., those following routes suggested by Google Maps) affects others, therefore it violates SUTVA~\cite{cox1958planning}. 
The variability of weather, strikes, and accidents introduces further confounding factors. 
For these reasons, we encounter only eight empirical studies \cite{schwieterman2019uber,santi2014quantifying,jalali2017emission,jiang2025carbon,hanna2017citywide,koh2019offline,edelman2014digital,zhang2021frontiers}, none of which are controlled. 

Among simulation studies, the majority of them (23 out of 32) adopts an observational approach \cite{johnson2017beautiful,mehrvarz2020optimal,barth2007environmental,colak2016understanding,cornacchia2023oneshot,maciejewski2016assignment,maciejewski2016largescale,Maciejewski2013SimulationAD,erhardt2019transportation,martinez2017assessing,zhu2017reducing,fagnant2014travel,fagnant2018dynamic,afeche2023ridehailing,agarwal2022impact,bokanyi2020understanding,alonsomora2017ondemand,mori2022developing,storch2021incentive,garcia2020short,sanchez2023bias,sanchez2021effects,mauro2025urban}, evaluating the city-wide effects of location, accommodation, routing, and ride suggestions.
Nine simulation studies employ controlled designs to examine how routing algorithms influence congestion, emissions, and other traffic-related outcomes \cite{arora2021quantifying,valdes2016eco,perezprada2017managing,cornacchia2022routing,cornacchia2023effects,thai2016negative,ahn2013eco,sabet2025exploring,cornacchia2024navigation}. 
While most simulations rely on standard routing algorithms such as Dijkstra or A*, a growing subset of studies integrates real navigation APIs into their experimental setups \cite{arora2021quantifying,cornacchia2022routing,cornacchia2024navigation}.

We group the main findings according to the four services that characterise this ecosystem -- location-based, accommodation, navigation and ride-hailing platforms.
Location-based platforms recommend points of interest in response to users’ queries \cite{sanchez2022point,pappalardo2023future}.
The surveyed studies reveal a key trade-off: while recommenders can increase individual diversity by spreading users' visits more evenly across locations, they also reinforce concentration, as these visits tend to converge on a small set of already popular venues \cite{sanchez2023bias, sanchez2021effects, mauro2025urban}.

The expansion of Airbnb and other accommodation platforms has been linked to rising rents and house prices~\citep{garcia2020short}, which we characterise as an increased item volume.  
Some studies document racial disparities in earnings, indicating that recommendations can disadvantage black hosts~\citep{edelman2014digital}; others find that recommendations can partially mitigate such discrimination~\citep{zhang2021frontiers} or foster the formation of echo chambers~\citep{koh2019offline}. 
The mechanisms through which discrimination may arise remain poorly understood, owing to the limited transparency of recommenders and the lack of controlled studies.

The impact of routing strategies \cite{johnson2017beautiful, mehrvarz2020optimal, barth2007environmental, colak2016understanding, cornacchia2023oneshot, valdes2016eco, perezprada2017managing, cornacchia2023effects, thai2016negative, ahn2013eco, sabet2025exploring} and commercial navigation services \cite{johnson2017beautiful, arora2021quantifying, cornacchia2022routing, cornacchia2024navigation} has received substantial attention in the literature. 
Research focuses predominantly on volume-related outcomes at the ecosystem level, such as reductions in emissions and travel time. 
A study by Google Maps -- the only conducted internally by an online platform -- employs a controlled simulation and finds that its routing suggestions reduce travel time and CO2 emissions in Salt Lake City \cite{arora2021quantifying}. 
Some works confirm beneficial effects of algorithmic routing on travel time, CO2 emissions, or fuel consumption \citep{valdes2016eco, ahn2013eco, barth2007environmental, mehrvarz2020optimal, colak2016understanding, sabet2025exploring}.
However, some other simulation studies show that, under certain circumstances (e.g., high service adoption), routing recommendations increase CO2 emission \citep{valdes2016eco, perezprada2017managing, cornacchia2022routing, cornacchia2024navigation}, reduce route diversity  \citep{cornacchia2024navigation}, and generate traffic concentration in some areas \citep{johnson2017beautiful,thai2016negative,cornacchia2022routing, cornacchia2024navigation}.
Overall, the impact of navigation services appears sensitive to algorithmic design and service adoption dynamics.

The urban mapping ecosystem also includes recommenders that match users with drivers within e-mobility, car-pooling and taxi platforms \citep{schwieterman2019uber, santi2014quantifying, jalali2017emission,jiang2025carbon, maciejewski2016assignment, maciejewski2016largescale, Maciejewski2013SimulationAD, erhardt2019transportation, martinez2017assessing, zhu2017reducing, fagnant2014travel, fagnant2018dynamic, afeche2023ridehailing, agarwal2022impact,bokanyi2020understanding, alonsomora2017ondemand, mori2022developing, storch2021incentive}. 
Research focuses primarily on volume-related outcomes at the item, individual, and ecosystem levels, considering elements like fares, travel time, idle time, vehicle utilisation, and CO2 emissions. 
Most of these studies show that recommenders reduce travel time, distance, fleet size, and emissions, and they make vehicle usage more efficient \citep{santi2014quantifying,jalali2017emission,jiang2025carbon,maciejewski2016assignment,maciejewski2016largescale,Maciejewski2013SimulationAD,martinez2017assessing,zhu2017reducing,fagnant2014travel,fagnant2018dynamic,schwieterman2019uber,agarwal2022impact,alonsomora2017ondemand,mori2022developing}. 
For example, ride-sharing can reduce travel time, cumulative trip length, and city-wide emissions \cite{santi2014quantifying, jalali2017emission}.
However, other studies document unintended outcomes, such as higher user costs \citep{schwieterman2019uber}, discrimination arising from matching and repositioning strategies \citep{afeche2023ridehailing,bokanyi2020understanding}, and increased traffic congestion \citep{erhardt2019transportation}. 
For instance, the widespread use of ride-hailing in San Francisco increased congestion, as a substantial share of vehicle miles were travelled without passengers \cite{erhardt2019transportation}.

\begin{table}[H]
\centering
\caption{Generative AI Ecosystem. Classification of selected papers based on methodology, outcomes and level of analysis.}
\vspace{-0.4cm}
{
\renewcommand{\arraystretch}{1.2}
\scriptsize
\begin{tabular}{@{}
l
l
P{2.425cm}
P{2.425cm}
P{2.425cm}
P{2.425cm}
@{}}

\toprule

\multicolumn{2}{c}{\multirow{2}{*}{\centering {\LARGE \textbf{Generative AI}}}} &
\multicolumn{2}{c}{\large \textbf{Empirical}} & 
\multicolumn{2}{c}{\large \textbf{Simulation}} \\

\cmidrule(lr){3-6}

\multicolumn{2}{c}{} & 
\multicolumn{1}{c}{\textbf{Observational}} & \multicolumn{1}{c}{\textbf{Controlled}} &
\multicolumn{1}{c}{\textbf{Observational}} & \multicolumn{1}{c}{\textbf{Controlled}} \\

\midrule

\multirow{2.5}{*}{\footnotesize \textbf{Individual}}

 & \textbf{Filter Bubble}  & & & & \\
 & \textbf{Radicalisation} & & & & \\

\midrule
\midrule

{\footnotesize \textbf{Item}}
 & \textbf{Content Degradation} & & & \cite{shumailov2023curse,shumailov2024ai,alemohammad2023self,guo2024curious,martinez2023combining,martinez2023towards,briesch2023large,dohmatob2024tale,dohmatob2024model,dohmatob2024strong,bohacek2023nepotistically, hataya2023will,seddik2024bad,herel2024collapse,wang2025llm,suresh2025rate,fu2025theoretical,feng2024beyond,drayson2025machine,gerstgrasser2024model,kazdan2024accumulating,ferbach2024self,gillman2024self,zhu2024synthesize,gambetta2025,bertrand2024stability,zhu2025matters,schaeffer2025position}        & \\

\midrule
\midrule

\multirow{6}{*}{\footnotesize \textbf{Ecosystem}}
 & \textbf{Concentration}  &  \cite{kobak2025delving}                    &  &        \cite{hou2024large, di2025addressing, spurlock2024chatgpt, zhang2021language, lichtenberg2024large}             &                \\
 & \textbf{Discrimination} & & & & \\
 & \textbf{Echo Chamber}  & & & & \\
 & \textbf{Homogenisation}  &       \cite{kobak2025delving}               &       \cite{agarwal2025ai, doshi2024generative, desdevises2025paradox}          &          \cite{tao2024cultural, bulte2025llms, rettberg2025ai}              &                 \\
 & \textbf{Polarisation} & & & & \\

\midrule
\midrule

\multirow{2}{*}{\begin{tabular}[c]{@{}l@{}}{\footnotesize \textbf{Individual}}\\ {\footnotesize \textbf{Item}}\\ {\footnotesize \textbf{Ecosystem}}\end{tabular}} 
& \textbf{Diversity} &                  &       individual: \cite{doshi2024generative, wan2025using},

ecosystem: \cite{wan2025using}

&     ecosystem: \cite{di2025addressing, deldjoo2024understanding}                    &              \\
& \textbf{Volume}     &             &  ecosystem: \cite{desdevises2025paradox}    &                 &                   \\
  
\bottomrule
\end{tabular}}
\label{tab:genAI}
\end{table}

\section{Generative AI ecosystem}
\label{sec:GenAI}
The generative AI ecosystem comprises online platforms that assist users in producing texts, images, codes, and other synthetic contents. 
It includes assistants based on Large Language Models (LLMs) such as ChatGPT, Claude, Gemini, DeepSeek, and Llama-powered services; image and video generation platforms such as Midjourney, DALL·E, Stable Diffusion, and Runway; and multimodal productivity suites integrated into applications like Microsoft Copilot and Google Workspace AI.
Figure \ref{fig:heatmaps}a summarises the distribution of methodologies
employed across studies in this ecosystem, while Table \ref{tab:genAI} presents an overview of the outcomes examined.
Appendix \ref{app:generative_AI} provides details about each study.

The analysis of the generative AI ecosystem includes 42 articles with a prevalence of simulation (37) over empirical studies \cite{agarwal2025ai, doshi2024generative, desdevises2025paradox, wan2025using, kobak2025delving}. 
This reflects the difficulty of conducting empirical experiments with generative AI platforms, which are costly, operationally demanding, and often limited by ethical and privacy constraints. 
In contrast, simulation studies offer greater experimental control, higher reproducibility, longitudinal perspective, and lower resource requirements. 
Within simulation work, all studies adopt observational designs~\cite{zhang2021language, tao2024cultural, bulte2025llms, rettberg2025ai, shumailov2023curse,shumailov2024ai,alemohammad2023self,guo2024curious,martinez2023combining,martinez2023towards,briesch2023large,dohmatob2024tale,dohmatob2024model,dohmatob2024strong,bohacek2023nepotistically, hataya2023will,seddik2024bad,herel2024collapse,wang2025llm,suresh2025rate,fu2025theoretical,feng2024beyond,drayson2025machine,gerstgrasser2024model,kazdan2024accumulating,ferbach2024self,gillman2024self,zhu2024synthesize,gambetta2025,bertrand2024stability,zhu2025matters,schaeffer2025position, hou2024large, di2025addressing, spurlock2024chatgpt, zhang2021language, lichtenberg2024large}. 

There are three main strands of literature: 
(1) the examination of how generative AI influences users’ content-creation capacity; (2) the assessment of the outcomes that arise when generative AI is used as item recommenders; and (3) the investigation of the impact of the self-consuming loop.

The first line of research employs controlled ~\cite{agarwal2025ai, doshi2024generative, desdevises2025paradox, wan2025using} and observational ~\cite{kobak2025delving, tao2024cultural, bulte2025llms, rettberg2025ai} studies to reflect on the impact of recommenders on the homogenisation of content creation capacity.
ChatGPT produces more ideas than humans, but most of these ideas are non-original, reinforcing homogenisation \cite{desdevises2025paradox}.
Western-centric AI models conform writing to Western norms \cite{agarwal2025ai}, and in general, the thinking and communication patterns of users adopting generative AI converge toward the dominant tendencies encoded in the models \cite{bulte2025llms, tao2024cultural}.
The use of generative AI for writing short stories boost individual creativity, but homogenises writing styles \cite{doshi2024generative}.
This homogenisation can be mitigated by introducing a set of generative-AI personas with distinct cultural backgrounds, cognitive styles, and interests \cite{wan2025using}. 
An increased uniformity is also observed in biomedical research outputs after the advent of LLMs \cite{kobak2025delving}.

The second line of research, exclusively based on observational simulation approaches~\cite{zhang2021language, lichtenberg2024large, deldjoo2024understanding, hou2024large, di2025addressing, spurlock2024chatgpt}, examines the use of generative AI for recommendations about movies~\cite{zhang2021language, deldjoo2024understanding, hou2024large, spurlock2024chatgpt, lichtenberg2024large}, songs~\cite{deldjoo2024understanding}, and software libraries~\cite{di2025addressing}.
With the exception of two studies \cite{deldjoo2024understanding, lichtenberg2024large}, this research consistently shows that recommendations favour popular items, increasing concentration \citep{hou2024large, zhang2021language, di2025addressing, spurlock2024chatgpt}. 
Some studies propose strategies (e.g., prompt engineering, fine-tuning, and post-processing) to mitigate this concentration, but reveal a trade-off between bias mitigation and recommendation accuracy~\cite{di2025addressing, spurlock2024chatgpt}. 

The third line of research investigates content degradation (a.k.a. model collapse \cite{schaeffer2025position}) that arises when generative AI models are repeatedly fine-tuned on their own outputs -- a process known as autophagy or self-consuming loop \cite{shumailov2024ai}.
This stream of research relies exclusively on observational simulation designs because it aims at investigating long-term impacts of generative AI.
Building on the seminal work of \citet{shumailov2024ai}, the literature consistently shows that autophagy leads to progressive content degradation across modalities, simulation setups, and content types \cite{alemohammad2023self, seddik2024bad, briesch2023large, martinez2023combining, martinez2023towards, gambetta2025}. 
As models are fine-tuned on their own outputs, frequent patterns (e.g., common words or visual features) become amplified and rare ones gradually suppressed, reducing content diversity and quality.
The literature also investigates how content degradation varies according to model characteristics, the proportion of synthetic data, and dataset construction strategies \citep{guo2024curious, herel2024collapse, suresh2025rate}. Other research focuses on mitigating content degradation, emphasising the need to maintain sufficient real data and preserve proximity to the true data distribution \citep{gerstgrasser2024model, kazdan2024accumulating, bertrand2024stability, dohmatob2024model, dohmatob2024tale, dohmatob2024strong, fu2025theoretical, ferbach2024self, gillman2024self, zhu2024synthesize, wang2025llm, zhu2025matters, bohacek2023nepotistically, feng2024beyond, hataya2023will, drayson2025machine}.
In this respect, a study shows that any text (human- or AI-generated) overly probable under the model’s own distribution may lead to content degradation \cite{gambetta2025}. 
Therefore, mitigating this outcome requires ensuring that training data remain sufficiently surprising relative to the generative AI model.

\section{Commonalities and differences across ecosystems}
\label{sec:commonalities_differences}
This section highlights commonalities and differences across the four human-AI ecosystems.
These concern data availability, the methodologies employed, and the outcomes observed.

\subsection{Data availability} An important difference among ecosystems lies in the availability of empirical data.
In the social media and online retail ecosystems, several empirical studies have been developed thanks to publicly accessible APIs, and collaborations between academy and industry (see \citet{huszar2022algorithmic} on Twitter and \citet{lee2014impact,lee2019How} on an online retail platform).
In contrast, empirical studies are scarce in the urban mapping and generative AI ecosystems because privacy constraints severely limit what platform APIs can reveal about user requests, choices, and the corresponding recommendations.

Across all ecosystems, the recommenders used by online platforms are typically inaccessible and often undocumented. 
As a result, researchers approximate them in simulations using prototypical or simplified models, which can only capture platform behaviour in stylised form. 

\subsection{Methodologies}
Observational studies dominate across all ecosystems, except for online retail where roughly 40\% of the studies are controlled (see Figure \ref{fig:heatmaps}).
This particularity is due to economic and technical reasons. 
At the economic level, online retail platforms aim to causally measure recommenders' impact on product sales to maximise their revenues. 
At a technical level, these platforms can create treatment and control groups with minimal cross-interaction (e.g., Amazon users do not directly interact with each other), approximating the SUTVA (discussed in Section \ref{sec:taxonomy}).
A canonical example of a controlled study is provided by \citet{lee2014impact,lee2019How}, who asked users to interact independently with an online retail platform with and without the aid of a recommender.

Weak cross-interactions also characterise controlled experiments in the generative AI ecosystem, where user studies can be designed to split participants into non-interacting groups. 
An example is \citet{doshi2024generative}'s study, in which hundreds of participants were asked to write short stories with or without assistance from a generative AI system. 
However, such field studies are resource-intensive and difficult to scale up.

In the social media and urban mapping ecosystems, interactions among individuals are intrinsic to the platform or physical environment. 
On social media, users influence one another through views, likes, reposts, and direct communication. 
In urban environments, users following navigation suggestions interact at road intersections, and exogenous events (storms, strikes, or accidents) affect the behaviour of drivers while travelling. 
All these factors can lead to the underestimation of the true effect of the recommender at any given moment, thereby violating the SUTVA.

We note that some online platforms conduct controlled studies internally, but do not disclose their results. 
The Facebook Files, a set of leaked internal reports by Meta, illustrates this point (https://facebookpapers.com/). 
The Wall Street Journal published these reports, revealing that Meta kept secret the results of their internally controlled studies, which showed negative impacts of Instagram on teenage users and of Facebook on fostering violence in developing countries.
Overall, controlled studies carried out independently by academics are scarce, largely because of limited accessibility to platform data and algorithms.
This not only limits knowledge accumulation and the replicability of studies, but also constrains our understanding the impact of recommenders.

\subsection{Outcomes}
While some outcomes are distinctive of specific ecosystems, others are
extensively investigated across multiple ones -- such as concentration, homogenisation, diversity and volume (see Figure \ref{fig:heatmaps}).
These outcomes all pertain to the ecosystem level of analysis, testifying a growing interest of the scientific community in the collective impact of recommenders. 
Yet, the absence of a common notation and research fragmentation across disciplines hinders the development of a comprehensive understanding of such systemic impact.

Filter bubbles, echo chambers and radicalisation are especially prominent in studies of the social media ecosystem. Outcomes regarding the volume of items, discrimination and polarisation have also attracted substantial attention in this literature.
In contrast, research in the online retail ecosystem centres on volume and diversity at both the individual and ecosystem levels, alongside concentration and homogenisation. 
A similar emphasis on individual and ecosystem volume is found in the urban mapping ecosystem. 
In this context, another key observed outcome is traffic congestion, defined here as concentration. 
Research in the generative AI ecosystem focuses only on the study of three outcomes (homogenisation, concentration, and content degradation), perhaps a reflection of the content or the youth of the research field.

Across all ecosystems, the outcomes reported in the literature exhibit substantial variability, even when studies address similar questions. 
This heterogeneity arises from differences in methodological approaches, disparities in data access, and the absence of standardised benchmarks or measurement practices. 
As a result, outcomes are highly sensitive to platform characteristics, recommender design, baseline choices, and the specific metrics employed. 
The difficulty of reproducing and replicating studies amplifies this variability: empirical and simulation studies, as well as controlled and observational designs, may yield different results under similar conditions. 
These issues are particularly pronounced in the urban mapping and generative AI ecosystems, where assumptions, data limitations, and modelling setups vary widely across studies.

\section{Open Challenges, Recommendations and Future Research Avenues}
\label{sec:open_challenges}
This section concludes the article, opening to present and future challenges related to user-recommender interactions.
It examines the challenge of limited data access and the ethical implications of these interactions. 
It then offers guidance for developers and policymakers on how to govern these impacts and design next-generation recommenders that better account for the observed outcomes. 
Finally, it identifies key gaps in the current literature and outlines promising directions for future research.

\subsection{Data Access and Reproducibility}
Two major obstacles to studying the impact of recommenders are the limited access to data by researchers external to platforms, and the lack of transparency about how recommenders are designed, deployed and updated. 
For example, the continual update of recommenders on online platforms may expose users to different algorithmic versions.
As a result, the findings of empirical studies become outdated quickly, limiting the reproduction of experiments, the comparability of results over time, and hindering large-scale empirical analyses. 

Article 40 of EU's DSA seeks to mitigate this issue, as it requires  VLOPs/VLOSEs to provide data access for research on the systemic risks they might cause. 
Since October 2025, vetted researchers can request such data through the EU Data Access Portal.\footnote{https://data-access.dsa.ec.europa.eu/home} However, uncertainties remain regarding the type, granularity, and completeness of the shared data, especially whether or not these data include the recommendations users receive.

Comprehensively studying the impact of recommenders at different scales requires the development of standardised datasets that capture the key elements of the human-AI feedback loop \cite{pedreschi2023human}: the recommendations users receive, their reactions and choices, and technical details of the recommender and its training. 
Such datasets would greatly enhance comparability across studies, improve the reproducibility of empirical findings, and provide a shared foundation for calibrating simulations \cite{knott2022transparency}.
Unlike fields such as computer vision and natural language processing, which benefit from large and standardised datasets, recommender systems research still relies on a few public datasets (e.g., MovieLens, Last.fm, Yelp) that capture only user choices and omit critical information covering the entire feedback loop. 
Constructing such public datasets requires collaboration between academy and online platforms. 

Another priority is the development of standardised digital twins and simulation frameworks. 
Digital twins are virtual replicas of online platforms that update continuously with real data. 
They allow researchers to test how algorithmic changes affect outcomes without placing users at risk. 
Although some major online platforms already employ digital twins internally, these tools remain inaccessible to academic researchers. Coordinated, cross-national legislative efforts (e.g., in the EU, US, and China) should therefore support the creation of open, research-oriented digital twins.
In parallel, simulation frameworks (which do not depend on live platform data) enable \emph{in silico} experiments that mimic user–recommender interactions. 
Several of such frameworks have been proposed across social media, online retail, urban mapping, and generative AI \cite{rossetti2024social,cornacchia2022routing,mauro2025urban,barlacchi2025simulation}, but the field still lacks a standardised framework that would allow cumulative and comparable scientific progress.

\subsection{Ethical concerns}
The impact of recommenders on users, items, and ecosystems pose ethical concerns, which stem from opacity, biases, and the misalignment between commercial objectives and societal well-being.
A key challenge concerns outcomes at the ecosystem level such as concentration, diversity, and volume. 
For example, concentration emerges in all ecosystems, illustrating how recommenders can amplify existing inequalities in society \cite{ferragina2025accumulation}.
In what follows, we examine this and other ethical concerns across the different human–AI ecosystems.

In social media, recommenders are optimised for user engagement, often privileging content that triggers strong emotional responses -- e.g., sensational, provocative, or divisive material. 
This can fuel political polarisation \cite{chitra2020analyzing,cinus2022effect,pansanella2022mean,pansanella2022modeling,peralta2021effect,peralta2021opinion,perra2019modelling,ramaciotti2021auditing,sirbu2019algorithmic,valensise2023drivers} and trap users into echo chambers \cite{bakshy2015exposure,gonzalez2023asymmetric,nyhan2023like,chitra2020analyzing,cinus2022effect,gausen2022using,peralta2021effect,perra2019modelling}, contributing to the erosion of the conditions for a healthy democratic discourse. 
Within this context, there are also ethical concerns related to autonomy and manipulation: users are subtly steered toward content that reinforces their biases without being aware of it. 
The lack of transparency regarding how content is prioritised and filtered can further erode trust and accountability, leaving users with little knowledge about how their informational environment is shaped.
Another concern is that recommenders concentrate attention on a small set of already prominent accounts \cite{ferragina2025accumulation,ferrara2022link,fabbri2022exposure, su2016effect, santini2023recommending, cakmak2024unveiling}, granting them disproportionate visibility and public influence.

In online retail, recommenders boost sales of online platforms by predicting what users are likely to purchase \cite{feng2024beyond,hosanagar2014will,lee2020,chen2004impact,lee2019How,donnelly2021longtail,long2022choice,li2022recommender,holtz2020engagement,pathak2010empirical,anderson2020algorithmic}. 
These recommenders have distributional consequences because they concentrate visibility around a small set of popular items or sellers \cite{yi2022recommendation,barlacchi2025simulation,matt2019factual,lee2014impact,lee2019How}, making it harder for niche producers and small businesses to gain exposure.
This amplifies economic inequality within digital marketplaces, a problem that received little attention in the recommender systems literature.

A similar inegalitarian dynamic arises in the urban mapping ecosystem: recommenders favour already popular points of interest \cite{sanchez2021effects, mauro2025urban}, further widening the visibility gap between venues.
Moreover, navigation services such as Google Maps or Waze optimise routes for individual travel time, which can create system-wide inefficiencies: when thousands of drivers are directed along the same fast path, congestion may grow, emissions can rise, and traffic may be funnelled through residential neighbourhoods \cite{cornacchia2022routing, cornacchia2024navigation, valdes2016eco, perezprada2017managing, johnson2017beautiful,thai2016negative}.
These side effects rise distributive justice questions: certain communities bear disproportionate burdens of noise and pollution, despite having no agency in the algorithms' decision-making processes. 
A related ethical concern is that recommenders behind accommodation platforms can contribute to discrimination -- e.g., widening the revenue gap between black and white hosts \cite{edelman2014digital}. 
There is limited ability to design effective interventions in this context because the mechanisms driving these outcomes are poorly understood.

In generative AI ecosystems, recommenders reduce the diversity of generated content \cite{shumailov2024ai, alemohammad2023self, seddik2024bad, briesch2023large, martinez2023combining, martinez2023towards, gambetta2025}, reinforce prevailing opinions and the popularity of items \cite{hou2024large, zhang2021language, di2025addressing, deldjoo2024understanding, spurlock2024chatgpt}, and increase homogenisation of users' styles \cite{agarwal2025ai, doshi2024generative, desdevises2025paradox}. 
Such dynamics hinder epistemic and cultural diversity, as mainstream styles and narratives are fostered and minority perspectives are sidelined. 
Moreover, increased homogenisation can be problematic for other reasons as well.
Consider CVs that converge toward the same structure and wording: while this may simplify evaluation on one level, it makes it harder for HR professionals to discern the distinctiveness of individual candidates. 

\subsection{Practical suggestions for developers and policy makers}

\subsubsection*{For developers}
Technical interventions to manage the impact of recommenders can target different elements: input data, recommender learning, and exposure to recommendations \cite{pedreschi2023human}.

The data used to train recommenders contain various forms of bias \cite{chen2023bias}. 
A way to manage them is to emphasise data that deviate from models' probability distributions \cite{gambetta2025}.
By exposing recommenders to underrepresented patterns, they can be partially nudged away from self-reinforcing loops --
a strategy that seems effective in mitigating content degradation in generative AI models \cite{gambetta2025}.

Most recommenders optimise short-term, individual-level objectives such as engagement or user satisfaction. 
Multi-objective optimisation can extend these goals through the incorporation of ecosystem-level considerations -- such as maximising ecosystem diversity or limiting concentration. 
The literature already proposes some interesting solutions: for example, relevance and diversity have been balanced through Bayesian scoring and submodular diversification \cite{Teo2016AdaptiveDiversity}; continual learning and customised losses have been used to correct accumulated model errors \cite{Cai2022ReLoop}; and unbiased pairwise learning helped mitigate recommenders' biases \cite{Ren2023UnbiasedPairwise}.  

The impact of recommenders can be also managed through post-hoc and online re-ranking.
Accuracy and intra-list diversity  can be improved by splitting user profiles into sub-profiles and re-ranking items accordingly \cite{Kaya2019Subprofiles}. 
Miscalibration -- when recommendation lists fail to mirror a user’s true taste proportions -- can be tackled by re-ranking items to better match the user’s preference distribution \cite{Steck2018Calibrated}. 
Personalised diversification that balances accuracy with novelty can help mitigate popularity bias \cite{Abdollahpouri2019PopularityBias}.

Overall, we suggest that user–recommender interaction should be conceived as a dynamic process unfolding within the human–AI feedback loop \cite{pedreschi2023human}.
Recommenders can be endowed with meta-adaptive capabilities to detect signs of degeneration (e.g., declining ecosystem diversity or excessive homogenisation) and trigger compensatory adjustments in data selection, model optimisation, or recommendation strategies. 
Together, these technical interventions outline a research agenda for managing the impact of recommenders. 

Managing this impact requires more than technical fixes. 
It calls for a rethinking of ethical governance in algorithmic ecosystems, one that integrates transparency, fairness, and accountability into the design of recommenders. 
An ethical future for recommenders depends on recognising that algorithmic optimisation is not neutral: it encodes values, distributes power, and shapes the fabric of society.

\paragraph{For policy makers.}
The regulation of risks generated by the impact of recommenders is a technical, social, economic, and political issue that extends beyond national jurisdictions.  
We identify three levels at which this regulation can be approached by policy makers \cite{ferragina2025accumulation}.

The first concerns the level of transparency at which recommenders operate. 
Enhancing transparency could enable users to understand the potential unintended consequences of their interactions with recommenders and make more informed decisions -- much like reading a product label before purchasing an item. 
Online platforms should be encouraged (or compelled) to make their algorithms auditable and transparent,
while policy makers should establish frameworks that treat algorithmic externalities with the same seriousness as
environmental or public health risks.
In this regard, the aforementioned DSA represents an important step as it requires VLOPs/VLOSEs to identify, analyse, and assess systemic risks stemming from the design and operation of algorithmic systems (including recommenders).\footnote{\url{https://www.eu-digital-services-act.com/Digital_Services_Act_Article_34.html}} 
However, while transparency alone empowers users against specific harms -- such as exposure to illegal content or violations of fundamental rights -- it does not necessarily address ecosystem outcomes such as concentration, discrimination, or reduction in ecosystem diversity.\footnote{For a critical reading of the DSA and its potential limitations, see \citet{terzis2024law}.}

A second level regards how governments could establish independent regulatory authorities to monitor ecosystem outcomes. 
For example, assess the extent to which recommenders exacerbate concentration by favouring already popular products in online retail and high-profile accounts in social media. 
The legitimacy of such authorities would rest on the recognition that VLOPs/VLOSEs provide services of public utility, and thus their operation should be subject to democratic scrutiny.
On the basis of this legitimacy, these authorities would be able to review platforms' recommenders and ask for revisions. 
Recent work has articulated this principle as a “right to constructive optimisation” \cite{naudts2025right}, i.e., the right of citizens to ensure that algorithmic systems are optimized toward public rather than purely commercial interests.

A third level would involve the development of public-interest alternatives to privately owned VLOPs/VLOSEs. 
Consider a scenario in which a major private platform refuses to comply with national transparency or oversight requirements. 
For regulatory enforcement to be credible, public institutions must be able to offer viable alternatives to citizens, echoing earlier debates on the governance of public goods \cite{mattei2012beni}. 
Historical precedents -- such as the post–World War II nationalization of electricity, railways, and telecommunications -- illustrate how states can reclaim strategic capacities in critical sectors \cite{ferragina2025accumulation}.
In today's context, measures of this kind could help counterbalance the market dominance of VLOPs/VLOSEs. 
Notably, national investment in AI and digital infrastructure, such as China's state-led initiatives \cite{roberts2021chinese}, reflects recognition of this strategic imperative.

\subsection{Avenues for future research}

\subsubsection*{Methodology} 
Advancing our knowledge about the impact of recommenders requires a greater investment in controlled experiments, as they are considered the gold standard for assessing causal relationships \cite{pedreschi2023human}. 
However, designing such studies for online platforms is far from straightforward, because it is hard to prevent interactions between users in the control and treatment groups -- a requirement of the SUTVA for causal inference. 
Minimising these cross-group interactions remains an open methodological challenge which requires the development of new strategies. 
For instance, one could envisage in the social media ecosystem to ensure that users in the control and treatment groups belong to distinct social communities, as inter-community interactions in social networks are fewer than intra-community ones \cite{rossetti2018community}.

A few works -- mainly simulation studies in the urban mapping ecosystem \cite{fagnant2018dynamic, storch2021incentive, mauro2025urban, arora2021quantifying, valdes2016eco, perezprada2017managing, cornacchia2022routing, cornacchia2023effects, thai2016negative, ahn2013eco, sabet2025exploring, cornacchia2024navigation, Maciejewski2013SimulationAD} -- examine the role of adoption rates, i.e., the fraction of the population who actually use the platform or activate the recommendations functionality.
This aspect is essential, as it can reveal threshold effects indicating whether certain outcomes emerge only once a critical mass of users relies on the recommender. 
For instance, navigation systems may trigger traffic congestion only when adoption rates exceed a given proportion of the population \cite{cornacchia2022routing,cornacchia2024navigation}.
Extending similar analyses to other ecosystems could offer valuable insights into whether and when threshold effects shape the outcomes of user–recommender interactions.

Another avenue for future research concerns understanding the impact of specific technological components used to implement recommenders.
We observe two contrasting situations in the literature. On the one hand, some studies analyse the impact of recommenders embedded within online platforms, yet the underlying algorithms are not disclosed.
This does not allow one to assess which components of the recommender contribute to its impact.
On the other hand, when scholars build recommenders for simulations or laboratory experiments, they typically employ simple and well-known techniques -- such as collaborative filtering or content-based filtering.
However, online platforms use far more sophisticated technological components that blend multiple paradigms and incorporate recent deep-learning techniques (e.g., transformers).
While the use of simpler models allows researchers to isolate and interpret specific mechanisms, it is essential to also assess how the different components of real recommenders (such as collaborative filtering, content-based filtering, and deep-learning architectures) shape the observed outcomes.

\subsubsection*{Outcomes} 
A comprehensive understanding of the impact of recommenders on users, items and ecosystems requires research on under-investigated outcomes.
For example, content degradation is only studied within the generative AI ecosystem, but could occur elsewhere.
Is recommenders' continual re-training causing a systematic avoidance of certain items or users?
At what rate do recommenders reduce the diversity of recommended content? 

In social media, empirical studies on polarisation are needed to validate simulation findings, and the question of whether user–recommender interactions foster radicalisation remains contested and warrants further investigation.
Are polarisation and other ecosystem outcomes mainly caused by algorithmic recommendations or users' behaviour? 
Under which conditions do recommenders contribute to radicalise users' opinions?  

More research on concentration and ecosystem diversity is also needed in the online retail ecosystem.
Many studies showed a decrease in ecosystem diversity, but this has not been linked to the inequality of visibility of brands, products, and product categories.
To what extent do recommenders promote popular brands and products while making others even less popular? How does the specific type of recommender (e.g., collaborative filtering vs personalised) influence this phenomenon? 
Do recommenders push people to buy specific categories of products (e.g., junk vs. healthy food)?
Research in this area is driven by the goal of increasing revenues for companies, while the broader impact of recommenders on the distribution of these revenues is not fully understood.

Studies within the urban mapping ecosystem often overlook the influence of recommenders on complex phenomena such as socio-economic segregation  \cite{gambetta2023mobility} and gentrification \cite{mauro2025dynamic}. 
To what extent do recommenders in this ecosystem steer users from different social classes to segregated urban areas? And do they impact on housing prices and availability in historically low-income areas? 
Do ride-hailing recommenders systematically avoid poorer neighbourhoods with less demand? 
Do route recommenders increase traffic in specific areas, leading to discomfort for residents and increased emissions? 

Research in the generative AI ecosystem focus on a limited set of outcomes -- concentration, content degradation, and homogenisation. 
However, it is unclear how these outcomes depend on the type of content generated (e.g., texts vs images) and how to effectively address them.
Is programming code generation more prone to content degradation than textual writing? 
How to counteract the homogenisation of users' style created by the increasingly pervasive use of generative AI?

\subsection*{Conclusion}
This survey offers the first taxonomy of outcomes and methodological categories aimed at addressing the fragmentation and lack of systematicity in the literature.
Our main objective is to facilitate a process of knowledge accumulation about the impact of recommenders at different scales. 
We provide a foundation for analysing outcomes at multiple levels, and our effort could be expanded by incorporating additional studies, outcomes, and human–AI ecosystems.

From this review, a key challenge emerges. 
Since recommenders are based on AI, and machine learning in particular, their interactions with users always give rise to the human-AI feedback loop \cite{pedreschi2023human}.
The literature does not investigate the feedback loop in its complexity: even rigorous controlled studies do not clarify which step of the loop is primarily responsible for an observed outcome. 
Outcomes in fact may depend on different phases of the loop -- from the data used for training to the behavioural patterns shaping user–item interactions, from the specific techniques underlying the recommendation algorithm to the periodicity in recommender retraining.
Understanding the human-AI feedback loop requires a holistic approach, one that accounts for the reciprocal impact of humans and recommenders.
This and other challenges will continue to shape the literature on human-AI interactions in the years to come.

\subsection*{Authors contributions}
LP and EF coordinated the taxonomy development, the categorisation of studies, and the overall writing process, and they prepared the final manuscript as well as Tables \ref{tab:outcomes} and \ref{tab:measures}.
VP, SC, VM, and GR collected and analysed the social media studies and contributed to the corresponding section and table.
ML and GB did the same for the online retail ecosystem; GC and GM for urban mapping; and DG and GG for generative AI.
VP created Figure \ref{fig:flowchart}, and GC produced Figure \ref{fig:heatmaps}.
All authors reviewed and approved the final manuscript.

\subsection*{Acknowledgements}
Luca Pappalardo has been supported by PNRR (Piano Nazionale di Ripresa e Resilienza) in the context of the research program 20224CZ5X4\_PE6\_PRIN 2022 “URBAI – Urban Artificial Intelligence” (CUP B53D23012770006), funded by European Union – Next Generation EU.
Emanuele Ferragina has been supported by a grant overseen by the French National Research Agency (ANR) as part of the ‘investissement d’avenir’ programme LIEPP (ANR-11-LBX-0091, ANR-11-IDEX-0005-02) and the université de Paris IdEX (ANR-18-IDEX-0001).
Dino Pedreschi has been supported by the Partnership Extended PE00000013 - “FAIR - Future Artificial Intelligence Research” - Spoke 1 “Human-centered AI”.
This work was partially conducted during a visiting stay supported by government funding managed by the National Research Agency under France 2030, reference ANR-22-EXES-0014.

We thank Daniele Fadda for the support on Figure \ref{fig:flowchart}. 
We also thank The Beatles for inspiring the hard task of writing. 
This paper would not have been possible without the medley of Abbey Road. 
When confronted with our abundant flaws (read animal spirits), there is consolation only in absolute beauty.

\bibliographystyle{ACM-Reference-Format}
\bibliography{bibliography}

\newpage
\appendix

\section{Social Media Ecosystem}
\label{app:social_media}

\subsection{Empirical studies}

\subsubsection*{\bf Observational studies}
\citet{ribeiro2020auditing} audit radicalisation pathways on YouTube's video and channel recommendations. By analysing users' migration patterns across 330k videos from 349 politically related channels, the study finds a recommendation flow from milder to more extreme (alt-right) content. \colorbox{black}{\color{white}\scriptsize Radicalisation}

\citet{santini2023recommending} examine YouTube's promotion of hyperpartisan content during Brazilian elections. 
They analyse the news sources recommended on the platform by simulating the browsing behaviour of new users. 
The findings highlight an increase in inequality with preferential treatment for right-wing media outlets over similar content from left-wing ones. \colorbox{black}{\color{white}\scriptsize Discrimination} \colorbox{black}{\color{white}\scriptsize Concentration}
\colorbox{black}{\color{white}\scriptsize Item Volume} \colorbox{black}{\color{white}\scriptsize Ecosystem Diversity}

\citet{haroon2023auditing} investigate YouTube's recommender tendency to generate filter bubbles, radicalisation pathways, and extremist content recommendations. 
They rely on a sock-puppet audit using 100k  accounts designed to represent various political leanings.
The study discovers a filter bubble effect, particularly pronounced for right-leaning users. 
It also finds an increase in recommendations from channels linked to extremist or conspiratorial content, particularly for users characterised by views of extreme right-wing content.
\colorbox{black}{\color{white}\scriptsize Filter Bubble}
\colorbox{black}{\color{white}\scriptsize Individual Diversity}
\colorbox{black}{\color{white}\scriptsize Ecosystem Volume}
\colorbox{black}{\color{white}\scriptsize Radicalisation}
\colorbox{black}{\color{white}\scriptsize Discrimination}

\citet{hosseinmardi2021examining} investigate the role of users' preferences on received recommendations. 
They retrieve browsing histories of 310k users and profile them based on viewing habits. 
The researchers then analyse on- vs. off-platform consumption habits of users, pathways to radical political content, and the effect of session length on content type exposure. 
In contrast with previous research, this study finds little evidence of the amplification of political content and radicalisation pathways.
\colorbox{black}{\color{white}\scriptsize Radicalisation}

\citet{ledwich2019algorithmic} examine the role that YouTube's recommender plays in encouraging online radicalisation.
By examining the recommendation patterns among 800 political channels, the research finds that the recommender amplifies views for mainstream media and politically neutral content rather than promoting radical or extremist content.
\colorbox{black}{\color{white}\scriptsize Radicalisation}
\colorbox{black}{\color{white}\scriptsize Filter Bubble}
\colorbox{black}{\color{white}\scriptsize Discrimination}
\colorbox{black}{\color{white}\scriptsize Concentration}
\colorbox{black}{\color{white}\scriptsize Ecosystem Volume}

\citet{ibrahim2023youtube} focus on YouTube's recommender propensity to create political filter bubbles.
The study collects video recommendations via a sock-puppet audit with 360 bots that represent six personas across the US political spectrum.
The findings show that the recommender steers users away from political extremes toward moderate content. 
This effect is more pronounced for far-right than for far-left content. 
\colorbox{black}{\color{white}\scriptsize Radicalisation}
\colorbox{black}{\color{white}\scriptsize Filter Bubble}
\colorbox{black}{\color{white}\scriptsize Item Volume}

\citet{hosseinmardi2024causally} estimate the causal impact of YouTube recommendations on the consumption of highly partisan and radical content. 
The study compares the behaviours of bots designed to mimic real users' viewing patterns with those of bots following predefined rule-based trajectories.
The findings show that the recommender does not steer users towards radical content. 
On the contrary, when users with strong political views start watching moderate content, the recommender shifts their recommendations after approximately 30 videos, breaking down users' filter bubbles.
\colorbox{black}{\color{white}\scriptsize Radicalisation}
\colorbox{black}{\color{white}\scriptsize Filter Bubble}

\citet{le2023modeling} examine the impact of YouTube's personalised recommendations on generating filter bubbles and rabbit holes.
The researchers conduct a sock-puppet audit to gather video recommendations and propose a straightforward theoretical model explaining why and how rabbit holes form on YouTube.
The results indicate that user interactions could influence recommendations, but users are not consistently led further into specialised content. 
In fact, after a certain number of interactions, YouTube's recommender may forget previous user preferences, breaking down users' filter bubbles.
\colorbox{black}{\color{white}\scriptsize Radicalisation}
 \colorbox{black}{\color{white}\scriptsize Homogenisation}
\colorbox{black}{\color{white}\scriptsize Filter Bubble}

\citet{zhou2010impact} investigate the impact of various YouTube features on video views, with a focus on the recommender's effectiveness in driving video popularity. 
By analysing metadata, related video lists, and view statistics for hundreds of thousands of videos, the study finds that recommendations increase video views and promote a wider variety of videos rather than just promoting the most popular ones.
\colorbox{black}{\color{white}\scriptsize Item Volume}
\colorbox{black}{\color{white}\scriptsize Ecosystem Diversity}

\citet{kirdemir2021assessing} inspects YouTube's recommendation biases across different topics, languages, and entry points. The study analyses the structure of video recommendation networks through PageRank distributions, covering 257k videos and 803k recommendations. Despite variations based on factors like video language, content topic, and the source of seed videos, findings reveal an increase in recommendations for a small fraction of videos. 
\colorbox{black}{\color{white}\scriptsize Ecosystem Diversity}

\citet{cakmak2024unveiling} investigate algorithmic bias in YouTube Shorts' recommender by analyzing how watch-time duration, topic sensitivity, and engagement metrics influence content visibility and drift. 
The results on three different domains (South China Sea dispute, 2024 Taiwan election, and general content) reveal a consistent drift away from politically sensitive content toward entertainment-focused videos. 
Emotion analysis shows a systematic preference for joyful or neutral content over negative emotions, while engagement patterns indicate that highly viewed and liked videos are disproportionately promoted, reinforcing popularity bias. 
Such findings demonstrate that YouTube Shorts exhibits systematic algorithmic bias that favors entertainment content and positive emotional tone while suppressing politically sensitive material.
\colorbox{black}{\color{white}\scriptsize Concentration}

\citet{yang2023bubbles} explore the dynamics of personalised search on Twitter using a sock-puppet audit. The findings indicate that factors such as following behaviour, cookies, and previous searches have a limited impact on personalisation, showing a tendency towards homogenisation of user feeds. 
However, when it comes to polarised searches, the results reveal a slight bias toward positive views, raising concerns about filter bubbles effects.
\colorbox{black}{\color{white}\scriptsize Homogenisation}

Using a similar sock-puppet audit methodology, \citet{chen2021neutral} evaluate the impact of Twitter's content curation mechanism on the creation of political echo chambers and filter bubbles, as well as radicalisation and partisan amplification. 
The study finds little evidence of the role of the recommender in generating these outcomes: in general, users receive content that is closely aligned with whatever their friends produce.
\colorbox{black}{\color{white}\scriptsize Echo Chamber} 
\colorbox{black}{\color{white}\scriptsize Filter Bubble} 
\colorbox{black}{\color{white}\scriptsize Radicalisation}

\citet{su2016effect} examine the impact of Twitter's who-to-follow recommender, comparing the social networks collected before and after the recommender was implemented. 
The findings reveal that there is a disproportionate increase in followers for popular users. 
However, the number of recommendations per follower decreases with popularity, so the recommender does not confer an obvious advantage to the most popular users.
\colorbox{black}{\color{white}\scriptsize Concentration} 
\colorbox{black}{\color{white}\scriptsize Item Volume}
\colorbox{black}{\color{white}\scriptsize Ecosystem Volume}

\citet{bouchaud2023algorithmic} explore how Twitter's engagement-maximising recommender affects the visibility of tweets by Members of Parliament (MPs) in users' timelines. 
The researchers use a Twitter dataset collected via a browser add-on installed by volunteers to simulate users' timelines. 
Using an adaptation of \citet{huszar2022algorithmic}'s metric, the study finds that, from the user's perspective, engagement-based timelines amplify right-wing MPs and reduce the ideological diversity of the content shown, contributing to a political filter bubble. 
From the MPs' perspective, the same algorithm narrows the ideological gap between MPs and the users who see their tweets, reducing the diversity of the audiences they reach.
\colorbox{black}{\color{white}\scriptsize Filter Bubble}
\colorbox{black}{\color{white}\scriptsize Discrimination}
\colorbox{black}{\color{white}\scriptsize Item Volume}
\colorbox{black}{\color{white}\scriptsize Item Diversity}

\citet{bartley2024impacts} focus on exposure bias, analyzing how interactions between social networks and recommenders can distort users' online experiences. 
Using archival Twitter data from before the platform introduced algorithmic feed construction, they reconstruct artificial user sessions to observe the prevalence of attributes across different simulated recommenders. 
Their findings show that users are likely to develop distorted views of their immediate network, with minority traits being overrepresented. 
Moreover, simpler algorithmic feeds lead to significantly higher exposure bias compared to chronological, popularity-based, and deep-learning recommender models.
\colorbox{black}{\color{white}\scriptsize Ecosystem Diversity}

\citet{ye2025auditing} investigate exposure bias in X's algorithmic recommendations during the 2024 U.S. Presidential Election.
They conduct a six-week sock-puppet audit with 120 monitoring accounts representing left- and right-leaning profiles to capture tweets from "For You" timelines.
The findings show that X's recommender disproportionately amplifies content from a small set of high-popularity users, with right-leaning accounts facing the highest inequality in exposure.
Both left- and right-leaning accounts experience reinforcement of their own political views and reduced exposure to opposing perspectives.
Additionally, newly created accounts encounter a right-leaning bias in their default timelines.
\colorbox{black}{\color{white}\scriptsize Item Volume}
\colorbox{black}{\color{white}\scriptsize Discrimination}
\colorbox{black}{\color{white}\scriptsize Filter Bubble}

\citet{bakshy2015exposure} explores the impact of Facebook's recommender on the formation of filter bubbles and echo chambers. The researchers examine the news consumption patterns of 10 million users in the US, focusing on how their political beliefs align with the content they encountered in their news feeds and through their social connections. The study finds that users' choices have a more significant impact than Facebook's recommender in limiting exposure to diverse political news. Additionally, the researchers emphasise that users' friend networks can serve as a potential source of diverse perspectives.
\colorbox{black}{\color{white}\scriptsize Filter Bubble}
\colorbox{black}{\color{white}\scriptsize Echo Chamber}

In a collaborative effort between Meta and a team of external researchers, \citet{gonzalez2023asymmetric} investigate the presence of ideological segregation in political news consumption on Facebook during the US 2020 election.
The researchers analyse news content and assess its ideological alignment with the content 208 million users visualise and interact with.
In contrast to \citet{bakshy2015exposure}, this study reveals that algorithmic curation worsens segregation, with conservative users exhibiting less diverse consumption patterns than liberals. Additionally, conservatives engage more with news ecosystems that feature misinformation.
\colorbox{black}{\color{white}\scriptsize Echo Chamber}
\colorbox{black}{\color{white}\scriptsize Item Diversity}
\colorbox{black}{\color{white}\scriptsize Filter Bubble}

\citet{boeker2022empirical} use a sock-puppet audit to investigate how user actions and their attributes impact recommendations displayed on TikTok's \say{For You} page. 
The research reveals that likes, follows, watch duration, as well as user language and location settings all play a role in shaping the volume and nature of content recommended to users.
Among these factors, follows, video view rate, and likes are the most influential in determining the content presented to users.
\colorbox{black}{\color{white}\scriptsize Item Volume}
\colorbox{black}{\color{white}\scriptsize Item Diversity}

Using sock-puppet audit, \citet{le2019measuring} examine whether Google News personalises search results based on a user's political browsing history. 
Sock-puppets, representing distinct political views (pro- and anti-immigration), browse related content and conduct identical searches on Google News. The findings show significant personalisation in Google News search results, indicating the presence of a filter bubble that reinforces the assumed political bias of sock-puppets. 
\colorbox{black}{\color{white}\scriptsize Filter Bubble}

\citet{whittaker2021recommender} investigate whether recommenders on YouTube, Reddit, and Gab promote radicalisation pathways. The researchers use a sock-puppet audit, exposing bots to varying levels of extreme content. The findings indicate a clear trend towards the promotion of more radical content on YouTube, especially after interacting with far-right material. Reddit and Gab do not show a significant algorithmic promotion of extremist content.
\colorbox{black}{\color{white}\scriptsize Item Volume}
\colorbox{black}{\color{white}\scriptsize Radicalisation}

\subsubsection*{\bf Controlled studies}

\citet{markmann2021youtube} investigate whether YouTube's autoplay recommender leads users to consume more radical and extreme content.
By using remote control of the browser, the researchers gather data from two groups of accounts: one with personalised recommendations and one without. 

Autoplay on YouTube narrows the diversity of suggested content, particularly at the channel level, by leaning heavily on users' immediate and overall watch history, leading to more homogeneous recommendations across users. 
On the other hand, there is weak evidence of radicalisation due to the autoplay function: mainstream videos are often preferred over controversial ones.
\colorbox{black}{\color{white}\scriptsize Radicalisation}
\colorbox{black}{\color{white}\scriptsize Homogenisation}

\citet{bartley2021auditing} examine how Twitter's recommender impacts users' information consumption habits. The researchers conduct a sock-puppet audit, dividing users into a treatment group that receives personalised tweet recommendations and a control group where recommendations are provided in inverse-chronological order. 
The study finds that personalised recommendations tend to prioritise popular content, leading to an increased visibility for a few accounts. 
As a result, there is a strong inequality in the visibility of content and users on the platform.
\colorbox{black}{\color{white}\scriptsize Ecosystem Diversity}

\citet{huszar2022algorithmic} employ proprietary data and a multi-year experiment with nearly two million users to analyse how Twitter's timeline algorithm affects the amplification of political content.
They compare a control group with a reverse-chronological feed against a treatment group with personalised feeds. 
Their findings reveal that mainstream right-leaning political content is consistently more amplified than left-leaning content. 
Additionally, the study indicates that algorithmic amplification generally increases the visibility of mainstream news sources in the US, while it does not disproportionately boost far-left or far-right groups compared to moderates. 
\colorbox{black}{\color{white}\scriptsize Discrimination}
\colorbox{black}{\color{white}\scriptsize Item Volume}

In a collaborative effort initiated in early 2020, Meta and a team of external researchers launched the US 2020 Facebook and Instagram Election Study\footnote{https://research.facebook.com/2020-election-research/} that resulted in the publication of four articles \cite{gonzalez2023asymmetric,guess2023social, guess2023reshares,nyhan2023like}, three of which are controlled empirical studies.
\citet{guess2023social} compare the behaviour of Instagram and Facebook users in a control group receiving chronologically ordered feeds to a treatment group of users with personalised recommendations. 
Chronologically ordered feeds show a decrease in time spent on the platform and engagement with content.
Over three months, these changes did not significantly influence affective and ideological polarisation levels, political knowledge, or other major attitudes, self reported by the users in the study.
\colorbox{black}
{\color{white}\scriptsize Item Volume}
\colorbox{black}{\color{white}\scriptsize Individual Volume}

\citet{guess2023reshares} extend this study by exploring the effects of reshared Facebook content on political news exposure and its impact on political polarisation and knowledge. 
By comparing a control group with a standard feed to a treatment group with reshared content removed, the researchers observe a significant reduction in exposure to political news, particularly from unreliable sources.
Also, this reduction does not affect political polarisation or attitudes but leads to a noticeable decrease in users' political knowledge.
\colorbox{black}{\color{white}\scriptsize Item Volume}
\colorbox{black}{\color{white}\scriptsize Individual Volume}

\citet{nyhan2023like} examines the effects of reducing Facebook users’ exposure to like-minded political content during the 2020 U.S. election by artificially removing the platform’s personalisation. 
They compare more than 20,000 treated users -- whose feeds were modified to show much less like-minded content -- to the rest of the population. 
The treatment confirms that users typically operate in echo-chamber-like environments, though less extreme than often claimed. Reducing like-minded exposure lowers total engagement with such content but increases the engagement rate for what remains, indicating a persistent preference for congenial information. 
The shift does not increase cross-cutting exposure but raises exposure to content from sources classified as neither like-minded nor cross-cutting. Finally, reduced like-minded exposure also leads to fewer uncivil posts and less misinformation.
\colorbox{black}{\color{white}\scriptsize Ecosystem Volume}
\colorbox{black}{\color{white}\scriptsize Echo Chamber}
\colorbox{black}
{\color{white}\scriptsize Filter Bubble}
\colorbox{black}{\color{white}\scriptsize Ecosystem Diversity}

\subsection{Simulation studies}
\label{subsec:simulation}
\subsubsection*{\bf Observational studies}
\citet{sirbu2019algorithmic} examine the impact of biasing interactions towards like-minded individuals in synthetic social networks. 
The researchers introduce a recommender parameter that influences the probability of interacting with users who hold similar opinions. 
By simulating the evolution of opinions on a fully connected network with bounded confidence, the study reveals that stronger semantic bias in the recommender leads to increased opinion polarisation.
\colorbox{black}{\color{white}\scriptsize Polarisation} 
\colorbox{black}{\color{white}\scriptsize Homogenisation} 

\citet{pansanella2022mean} builds upon \citet{sirbu2019algorithmic}'s research by exploring various network topologies, including random, scale-free, and clustered networks. The study reveals that opinion polarisation persists across different network topologies. Additionally, introducing a certain degree of sparsity in the network amplifies the divisive impact of recommenders on the distribution of opinions within the population. Furthermore, the researchers indicate that the presence of homophilic communities, combined with cognitive biases, leads to the formation of echo chambers.
\colorbox{black}{\color{white}\scriptsize Polarisation} 
\colorbox{black}{\color{white}\scriptsize Homogenisation} 

Expanding on \citet{sirbu2019algorithmic}, \citet{pansanella2022modeling} explores the impact of adaptive topologies, which allow connections to be changed from conflicting agents to those with similar views. The study finds that recommenders may intensify polarisation, but at the same time hinder the formation of echo chambers. This is due to the homophilic rewiring process and the evolution of opinions. 
\colorbox{black}{\color{white}\scriptsize Polarisation} 
\colorbox{black}{\color{white}\scriptsize Homogenisation} 
\colorbox{black}{\color{white}\scriptsize Echo Chamber} 

Building on a different opinion evolution model, \citet{valensise2023drivers} simulate social network sessions exposed to a feed algorithm that adjusts the range of opinions viewed by users. The simulation accounts for bounded confidence and adaptive topologies.
The study finds that a strong filtering algorithm increases polarisation, while milder personalisation is necessary for echo chamber formation.
\colorbox{black}{\color{white}\scriptsize Polarisation}
\colorbox{black}{\color{white}\scriptsize Echo Chamber}

\citet{chitra2020analyzing} explore the impact of recommenders on social network polarisation using an opinion dynamics model. 
A recommender encourages connections among users with similar viewpoints, thus creating a similarity bias. 
The findings show that a greater bias results in increased polarisation and the creation of echo chambers within clustered networks.
\colorbox{black}{\color{white}\scriptsize Polarisation}
\colorbox{black}{\color{white}\scriptsize Echo Chamber}

\citet{perra2019modelling} examine the impact of different network topologies and timeline filtering strategies, such as random, chronological, reverse chronological, semantic ordering, and nudging. 
They represent users' opinions as binary variables, simulating a two-party system, and find that algorithmic filtering exacerbates inequalities and reduces the visibility of minority opinions. 
The study also highlights that semantic or temporal biases in highly clustered networks lead to opinion polarisation and the formation of echo chambers. 
If there is a majority opinion from the beginning, the filtering algorithm amplifies it, gradually pushing the minority view out of visibility and creating an artificial consensus.
Additionally, combining semantic filtering and nudging in networks with spatial correlations impedes convergence, reinforcing echo chambers that resist nudged opinions.
\colorbox{black}{\color{white}\scriptsize Echo Chamber}
\colorbox{black}{\color{white}\scriptsize Polarisation}
\colorbox{black}{\color{white}\scriptsize Discrimination}
\colorbox{black}{\color{white}\scriptsize Homogenisation}

\citet{peralta2021effect} investigate the interactions between semantic filtering and network topology. Semantic filtering is adjusted using a bias parameter that hides a portion of the population from the recommender. 
The stronger the bias, the more contrasting opinions are hidden. 
The study employs mathematical analyses and simulations of extended binary opinion models considering pairwise and group interactions.
The findings show that semantic bias drives opinion polarisation and echo chamber formation in modular networks, while it fosters polarisation in non-modular networks. 
When the bias is below a threshold, it encourages consensus around a single opinion after pairwise or small-group interactions, whereas interacting in larger groups encourages polarisation.
 \colorbox{black}{\color{white}\scriptsize Polarisation}
 \colorbox{black}{\color{white}\scriptsize Homogenisation}
\colorbox{black}{\color{white}\scriptsize Echo Chamber}
\colorbox{black}{\color{white}\scriptsize Discrimination}
\citet{peralta2021opinion} expand this model to include algorithmic nudging, where the algorithm exhibits a bias towards one of two opinions. 
The simulations reveal that if the platform favours the opinion of the minority group, it promotes polarisation. 
Conversely, if the visibility of the minority opinion is hindered, it leads the population towards consensus.
\colorbox{black}{\color{white}\scriptsize Homogenisation}
\colorbox{black}{\color{white}\scriptsize Discrimination}
\colorbox{black}{\color{white}\scriptsize Polarisation}
\colorbox{black}{\color{white}\scriptsize Ecosystem Diversity}

\citet{gausen2022using} investigate the impact of different recommenders on the spread of information in news feeds. Using an agent-based model to simulate information diffusion and opinion evolution, the researchers compare a random recommender with three filtering strategies: chronological, belief-based, and popularity-based. The findings reveal that belief-based and popularity-based recommenders increase the spread of information, while the random recommender decreases the amount of content shared. Additionally, belief-based recommenders lead to a higher belief purity of agents' feeds, decreasing content diversity.
\colorbox{black}{\color{white}\scriptsize Ecosystem Volume} 
\colorbox{black}{\color{white}\scriptsize Filter Bubble} 
\colorbox{black}{\color{white}\scriptsize Individual Diversity} 
\colorbox{black}{\color{white}\scriptsize Echo Chamber}

\citet{cinus2022effect} look into the long-term evolution of opinions when people recommenders are used on synthetic networks with tunable levels of homophily and segregation. The findings reveal that when initial network conditions are homophilic and non-modular, following link recommendations leads to the formation of echo chambers. This effect becomes absent or reversed if networks are already segregated or heterophilic. 
Additionally, when link recommendations are biased toward connecting users with similar opinions \cite{sirbu2019algorithmic}, they generate far stronger echo chambers than recommenders that do not rely on user similarity. This result contrasts sharply with \citet{pansanella2022modeling} and \citet{valensise2023drivers}, where the recommender biases interactions directly rather than shaping the network's structure through link formation.
\colorbox{black}{\color{white}\scriptsize Echo Chamber}
\colorbox{black}{\color{white}\scriptsize Polarisation}

Similarly, \citet{ramaciotti2021auditing} explore how the evolution of links, combined with an opinion evolution model, impacts polarisation. 
The study shows that when there is no biased assimilation (i.e., the tendency to be more influenced by similar opinions), some recommenders reduce polarisation, while others slightly increase it. 
However, with high levels of biased assimilation, all recommenders lead to smaller increases in polarisation compared to what is caused by sole cognitive biases on a fixed population. This indicates that recommenders often expose users to more diverse connections, mitigating polarisation compared to what would be achieved through user choice alone.
\colorbox{black}{\color{white}\scriptsize Polarisation}

\citet{fabbri2022exposure} examine the long-term impacts of user-recommender feedback loops. The researchers compare random link prediction with link prediction algorithms based on network topology, random walk, and collaborative filtering. 
They then create recommendations in a social network divided into a majority cluster and a minority cluster. 
The study shows the emergence of  discrimination, whereby minority visibility increases only under homophilic initial conditions and otherwise the majority dominates, and concentration, as exposure becomes increasingly skewed toward a small set of high-degree minority or majority hubs, exacerbating existing centrality inequalities.
\colorbox{black}{\color{white}\scriptsize Discrimination}
\colorbox{black}{\color{white}\scriptsize Concentration}
\colorbox{black}{\color{white}\scriptsize Ecosystem Diversity}

\citet{ferrara2022link} investigate the impact of who-to-follow recommenders on networks with two distinct groups, one being the minority category. 
The researchers consider various recommenders suggesting new connections while removing existing random links. 
These include personalised page rank (PPR), egocentric random walks (WTF), friends-of-friends recommender (2H), common-following (CF), and Node2Vec (N2V). 
The findings indicate that networks become more closely connected with repeated recommendations, regardless of the recommender used. 
However, not all tested recommenders exhibit an increase in concentration: Node2Vec prevents the network from increasing inequalities. 
Overall, CF can increase or decrease the visibility of the minority, N2V maintains a balanced impact, and PPR, WTF, and 2H generally maintain the status quo but may decrease minority visibility.
\colorbox{black}{\color{white}\scriptsize Discrimination}
\colorbox{black}{\color{white}\scriptsize Concentration}
\colorbox{black}{\color{white}\scriptsize Ecosystem Diversity}

\citet{rossi2021closed} examine the influence of recommenders on news platforms on user opinions and engagement. In the simulations, users interact with a popularity-based recommender with random exploration, which suggests articles supporting or opposing a topic. 
The findings show that personalisation drives users towards more extreme opinions.

\colorbox{black}{\color{white}\scriptsize Radicalisation}

\citet{ribeiro2023amplification} examine YouTube's amplification paradox. 
This refers to the discovery that sock-puppet audits reveal amplification of problematic content due to recommenders, while user data suggest recommenders are not the primary driver of attention towards this content. 
The researchers build a recommender based on collaborative filtering to simulate recommendations. 
Moreover, they develop an agent-based model where users consume content based on their preferences. 
The results help explain the paradox: when recommendations are followed indiscriminately, extreme content (Far Left/Right) becomes increasingly amplified, confirming trends reported by \citet{haroon2023auditing} and similar auditing work. 
However, when users select content according to their preferences, extreme items are instead de-amplified: users consume less of them than they would without algorithmic guidance. 
This aligns with empirical evidence based on real navigation logs \cite{hosseinmardi2021examining}, suggesting that recommender systems are not necessarily the primary force driving exposure to extreme content.
\colorbox{black}{\color{white}\scriptsize Radicalisation}

\section{Online Retail Ecosystem}
\label{app:online_retail}
\subsection{Empirical studies}
\subsubsection*{\bf Observational studies}
 
\citet{nguyen2014exploring} explore the impact of collaborative filtering on MovieLens users. 
The findings reveal an overall diversity decrease both in the movies recommended and consumed. However, this effect is less pronounced for users who follow recommendations, as they tend to consume a wider variety of movies compared to those who ignore recommendations. Additionally, the recommendation-following users actively seek out diverse movies, which helps reduce the risk of creating filter bubbles. These users also tend to give more positive ratings to the recommended items.
\colorbox{black}{\color{white}\scriptsize Individual Diversity}
\colorbox{black}{\color{white}\scriptsize Filter Bubble}

\citet{ge2020understanding} analyse clicking and purchasing behaviours using real-world data consisting of user clicks, purchases and browse logs from Alibaba Taobao. 
To measure the impact of recommenders on users, the researchers follow the strategy proposed by~\citet{nguyen2014exploring} and separate all users into ``following'' and ``ignoring'' groups.   
The study shows that personalised recommendations reinforce cluster formation in click-behaviors (echo chambers), i.e., there is a strengthening trend over time for the ``following'' group of users. 
Moreover, the set of suggested products is less diverse for the ``following'' group in comparison to the ``ignoring'' group.
This is because personalised recommendations shrink the scope of the offered content, and therefore, the gap further enlarges over time.
\colorbox{black}{\color{white}\scriptsize Echo Chamber}
\colorbox{black}{\color{white}\scriptsize Individual Diversity}

\citet{anderson2020algorithmic} investigate how Spotify's recommender impacts the diversity of streaming content users listen to. 
The researchers split user streaming behaviour into two categories: user-driven listening, where users actively seek out specific music or listen to playlists created by other users, and algorithm-driven listening, where users listen to algorithmically personalised playlists (e.g., Discover Weekly) or radio stations generated by Spotify's algorithm. 
The study finds that each user listens to a more constrained set of songs when relying on algorithmic suggestions, compared to when they explore themselves, thereby signalling a diversity decrease effect of the recommender on individual consumption behaviour. Furthermore, users who listen to a diverse range of songs are less likely to leave the platform and more likely to become paying subscribers.
\colorbox{black}{\color{white}\scriptsize Individual Diversity}
\colorbox{black}{\color{white}\scriptsize Ecosystem Volume}

\citet{chen2004impact} analyse a dataset sourced from Amazon to examine the effects of recommendations and consumer feedback on sales. The findings indicate that more recommendations are associated with high sales volume, but consumer ratings do not have a significant impact on sales. However, the number of consumer reviews positively correlates with sales volume. The study also finds that recommendations work better for less-popular books than for more-popular books.
\colorbox{black}{\color{white}\scriptsize Ecosystem Volume}
\colorbox{black}{\color{white}\scriptsize Item Volume}

\citet{pathak2010empirical} analyse a dataset from Amazon and Barnes \& Noble to explore how the strength of recommendations (i.e., the number of books pointing to a particular book and their popularity) impacts book sales and prices. The study finds that stronger recommendations lead to increased sales volume and higher prices. Additionally, the recommender may contribute to increased diversity in book sales, a phenomenon referred to in the paper as a long-tail effect.
\colorbox{black}{\color{white}\scriptsize Ecosystem Diversity}
\colorbox{black}{\color{white}\scriptsize Ecosystem Volume}
\colorbox{black}{\color{white}\scriptsize Concentration}

\citet{hosanagar2014will} analyse consumer behaviour in an online music store. The store uses a free software add-on to Apple’s iTunes to provide personalised recommendations to registered users through collaborative filtering. 
The findings show that recommendations lead to an increase in commonality among consumers. 
This is because, after receiving the recommendations, individual consumers purchase a greater volume and more similar mix of songs.
\colorbox{black}{\color{white}\scriptsize Individual Volume}
\colorbox{black}{\color{white}\scriptsize Ecosystem Volume}
\colorbox{black}{\color{white}\scriptsize Homogenisation}

\citet{FENG2024103905} investigate how the recommender on a kitchen-sharing platform  affects consumer clickstream and purchase data. They find that the recommender increases user engagement (15$\%$ increase in session count and a 2$\%$ increase in purchase intensity) and decreases (by 29$\%$) the process efficiency (extended decision time). 

\colorbox{black}{\color{white}\scriptsize Individual Volume}

\subsubsection*{\bf Controlled studies}

\citet{yi2022recommendation} conduct a laboratory experiment to explore the impact of product recommendations on search (e.g., computers) and experience goods (e.g., music). 
The participants are divided into a treatment group that receives recommendations and a control group that does not. 
The findings show that users under treatment visualise a wider range of products (either in average and at aggregate) but end up purchasing fewer products and concentrating their purchases on the most popular items. 
This effect is more pronounced for search goods.
\colorbox{black}{\color{white}\scriptsize Individual Diversity}
\colorbox{black}{\color{white}\scriptsize Ecosystem Diversity}
\colorbox{black}{\color{white}\scriptsize Concentration}

\citet{matt2013} split users on an online music store into four treatment groups exposed to different recommenders: content-based filtering, collaborative filtering, bestseller recommender, and random recommender. A control group receives no recommendations. The results indicate that, compared to the baseline, all recommenders (except content-based filtering) lead to an increase in sales diversity.
In a subsequent study, \citet{matt2019factual} introduce two additional randomised treatment groups. These groups are exposed to variants of collaborative filtering and bestseller recommender trained on data describing other users' ratings and purchases.
The perceived recommendation diversity for individuals was highest for collaborative filtering.
At the ecosystem level, recommenders have varying effects on sales diversity: the random recommender increases it, collaborative filtering decreases it, and other recommenders (including the collaborative filtering variant) have no effect.
Neither of the bestseller recommender variants differs from the baseline. The other recommenders have no noticeable effect on sales diversity.
\colorbox{black}{\color{white}\scriptsize Individual Diversity}
\colorbox{black}{\color{white}\scriptsize Ecosystem Diversity}
\colorbox{black}{\color{white}\scriptsize Concentration}

\citet{lee2014impact} investigate the impact of different recommenders on movie sales at a top retailer in North America. Customers are randomly assigned to one of four groups: a control group with no recommendations, three treatment groups exposed to purchase-based collaborative filtering (who bought this also bought that), view-based collaborative filtering (who viewed this also viewed that), and recently-viewed recommender (recently viewed items).
The study finds that purchase-based collaborative filtering significantly increases the average number of views per individual and the average number of purchases compared to the control group. In contrast, the effects of view-based collaborative filtering and recent-views-based recommenders are not statistically significant. 
Both collaborative filtering algorithms increase sales diversity at the individual level but decrease aggregate sales diversity. 
This indicates that both algorithms encourage users to purchase the same products, leading to a concentration effect at the ecosystem level. 
The recently-viewed recommender decreases sales diversity at the ecosystem level but has no effect at the individual level.
\colorbox{black}{\color{white}\scriptsize Individual Volume}
\colorbox{black}{\color{white}\scriptsize Item Volume}
\colorbox{black}{\color{white}\scriptsize Ecosystem Volume}
\colorbox{black}{\color{white}\scriptsize Individual Diversity}
\colorbox{black}{\color{white}\scriptsize Ecosystem Diversity}

\citet{liang2022exploring} examine the behaviour of four random groups of Spotify users over six weeks. 
These groups are composed on the basis of algorithm personalisation and the visual presentation of recommendations. 
In their first session, users are randomly assigned to either a representative or a personalised initial playlist. Then, they are further assigned to one of two visual presentations.  
The study reveals an initial increase in the diversity of music exploration within the playlist, driven by nudging techniques such as default initial playlists and visual anchors. However, this heightened exploration gradually diminishes over time. 
The residual effect on the change in users’ profiles indicates the potential (long-term) benefits of combining nudging with personalisation in exploration tools.
\colorbox{black}{\color{white}\scriptsize Individual Diversity}

\citet{long2022choice}  employ data from 1.6 million Alibaba customers to examine how the quantity of recommended products impacts on consumers' search and purchase behaviours. The researchers leverage Alibaba's recommendation technology and randomly assign consumers to one of four treatment groups, each receiving a different number of recommended products. The findings reveal that increasing the number of recommended products boosts the probability of purchasing those items. However, this probability decreases as the number of recommended products continues to rise.
Purchase probability declines mainly because consumers reduce the number of searches as a consequence of choice overload. 
\colorbox{black}{\color{white}\scriptsize Individual Volume}
\colorbox{black}{\color{white}\scriptsize Ecosystem Volume}

\citet{lee2020} investigate the impact of recommendations on views and sales of cosmetics and clothing on mobile and personal computer (PC) channels. The experiment was carried out on an online retailing firm’s web application in South Korea. The researchers split customers into a treatment group exposed to collaborative filtering trained on users' recent views and a control group exposed to best-selling items. The study finds that collaborative filtering increases views and sales volume when users access the platform through mobile devices. When users access the platform through PCs, the recommender only increases the volume of views. These outcomes are particularly pronounced for the most expensive items. Moreover, collaborative filtering increases view diversity on both mobile and PC platforms, but it has no significant impact on sales diversity. 
\colorbox{black}{\color{white}\scriptsize Item Volume}
\colorbox{black}{\color{white}\scriptsize Ecosystem Volume}
\colorbox{black}{\color{white}\scriptsize Ecosystem Diversity}

\citet{donnelly2021longtail} investigate the impact of personalised recommendations generated by Wayfair's collaborative filtering on consumption patterns in the context of online furniture shopping. 
In the experiment, a treatment group of 95\% of customers exposed to Wayfair's recommender is compared to a control group of 5\% of the customers exposed to popularity-based recommendations. The study finds that the recommender encourages users to engage in more searches, increasing the number of clicks and positively influencing purchase probability at the individual level. 
Furthermore, the findings indicate that Wayfair's recommendations increase diversity in searches and sales at the ecosystem level. 
\colorbox{black}{\color{white}\scriptsize Individual Volume}
\colorbox{black}{\color{white}\scriptsize Ecosystem Diversity}

\citet{holtz2020engagement} examine the impact of recommendations on podcast consumption among approximately 900,000 Spotify premium users across seventeen countries. Users in the treatment group are exposed to personalised recommendations based on their historical listening behaviour, while those in the control group are exposed to the most popular podcasts. 
The study finds that at the individual level, personalised recommendations lead to an increased volume of podcasts listened to, but a decrease in podcast streaming diversity. However, at the systemic level, personalised recommendations increase podcast streaming diversity \textit{across} users.
\colorbox{black}{\color{white}\scriptsize Individual Volume}
\colorbox{black}{\color{white}\scriptsize Individual Diversity}
\colorbox{black}{\color{white}\scriptsize Homogenisation}

\citet{chen2022more} examine how recommendations affect the relationship between filter bubbles and consumers' preferences on the e-commerce platforms Jingdong and Taobao. 
The study distinguishes between personalised recommendations for users with personal accounts and non-personalised recommendations for users without them. 
The findings show that recommendations reinforce individual consumer preferences, creating a filter bubble effect. 
\colorbox{black}{\color{white}\scriptsize Filter Bubble}
\colorbox{black}{\color{white}\scriptsize Ecosystem Diversity}

\citet{lee2019How} explore the impact of collaborative filtering on sales diversity using data from a randomised field experiment conducted over a top retailer application in North America. 
The researchers split users into a control group with no recommendations and two treatment groups exposed to view-based (who viewed this also viewed that) and purchase-based collaborative filtering (who purchased this also purchased that). 
The study finds that the two recommenders increase sales diversity at the individual level, leading to a decrease in views and sales diversity at the ecosystem level. 
As similar users explore the same products, this results in a concentration effect. 
At the item level, both recommenders generate an increase in views and sales.
\colorbox{black}{\color{white}\scriptsize Item Volume}
\colorbox{black}{\color{white}\scriptsize Ecosystem Volume}
\colorbox{black}{\color{white}\scriptsize Ecosystem Diversity}
\colorbox{black}{\color{white}\scriptsize Individual Diversity}
\colorbox{black}{\color{white}\scriptsize Concentration}

\citet{li2022recommender} conduct three experiments in a laboratory and a real-life online bookstore. Each experiment involves a control group of users receiving no recommendations and treatment groups exposed to recommendations, based on a basket value. Here, basket value refers to the total dollar amount of a user’s shopping basket. 
In the first laboratory experiment, three treatment groups are exposed to collaborative filtering, best-selling, and random recommendations.
The researchers find that collaborative filtering provides the highest basket value. 
In a subsequent laboratory experiment based on collaborative filtering only, the study finds that recommending three products of the same type is the most effective way to increase basket value.
In the real-life experiment, the treatment group receives recommendations about three different products from memory-based collaborative filtering.  
The findings show that this recommender leads to an increase in diversity in consumers' consideration sets, as well as an increase in views and sales.
\colorbox{black}{\color{white}\scriptsize Individual Volume}
\colorbox{black}{\color{white}\scriptsize Individual Diversity}

\subsection{Simulation studies}

\subsubsection*{\bf Observational studies}
\citet{noordeh2020echo} measure the impact of collaborative filtering on content consumption on MovieLens. 
The study reveals that prolonged exposure to recommendations decreases content diversity and fosters the emergence of filter bubbles. Furthermore, once a filter bubble is established, it becomes challenging for users to break out of it.
\colorbox{black}{\color{white}\scriptsize Individual Diversity}
\colorbox{black}{\color{white}\scriptsize Filter Bubble}

\citet{barlacchi2025simulation} examine how the user-recommender feedback loop affects ecosystem diversity, purchase concentration, and user homogenisation over time. 
The results show that the feedback loop increases individual diversity, but it reduces ecosystem diversity across the user population. 
In addition, homogenisation effects emerge as users purchase increasingly similar items over time.
\colorbox{black}{\color{white}\scriptsize Individual Diversity}
\colorbox{black}{\color{white}\scriptsize Ecosystem Diversity}
\colorbox{black}{\color{white}\scriptsize Concentration}
\colorbox{black}{\color{white}\scriptsize Homogenisation}

\citet{hazrati2022recommender} employ log data from three Amazon services (Kindle, Games, and Apps) to analyse the effects of recommendations on the evolution of users' choices over time. 
The simulation combines a choice model with five recommenders. Three recommenders offer personalised recommendations: popularity-based collaborative filtering, low popularity-based collaborative filtering (penalising the score with the inverse popularity), and factor model (mapping users and items into a common latent factor space). 
Additionally, the study includes two non-personalised recommenders, namely popularity-based and average rating, as well as a baseline case with no recommendations.
The study finds that personalised recommendations lead to a greater increase in sales diversity compared to non-personalised recommendations, both at the item and ecosystem levels. 
Furthermore, at the ecosystem level, the low popularity-based collaborative filtering and the factor model increase sales diversity for the Kindle dataset. 
However, for the Games dataset, only the low popularity-based collaborative filtering increased sales diversity compared to the baseline case. In addition, the average popularity of recommended products is larger for non personalised algorithms with respect to all personalised ones and to the empirical data.
Surprisingly, non-personalised recommenders result in choices for items that have a larger predicted rating compared to
personalised recommenders.
\colorbox{black}{\color{white}\scriptsize Ecosystem Diversity}
\colorbox{black}{\color{white}\scriptsize Concentration}

\citet{mansouryAbdollahpouri2020} simulate the feedback loop of user-recommender interactions by analysing the progressive effects of three different recommenders: user-based collaborative filtering, Bayesian personalised ranking, and a recommender suggesting the most popular items.
The findings reveal that all recommenders lead to a progressive reduction in ecosystem diversity (catalog coverage) and increased concentration. 
Moreover, they find a general homogenisation of users' preferences towords the tastes of a majority group.
\colorbox{black}{\color{white}\scriptsize Concentration}
\colorbox{black}{\color{white}\scriptsize Ecosystem Diversity} 
\colorbox{black}{\color{white}\scriptsize Homogenisation} 
\colorbox{black}{\color{white}\scriptsize Discrimination}

\citet{wu2011} compare various recommenders trained on MovieLens data (a collaborative filtering and a content-based recommender) with a baseline condition with no recommendations.
The findings reveal that the content-based recommender increases sales diversity, whereas collaborative filtering decreases it. Moreover, the impact of these effects depends on how well the recommendations align with consumer awareness. For instance, suggesting popular products to consumers already aware of them has little impact. 
Recommending niche products could significantly influence consumer behaviour.
\colorbox{black}{\color{white}\scriptsize  Ecosystem Diversity}

\citet{aridor2020deconstructing} design a model in which products have both intrinsic and user-specific values.
In this model, users (unaware of item values) make choices on the basis of their beliefs and risk aversion. 
This baseline condition is compared to one where users are exposed to recommendations that allow them to combine their value with the intrinsic value of items. 
The study shows that the more users become risk-averse, the more they consume items similar to those they previously considered valuable.
This leads to filter bubbles that narrow their consumption patterns. Recommendations help reduce these filter bubbles, but they also leads to an increase in homogeneity across users.
\colorbox{black}{\color{white}\scriptsize Individual Diversity}
\colorbox{black}{\color{white}\scriptsize Filter Bubble}
\colorbox{black}{\color{white}\scriptsize Homogenisation}

\citet{chaney2018algorithmic} explore how training recommenders using data from users influenced by automatic recommendations can lead to algorithmic confounding in a fully synthetic simulation. 
The researchers compare the effects of six recommenders (popularity-based, content filtering, matrix factorisation, social filtering, and random) with an ideal benchmark that recommends items based on the true utility of users. 
The study finds that a single training session leads to a small homogenisation in user behaviour, which then reverts to the ideal case. 
However, repeated training causes a greater homogenisation of user behaviour, with the effect becoming more pronounced with each cycle through the loop. 
This homogenisation occurs both at the local level (users behave more like their nearest neighbours) and population level (users become more similar on average) for all recommenders (except for the random recommender). 
In addition, they find that repeated training amplifies the impact of recommenders on the distribution of item consumption, irrespective of homogenisation effects. For example, matrix factorization and content filtering have comparable homogenising effects, but the formers reduce the diversity of consumed items.
\colorbox{black}{\color{white}\scriptsize Homogenisation}
\colorbox{black}{\color{white}\scriptsize Ecosystem Diversity}

\citet{fleder2007recommender} perform a simulation where users are exposed to collaborative filtering and have a certain probability of accepting the recommender's suggestion. 
The outcome is compared to that resulting from the same process, except for that when recommendations are not enabled. 
The study reveals a concentration effect towards a few items. 
A subsequent study \cite{flederHosanagar2009blockbuster} employs the same simulation settings to suggest that recommendations can increase sales diversity at the individual level, but decrease it at the ecosystem level.
\colorbox{black}{\color{white}\scriptsize Ecosystem Diversity}

\citet{de2023recommender} investigate how different implementations of user-based collaborative filtering influence users’ ability to discover novel content.
The study shows that recommenders increase the rate of individual novelty discovery. 
Under collaborative filtering, there is a faster growth in the number of distinct items encountered.
The study finds that all recommenders tend to enhance individual diversity, but some of them simultaneously promote polarised audience segmentation.
\colorbox{black}{\color{white}\scriptsize Individual Diversity}
\colorbox{black}{\color{white}\scriptsize Homogenisation}

\section{Urban Mapping}
\label{app:urban_mapping}

\subsection{Empirical studies}
\subsubsection*{\bf Observational studies}

\citet{schwieterman2019uber} shows that transportation network companies (e.g., Uber and Lyft) in Chicago contribute to reducing travel times compared to public transit, but are also slightly more costly on average for users. Moreover, during peak weekday hours, the prices are marginally higher than at other times, suggesting that transportation network companies may use surge pricing to respond to mobility demand.
\colorbox{black}{\color{white}\scriptsize  Individual Volume}

\citet{santi2014quantifying} employ a large dataset of taxi trips in New York City to model the collective benefits of ride-sharing as a function of prolonged travel time. They find that ride-sharing reduces users' travel time, cumulative trip length, and service cost. 
However, it entails an increase in the number of taxi passengers. \colorbox{black}{\color{white}\scriptsize Individual Volume} \colorbox{black}{\color{white}\scriptsize Item Volume}

\citet{jalali2017emission} use GPS trajectories from private vehicles to investigate the potential impact of ride-sharing in a Chinese city. 
They discover that ride-sharing reduces the number of trips, drivers' total travelled distance, and emissions, especially if users are willing to walk to drivers within 3 km. 
\colorbox{black}{\color{white}\scriptsize Ecosystem Volume}

\citet{jiang2025carbon} conduct a large-scale empirical analysis of ride-pooling in Suzhou, China, combining regulatory trajectory data with route reconstructions from the Amap API. Results reveal a positive relationship between route overlap and carbon emission reductions. Across 12.5 million rides, of which 3.5\% were pooled, pooling reduced emissions by an average of 22.5\% compared to the equivalent non-pooled rides. A scenario featuring a full ride-pooling adoption could cut Suzhou’s annual emissions by 30,000 tons, representing a 19.14\% reduction.
\colorbox{black}{\color{white}\scriptsize Ecosystem Volume}

\citet{hanna2017citywide} analyse the impact of lifting Jakarta's "three-in-one" high-occupancy vehicle policy (HOV), which restricted certain roads at specific hours to vehicles with a minimum of three occupants. 
By gathering data on road travel times from Google Maps before and after the policy lifting, the researchers uncover noticeable effects of HOV on traffic congestion: lifting the policy increased travel times both on high-occupancy roads and alternative routes, and both during and outside HOV periods.
\colorbox{black}{\color{white}\scriptsize Concentration}
\colorbox{black}{\color{white}\scriptsize Item Volume}

A few works \cite{koh2019offline, edelman2014digital, zhang2021frontiers} focus on the empirical analysis of data from Airbnb. 
Koh et al. \cite{koh2019offline} analyse the diversity of the user base on the Airbnb platform across five cities in three continents. The study observes a predominantly young, female, and white user base, even in cities with a diverse racial composition. This creates an echo chamber effect where similar demographics tend to cluster. The authors also observe a similar homophily tendency between female hosts and guests and a relevant homophily tendency regarding race, while no tendency is highlighted in age. 
\colorbox{black}{\color{white}\scriptsize Echo Chamber}

Similarly, \citet{edelman2014digital} analyse pictures of New York City landlords on Airbnb and observe revenue inequalities: non-black hosts' houses are about 12\% more expensive than those of black hosts, even when the houses have similar attributes like the number of bedrooms, type of room, and user ratings. 
\colorbox{black}{\color{white}\scriptsize Discrimination}

\citet{zhang2021frontiers} investigate the impact of the smart-pricing algorithm of Airbnb on racial disparities in daily host revenue. The researchers collect data on venue prices, host race (inferred from profile pictures), host revenues, and venue occupancy rates before and after hosts adopt the smart-pricing algorithm. They find that the algorithm reduces venue prices, increases host revenues, and decreases the revenue gap between white and black hosts. 
\colorbox{black}{\color{white}\scriptsize Discrimination}
\colorbox{black}{\color{white}\scriptsize Individual Volume}
\colorbox{black}{\color{white}\scriptsize Item Volume}

\subsection{Simulation studies}
\subsubsection*{\bf Observational studies}

\citet{johnson2017beautiful} investigate the impact on urban traffic of three routing criteria: scenic routing optimises routes for aesthetic enjoyment; safety routing avoids areas with higher rates of accidents or crime; and simplicity routing, where route complexity is reduced on the basis of the number of intersections and actions needed to traverse it (i.e., going straight or turning).
Simulations in San Francisco, New York City, London, and Manila show that scenic routing leads to more complex routes, potentially increasing the risk of accidents and negatively affecting driver safety. Additionally, it diverts traffic from highways to parks, popular areas, tourist destinations, and slower roads. Safety routing, though to a lesser degree than scenic routing, also generates more complex routes and redirects traffic away from identified unsafe zones.
Simplicity routing amplifies traffic on highways but does not explicitly favour or avoid any particular region.
\colorbox{black}{\color{white}\scriptsize Concentration}
\colorbox{black}{\color{white}\scriptsize Discrimination}

\citet{mehrvarz2020optimal} compare the impact of vehicle routing incorporating sustainability variables (e.g., fuel consumption, engine load, acceleration rate, speed, road slope) with traditional routing that prioritises travel time or distance. The study finds that fastest routes are not necessarily the most sustainable and that sustainable routing might reduce fuel consumption by about 5\%.
\colorbox{black}{\color{white}\scriptsize Ecosystem Volume}

\citet{barth2007environmental} introduce a method for reducing energy consumption and emissions in navigation services. The method combines mobile-source energy and emission models with advanced route optimisation algorithms. The study applies this method in several case studies across Southern California, showing substantial energy savings and reduced emissions compared to navigation services that minimise distance or travel time.
\colorbox{black}{\color{white}\scriptsize Ecosystem Volume}

\citet{colak2016understanding} introduce a centralised strategy that optimises route choices to alleviate urban congestion while considering varying levels of social good awareness. The study shows that routing solutions mimicking socially optimal configurations decrease time lost in congestion by up to 30\%, with individual travel time reduction ranging between one and three minutes. 
\colorbox{black}{\color{white}\scriptsize Concentration}
\colorbox{black}{\color{white}\scriptsize Ecosystem Volume}
\colorbox{black}{\color{white}\scriptsize Individual Volume}

Cornacchia et al. \cite{cornacchia2023oneshot} introduce METIS, a traffic assignment algorithm designed to optimise vehicle routing by offering diverse alternatives. 
The study conducts a simulation across Florence, Rome, and Milan, evaluating the impact of METIS on various urban metrics, including CO2 emissions and road coverage.
The study reveals that METIS produces a more equitable distribution of traffic on the road network than other state-of-the-art routing algorithms, increases road coverage and mitigates CO2 emissions considerably. 
\colorbox{black}{\color{white}\scriptsize Ecosystem Diversity}
\colorbox{black}{\color{white}\scriptsize Ecosystem Volume}

Maciejewski et al. \cite{maciejewski2016assignment, maciejewski2016largescale} employ floating car data to simulate the impact of taxi fleets on traffic in Berlin and Barcelona.
The study evaluates two dispatching strategies: the ``nearest-idle-taxi" approach, where the closest available taxi is dispatched to the first available request; and the ``demand-supply balancing strategy," which classifies system states into oversupply and undersupply conditions.
The demand-supply balancing strategy outperforms the nearest-idle-taxi approach, considerably reducing waiting time for both drivers and passengers. 
\colorbox{black}{\color{white}\scriptsize Individual Volume}
\colorbox{black}{\color{white}\scriptsize Item Volume}

In another work, \citet{Maciejewski2013SimulationAD} evaluates three taxi dispatching strategies: a ``no-scheduling strategy'' (NOS) that assigns the nearest empty taxi to each request; a ``one-time schedule strategy'' (OTS) that assigns new customers to the taxi soonest available after current trips; and a ``re-scheduling strategy'' (RES) that recalculates assignments after each drop-off. 
Although NOS performs well under light system loads, slightly outperforming the other strategies in reducing passenger waiting times, RES is more effective as demand increases. 
\colorbox{black}{\color{white}\scriptsize Individual Volume}

\citet{erhardt2019transportation} simulate the effects of ride-hailing on San Francisco's traffic congestion. 
They compare traffic volumes in 2010, before significant ride-hailing activity, with those in 2016 when such services were available. Findings highlight that ride-hailing increases congestion, mainly because about 50\% of vehicles' miles travelled are with no passengers. The study also finds a 62\% increase in weekday vehicle hours of delay from 2010 to 2016 versus a 22\% increase under a hypothetical scenario without ride-hailing.
\colorbox{black}{\color{white}\scriptsize Concentration}
\colorbox{black}{\color{white}\scriptsize Ecosystem Volume}

\citet{martinez2017assessing} develop an agent-based model to examine the impact of moving from private transportation to a shared and self-driving vehicle fleet (taxis and mini-buses) in Lisbon.
Their study reveals that implementing the full-sharing scenario could substantially reduce CO2 emissions, congestion levels, and travel distances.
Sharing vehicles leads to more intensive vehicle utilisation, significantly increasing vehicles' daily usage and travel distances. 
\colorbox{black}{\color{white}\scriptsize Concentration} \colorbox{black}
{\color{white}\scriptsize Item Volume}
\colorbox{black}{\color{white}\scriptsize Ecosystem Volume}

\citet{zhu2017reducing} introduce a combinatorial optimisation model for long-term taxi trip assignment to minimise idle time and the number of taxis required. 
Simulations in New York City demonstrate that the model reduces by 28\% the taxi fleet size needed to complete all trips and cuts by 32\% per taxi average idle time.
\colorbox{black}{\color{white}\scriptsize Ecosystem Volume}
\colorbox{black}{\color{white}\scriptsize Item Volume}

\citet{fagnant2014travel} use agent-based modelling to evaluate the environmental impact of shared autonomous vehicles (SAV) compared to conventional vehicle ownership and usage patterns. Their simulations indicate that a single SAV can replace eleven traditional vehicles. Despite a projected 10\% increase in travel distances, the overall impact of SAVs remains favourable for reducing emissions compared to non-SAV trips. 
Additionally, the study suggests that centralised global strategies for SAV relocation are more effective in mitigating environmental impacts than localised approaches. 
\colorbox{black}{\color{white}\scriptsize Ecosystem Volume}

In subsequent research, \citet{fagnant2018dynamic} investigate the impact of SAV on travel costs and service times in Austin employing a Dynamic Ride-Sharing strategy (DRS). DRS brings together multiple users with similar origin and destination points at the same time.
The findings show that DRS reduces average service times and travel costs for SAV users, presenting potential benefits for both autonomous taxis and travellers. 
\colorbox{black}{\color{white}\scriptsize Individual Volume}

\citet{afeche2023ridehailing} use a game-theoretic model to analyse ride-hailing services, focusing on passenger-driver matches in a spatial network. They assess the impact of admission control (accepting or rejecting ride requests based on destination) and positioning control (relocating drivers to high-demand areas). 
The study compares three approaches: centralised control with strict admission and repositioning; minimal control with open admission and decentralised repositioning optimal admission control, which combines centralised and decentralised repositioning. 
Results show that while decentralised repositioning can result in drivers idling in low-demand zones, admission control reduces such inefficiencies. This approach can also lead to rejecting requests from less busy areas.
This may exacerbate inequality in service access and potentially decrease driver satisfaction. 
\colorbox{black}{\color{white}\scriptsize Discrimination}

\citet{agarwal2022impact} explore the interaction between ride-hailing apps' dynamic pricing and taxi bookings in Singapore. They find that a 10\% increase in ride-hailing surge prices results in a 2.6\% increase in taxi bookings within the same region and time interval. 
\colorbox{black}{\color{white}\scriptsize Ecosystem Volume}

\citet{bokanyi2020understanding} conduct an agent-based simulation to study the impact of ride-hailing matching algorithms. The study finds that the ``nearest algorithm'', which assigns passengers to the closest vehicle, exacerbates inequality among drivers' gains and is affected by the spatial location of pick-ups and drop-offs. 
Conversely, a ``poorest algorithm'', which prioritises drivers with lower earnings, reduces gain disparities.
Moreover, with outward flows, it also boosts average driver gains. \colorbox{black}{\color{white}\scriptsize Discrimination}

\citet{alonsomora2017ondemand} present a ride-sharing algorithm for assigning passenger requests to a fleet of vehicles of varying capacity (i.e., number of passengers), validating its performance in New York City.  
The results show that 2000 vehicles (15\% of the taxi fleet) of capacity ten or 3000 of capacity four can serve 98\% of the demand within a mean waiting time of 2.8 min and a mean trip delay of 3.5 min. Moreover, the study finds that increasing vehicle capacity improves service rate and reduces the mean distance travelled by vehicles.
\colorbox{black}{\color{white}\scriptsize Item Volume}
\colorbox{black}{\color{white}\scriptsize Ecosystem Volume}

Mori et al. \cite{mori2022developing} explore the advantages of integrating ride-sharing taxis with traditional taxi services through traffic simulation and dynamic vehicle allocation. The findings indicate that increasing the number of vehicles decreases the average time from booking to arrival. This effect is especially pronounced for ride-sharing taxis, although it reduces vehicle occupancy rates. 
\colorbox{black}{\color{white}\scriptsize Individual Volume}
\colorbox{black}{\color{white}\scriptsize Item Volume}

\citet{storch2021incentive} employ a game-theoretic approach to investigate the incentives (financial discounts, expected detours and trip uncertainty, and the inconvenience of sharing a vehicle with strangers) affecting ride-sharing adoption. The study identifies two distinct adoption regimes: one characterised by decreased sharing as demand rises and another by consistent sharing regardless of demand levels. The simulation reveals a discontinuous transition between these regimes, suggesting that even modest increases in financial incentives could significantly boost ride-sharing adoption. This pattern is also observed in empirical data about ride-sharing adoption in New York City and Chicago. 
\colorbox{black}{\color{white}\scriptsize Ecosystem Volume}

Garcia-López et al. \cite{garcia2020short} investigate the influence of Airbnb on housing rents and prices in Barcelona. The study proposes a model that uses rent data, transaction prices, and posted prices in which Airbnb house owners decide between long-term or short-term rentals. 
The findings reveal that, on average, neighbourhoods experience a 1.9\% increase in rents and a 5.3\% rise in transaction prices after the introduction of the Airbnb service. 
This impact is strongest in areas with high Airbnb activity, with estimated rent hikes of 7\% and transaction price increases of 19\%.
\colorbox{black}{\color{white}\scriptsize Item Volume}
\colorbox{black}{\color{white}\scriptsize Concentration}

\citet{sanchez2023bias} investigate bias in location-based recommenders, focusing on how algorithms may skew recommendations toward popular venues, certain categories, or geographically close places. 
Their evaluation across different recommendation models shows that such biases are systematic and vary by algorithm family. 
The authors also test two mitigation strategies, showing that they can reduce concentration without compromising recommendation accuracy. \colorbox{black}{\color{white}\scriptsize Concentration} 
\colorbox{black}
{\color{white}\scriptsize Diversity Individual} 

\citet{mauro2025urban} propose a simulation framework to study how location-based recommenders influence co-location and venue visitation patterns. 
The simulations show that, as recommender adoption increases across the population, users visit a wider range of venues. At the ecosystem level, the recommendations drive visits toward a small set of popular places, and the similarity of the visitation patterns among users increases.
\colorbox{black}{\color{white}\scriptsize Concentration} 
\colorbox{black}{\color{white}\scriptsize Homogenisation}
\colorbox{black}{\color{white}\scriptsize Individual Diversity}
\colorbox{black}{\color{white}\scriptsize Ecosystem Diversity}
  
\citet{sanchez2021effects} investigate whether augmenting the training set of a target city by adding check-in data from other source cities improves the accuracy of next-venue recommendations. They select source cities by interaction volume or geographic proximity and evaluate the effects on locals and tourists using collaborative-filtering models. Although augmentation increases overall accuracy and coverage, the gains are more pronounced for tourists, whose globally consistent mobility patterns are better supported by the added data. For locals, whose preferences are more city-specific, the augmented data amplify popularity bias and reduce the ability of the model to represent their niche behaviours, leading to systematically poorer recommendation quality for this group.
\colorbox{black}{\color{white}\scriptsize Concentration} \colorbox{black}{\color{white}\scriptsize Discrimination}

\subsubsection*{\bf Controlled studies}
\citet{arora2021quantifying} use Google Maps data and to simulate the impact of Google Maps' real-time navigation on travel time and CO2 emissions in Salt Lake City, targeting the subset of vehicles that use Google Maps' suggestions.
The study reveals that, on average, Google Maps users reduce CO2 emissions by 1.7\% and travel time by 6.5\%. 
For users whose routes differ from their original plans due to Google Maps' suggestions, there is a reduction of 3.4\% in CO2 emissions and 12.5\% in travel time. 
\colorbox{black}{\color{white}\scriptsize Ecosystem Volume}

\citet{valdes2016eco} and \citet{perezprada2017managing} examine the impact of adopting eco-friendly routes on traffic and emissions under various traffic conditions in the region of Madrid.
\citet{valdes2016eco} show that eco-routing reduces CO2 emissions in low and high traffic conditions. 
These gains grow with the share of green drivers, though with diminishing returns at higher penetration levels.
However, the environmental benefits come at the cost of increased travel times, as greener routes tend to favor shorter but slower paths.
\colorbox{black}{\color{white}\scriptsize Ecosystem Volume}
\citet{perezprada2017managing} discover that, in situations of congested traffic, a 90\% adoption rate of eco-friendly routes leads to reductions of up to 10\% in CO2 and 13\% in NOx emissions. However, NOx exposure for the population increases by up to 20.2\% and travel times by 28.7\%. Additionally, while traffic volumes decrease by 13.5\% for the entire region, downtown areas experience an increase in vehicle concentration of up to 16.4\%.
\colorbox{black}{\color{white}\scriptsize Concentration}
\colorbox{black}{\color{white}\scriptsize Ecosystem Volume}

\citet{cornacchia2022routing} propose a simulation framework based on SUMO to assess the impact of navigation services on urban CO2 emissions. 
Using route suggestions collected through APIs provided by TomTom and OpenStreetMap, the authors set up scenarios with different service adoption rates in Milan. 
When the adoption rate is high or low, CO2 emissions are higher than the baseline scenario where vehicles do not follow recommendations. 
In contrast, when the adoption rate is around 50\%, there is a reduction in the overall CO2 emissions over the road network. Furthermore, the higher the adoption rate, the fewer emissions are in the city centre and the more in the external ring road. \colorbox{black}{\color{white}\scriptsize Concentration}
\colorbox{black}{\color{white}\scriptsize Ecosystem Volume}
In an extension of this work, \citet{cornacchia2024navigation} show that navigation services may reduce emissions when a small share of drivers use them.
However, once the service adoption rate exceeds a city- and service-specific threshold, these benefits flatten or reverse. 
At high adoption rates, navigation services also reduces route diversity, as many drivers follow the same suggested paths.
This concentrates traffic and emissions in specific neighbourhoods,  increasing their environmental burden.
\colorbox{black}{\color{white}\scriptsize Concentration}
\colorbox{black}{\color{white}\scriptsize Ecosystem Volume}
\colorbox{black}{\color{white}\scriptsize Ecosystem Diversity}
In subsequent research, \citet{cornacchia2023effects} demonstrate that introducing randomness into recommended routes can reduce travel time and CO2 emissions.
\colorbox{black}{\color{white}\scriptsize Ecosystem Volume}

\citet{thai2016negative} analyse the impact of navigation services on road usage from a theoretical perspective, modelling it as a heterogeneous routing game. The model distinguishes between routed users, who utilise real-time navigation data to follow the shortest routes, and non-routed users, who rely on highways due to limited knowledge of low-capacity roads. Simulations in Los Angeles reveal that navigation services can reduce average travel times and total vehicle miles travelled. However, they also transfer significant traffic from highways to city streets, exacerbating urban congestion. 
\colorbox{black}{\color{white}\scriptsize Concentration}
\colorbox{black}{\color{white}\scriptsize Ecosystem Volume}

\citet{ahn2013eco} investigate the impacts of dynamic eco-routing on the transportation network. The researchers consider various adoption rates and traffic conditions in downtown Cleveland and Columbus. The study finds that dynamic eco-routing may lead to considerable fuel savings and emissions reduction compared to traditional routing methods. Additionally, dynamic eco-routing reduces travel distance, but does not always result in shorter travel times. The findings also reveal that as the proportion of dynamic eco-routing increases, fuel consumption decreases at the ecosystem level.
\colorbox{black}{\color{white}\scriptsize Ecosystem Volume}

\citet{sabet2025exploring} examine the combined effects of eco-routing, eco-driving, and alternative fuel technologies (Battery Electric Vehicles, Hybrid Electric Vehicles, and e-fuels) in downtown Toronto, under varying levels of Connected and Automated Vehicle (CAV) penetration. 
The results show that eco-routing and eco-driving reduce emissions and travel time, with the strongest gains achieved when integrated with CAVs. At the same time, the study observes modal shifts from active and public transport to private electric vehicles, which may increase vehicle kilometers traveled and congestion and offset part of the environmental benefits.
\colorbox{black}{\color{white}\scriptsize Ecosystem Volume}
\colorbox{black}{\color{white}\scriptsize Concentration}

\section{Generative AI}
\label{app:generative_AI} 
\subsection{Simulation studies}
\subsubsection*{\bf Observational studies}

\citet{bulte2025llms} examine how prompt language and explicit cultural framing align LLM responses to cultural values. 
The study employs items from standardized cultural surveys across 11 languages, and formulates prompts with and without explicit different cultural perspectives. 
The findings show that both prompt language and explicit cultural framing influence model outputs, but explicit cultural framing is more effective for aligning with human values. \colorbox{black}{\color{white}\scriptsize Homogenisation}

\citet{rettberg2025ai} show that stories generated by GPT-4o-mini consistently favor stability over change, irrespective of national context. Prompting GPT-4o-mini to produce more than 11{,}800 stories (50 for each of 236 countries) the authors find that story protagonists resolve minor conflicts by returning to tradition, organizing community events, and avoiding romance or major tension in favor of nostalgia and reconciliation. 
The study concludes that this standardisation constitutes a distinctive form of AI narrative bias, promoting global homogeneity and reinforcing cultural stereotypes in generative storytelling.
\colorbox{black}{\color{white}\scriptsize Homogenisation}

\citet{tao2024cultural} evaluate cultural bias and alignment in five GPT-based LLMs by comparing their outputs against standardized cultural surveys and cross-national cultural benchmarks. Across models, the expressed values align most closely with those of English-speaking and Protestant European countries, with substantial misalignment observed for many other regions. The authors further show that cultural prompting (i.e., specifying a cultural identity in the prompt) improves alignment with local cultural values for 71–81\% of countries and territories, although notable biases remain for underrepresented regions. 

\citet{zhang2021language} evaluate both encoder-only and decoder-only LLMs for movie recommendation tasks using zero-shot prompting and fine-tuning. 
In the zero-shot setting, they find that LLMs outperform a random recommendation baseline, yet exhibit strong linguistic biases toward frequent or popular items. With fine-tuning, these biases are mitigated given sufficient training data.
\colorbox{black}{\color{white}\scriptsize Concentration}

\citet{hou2024large} show that prompting LLMs with ranking tasks make them effective recommenders. 
They find that LLM outputs are susceptible to biases induced by item popularity and prompt position. These limitations can be mitigated through carefully designed prompting techniques and bootstrapping strategies. \colorbox{black}{\color{white}\scriptsize Concentration}

\citet{spurlock2024chatgpt} introduce a rigorous conversational recommendation pipeline using ChatGPT, where users iteratively reprompt the model with feedback to refine its recommended items. 
The results show that this reprompting improves recommendation relevance and help mitigate popularity bias. 
\colorbox{black}{\color{white}\scriptsize Concentration}

\citet{lichtenberg2024large} investigate whether LLMs reinforce popularity bias. 
They compare an LLM-based movie recommender against traditional recommenders. The results show that LLM-based recommenders exhibit lower popularity bias, indicating a reduction in concentration. 
However, accuracy decreases as the models recommend increasingly niche or less frequently consumed items. 
\colorbox{black}{\color{white}\scriptsize Concentration}

\citet{di2025addressing} investigate the ability of Llama-based LLMs to address popularity bias in third-party library recommenders. 
The study uses zero-shot and few-shot settings, fine-tuning, and post-processing penalty mechanisms. 
The findings show that while these strategies can increase recommendation diversity, LLMs struggle to mitigate popularity bias in software library recommendations. 
The paper also highlights the trade-off between relevance and diversity. 
\colorbox{black}{\color{white}\scriptsize Concentration}
\colorbox{black}{\color{white}\scriptsize Ecosystem Diversity}

\citet{deldjoo2024understanding} compare a GPT-based movie recommender with collaborative filtering. The study finds that the GPT-based recommender does not consistently outperform collaborative filtering in accuracy, but it tends to recommend newer and more diverse movies, particularly those released after 2000, indicating increased ecosystem diversity. 
The authors further observe distinct genre preferences: the GPT-based model favours drama, comedy, and romance, whereas collaborative filtering more often recommends action and adventure titles.
\colorbox{black}{\color{white}\scriptsize Ecosystem Diversity}

\citet{shumailov2023curse, shumailov2024ai} investigate the effects of incorporating model-generated content into the fine-tuning of generative models like Variational Autoencoders (VAEs), Gaussian Mixture Models (GMMs), and LLMs. They document the emergence of model collapse, characterized by the progressive overestimation of frequent events and underestimation of rare ones. 
\colorbox{black}{\color{white}\scriptsize Content Degradation}

\citet{alemohammad2023self} investigate self-consuming loops in generative image models, focusing on StyleGAN2~\cite{karras2020analyzing} and diffusion models such as DDPM~\cite{ho2020denoising}. They analyze three families of autophagous loops that vary in the availability of real versus synthetic data across training generations: fully synthetic loop, synthetic-augmentation loop, and fresh-data loop. Their results show that, in the absence of fresh real data, models exhibit progressively declining precision across generations, and while fixed real datasets can delay this degradation, they cannot ultimately prevent it. 
\colorbox{black}{\color{white}\scriptsize Content Degradation}

\citet{guo2024curious} examine linguistic diversity in self-consuming loops using the OPT-350M model~\cite{zhang2022opt} and introduce metrics for quantifying lexical, syntactic, and semantic diversity across model generations. They evaluate three use cases -- news summarisation, scientific abstract generation, and story generation -- and find that diversity degradation is most pronounced in high-entropy, more creative tasks.
\colorbox{black}{\color{white}\scriptsize Content Degradation}

\citet{martinez2023combining} replicate the experiment of \citet{guo2024curious} in the domain of image generation, employing denoising diffusion implicit models (DDIMs)~\cite{songdenoisingdiff}. In their simulation, the original training dataset is supplemented with AI-generated images, and a new model is trained within an autophagous loop, whereby the model learns from its own synthetic outputs. This process results in a decline in the quality of images generated by successive models. 
In a related effort, \citet{martinez2023towards} establish an autophagous loop where training uses an equal mix of authentic and self-generated data. They observe model collapse characterized by increased similarity to real data alongside reductions in precision and recall, indicating a degradation in generative performance.
\colorbox{black}{\color{white}\scriptsize Content Degradation}

\citet{briesch2023large} investigate autophagous loops using nanoGPT~\cite{Karpathy2022}, comparing a fully synthetic data cycle -- in which training data are entirely replaced by model -- generated outputs each generation—with three data-augmentation cycles that vary the proportion of real and synthetic data (balanced, incremental, and expanding). They find that the fully synthetic cycle leads to model collapse, while both the incremental and balanced cycles reduce diversity, with the incremental setup exhibiting the strongest degradation. Only the expanding cycle, in which new data are added rather than replaced, maintains diversity with no observable decline over 50 simulation steps.
\colorbox{black}{\color{white}\scriptsize Content Degradation}

\citet{hataya2023will} examine the impact of incorporating AI-generated images into future training datasets for downstream vision tasks. They simulate dataset contamination by progressively replacing portions of real images with synthetic images produced by state-of-the-art diffusion models. 
Their results show that synthetic images substantially degrade model performance, with the severity of degradation increasing with the level of contamination, underscoring the need for caution when using generated images for data augmentation. The authors further highlight the importance of reliable watermarking techniques for detecting AI-generated images. To this end, they propose a self-supervised detection method based on a masked autoencoder~\cite{he2022masked}, demonstrating that it can mitigate degenerative effects even when training datasets contain significant amounts of synthetic content.
\colorbox{black}{\color{white}\scriptsize Content Degradation}

\citet{bohacek2023nepotistically} also focus on a diffusion model. Starting with an image dataset containing 70000 faces, they automatically classify each face based on gender (man/woman), race (asian, black, hispanic, indian, and white), and age (young, middle-aged, old). 
These real images were used as input to image-to-image synthesis of the diffusion model to generate 900 images consistent with the demographic prompt “a photo of a [age] [race] [gender].” These 900 generated faces are used to seed the iterative model trainiing. This entire process is repeated with different proportions of generated faces and real faces. Regardless of the mixture ratio, the iterative retraining eventually leads to collapse by the fifth iteration, at which point the generated images are highly distorted. Lastly, authors retrained collapsed model for other five iterations on only real images, showing that in this way the model can be partially healed. \colorbox{black}{\color{white}\scriptsize Content Degradation}

\citet{dohmatob2024model, dohmatob2024tale, dohmatob2024strong} conduct experiments with Llama-2 and develop a mathematical formalisation of model collapse. They show that models trained on mixtures of real and AI-generated data exhibit initial performance gains but ultimately collapse as training continues. Their findings underscore the importance of early detection mechanisms to prevent or mitigate collapse. 
\colorbox{black}{\color{white}\scriptsize Content Degradation}

\citet{bertrand2024stability} present a theoretical framework for iterative retraining of generative models and show that stability can be achieved when the initial model sufficiently approximates the real data distribution and when each iteration retains a high enough proportion of genuine data. Otherwise, models risk collapsing toward degenerate, nearly identical outputs. Their empirical results on synthetic and natural image datasets corroborate these conditions, highlighting that stability depends more on a strong initial model and adequate preservation of real data than on injecting large amounts of new data.
\colorbox{black}{\color{white}\scriptsize Content Degradation}

\citet{seddik2024bad} study model collapse through a statistical model and conducting experiments in which they consider fully synthetic and partially synthetic setting, finding that total collapse is unavoidable when training solely on synthetic data and that it can be avoided if enough real data is injected into the recursive training process. \colorbox{black}{\color{white}\scriptsize Content Degradation}

\citet{herel2024collapse} place pre-trained GPT-2 architectures of varying sizes into an autophagous loop, running each simulation until model collapse is detected or until 1,000 iterations are reached. They find that collapse occurs more rapidly when using higher learning rates and larger-parameter models.
\colorbox{black}{\color{white}\scriptsize Content Degradation}

\citet{gerstgrasser2024model} examine the behavior of generative models in autophagous loops when synthetic content is accumulated across generations. Using both an analytical framework and empirical experiments with diffusion models, VAEs, and GPT- and Llama-based LLMs, they find that data accumulation can mitigate degradation. Their results align with \citet{briesch2023large}, confirming that accumulating generated data -- rather than repeatedly replacing the dataset -- can help prevent model collapse.
\colorbox{black}{\color{white}\scriptsize Content Degradation}

\citet{wang2025llm} highlight that most prior work examines model collapse in isolated, single-model settings and introduce LLM Web Dynamics (LWD), a framework for studying collapse at the network level by simulating an Internet-like environment. Using this framework, they observe convergence patterns in model outputs, revealing how collapse dynamics can propagate across interconnected models.
\colorbox{black}{\color{white}\scriptsize Content Degradation}

\citet{suresh2025rate} theoretically characterize the rate of collapse under recursive training, highlighting fundamental risks associated with relying exclusively on generated data. 
For discrete distributions, they show that the model “forgets’’ original data elements at a rate approximately proportional to their initial frequency, with rare elements disappearing most quickly across iterations. For Gaussian models, the standard deviation of the learned distribution collapses toward zero after roughly $n$ iterations -- where $n$ is the number of samples used at each step -- signaling a rapid loss of diversity.
\colorbox{black}{\color{white}\scriptsize Content Degradation}

\citet{fu2025theoretical} introduce the notion of recursive stability to quantify how a generative model’s outputs change over multiple rounds of recursive training when small perturbations are applied to the initial real dataset, demonstrating that model stability -- alongside maintaining a non-negligible proportion of real data -- is a critical condition for preventing model collapse.
\colorbox{black}{\color{white}\scriptsize Content Degradation}

\citet{feng2024beyond} investigate how data selection can prevent model collapse, providing a theoretical characterisation based on Gaussian mixtures, linear classifiers, and linear verifiers to determine when a verifier can effectively identify synthesized data that preserve optimal performance. They evaluate their framework on two practical tasks -- computing matrix eigenvalues with transformers and news summarisation with LLMs -- and show that even moderately effective selection mechanisms can prevent collapse.
\colorbox{black}{\color{white}\scriptsize Content Degradation}

\citet{drayson2025machine} train a machine-generated text detector and propose a resampling approach to prevent model collapse by up-sampling likely human content in the training data.  Validating the method on four LLMs from two different families and across a range of model sizes, they prove that it not only prevents model collapse but also improves performance compared to training on purely human data.
\colorbox{black}{\color{white}\scriptsize Content Degradation}

\citet{kazdan2024accumulating} analyze the effects of various strategies for incorporating synthetic data during training, demonstrating across three case studies that data replacement leads to collapse, whereas data accumulation avoids it. Moreover, for certain combinations of datasets and sample sizes per iteration, training on a mixture of real and accumulated synthetic data yields lower loss on real test sets than training solely on real data.
\colorbox{black}{\color{white}\scriptsize Content Degradation}

\citet{gillman2024self} introduce a technique for stabilizing self-consuming generative model training by employing self-correction functions that remap synthesized data to be more probable under the true data distribution. They demonstrate that this approach effectively prevents model collapse, even when synthetic data constitute up to 100\% of the training set.
\colorbox{black}{\color{white}\scriptsize Content Degradation}

\citet{zhu2024synthesize} examine how different proportions of synthetic data in pre-training affect language model performance, showing that increasing the share of synthetic data leads to predictable model collapse, manifested as over-concentration of n-grams. To mitigate this effect, they propose a token-level editing strategy that replaces tokens assigned high model probability given their context, and demonstrate that this approach effectively prevents collapse.
\colorbox{black}{\color{white}\scriptsize Content Degradation}

\citet{ferbach2024self} examine how data curation -- such as users selecting preferred outputs from generative models -- shapes the dynamics of iterative retraining on synthetic data. They show that curation implicitly performs preference optimisation, despite the model never observing explicit rewards or comparisons. The authors theoretically prove that curated retraining maximizes expected reward and analyze the conditions under which the retraining loop remains stable when real data is incorporated. \colorbox{black}{\color{white}\scriptsize Content Degradation}

\citet{zhu2025matters} investigate how different levels of diversity in LLM-generated data influence model performance, examining the effects of mixing varying proportions of synthetic data. They find that moderately diverse generated data can improve performance in settings with limited labeled data, whereas highly diverse synthetic data has a detrimental effect.
\colorbox{black}{\color{white}\scriptsize Content Degradation}

\citet{gambetta2025} show that model collapse can be mitigated by fine-tuning the generative AI model using documents that contain a high degree of "surprise", i.e., those documents that do not align with the token probability distribution of the generative AI model.
\colorbox{black}{\color{white}\scriptsize Content Degradation}

\citet{schaeffer2025position} highlight the conceptual ambiguity surrounding the term model collapse, noting that the literature employs multiple, sometimes conflicting, definitions. They propose a taxonomy of eight collapse types and argue that clarifying terminology is essential for developing a unified theoretical framework for the phenomenon.
\colorbox{black}{\color{white}\scriptsize Content Degradation}

\subsection{Empirical studies}
\subsubsection*{\bf Observational studies}
\citet{kobak2025delving} analyse over 15 million PubMed biomedical abstracts from 2010–2024 and show that the emergence of LLMs coincides with an increase in the frequency of certain terms. The study also estimates that at least 13.5\% of 2024 biomedical abstracts contain LLM-associated vocabulary, suggesting an increasing standardisation of scientific writing.
\colorbox{black}{\color{white}\scriptsize Concentration}
\colorbox{black}{\color{white}\scriptsize Homogenisation}

\subsubsection*{\bf Controlled studies}

\citet{agarwal2025ai} investigate the cultural impact of Western-centric AI writing models in a controlled study involving 118 participants who completed culturally grounded writing tasks both with and without AI assistance. The authors find that AI-generated suggestions systematically nudge non-Western users toward Western stylistic norms and cultural framings. This shift leads to more homogenised writing and a measurable reduction in culturally distinctive expression.
\colorbox{black}{\color{white}\scriptsize Homogenisation}

\citet{desdevises2025paradox} set up a task in which participants use ChatGPT-4o to generate as many uses or ideas as possible for a common object (an egg). 
The outputs generated by ChatGPT-4o are evaluated against (i) those of 47 human participants and (ii) aggregated results from eight prior studies employing the same protocol.
The findings show that ChatGPT-4o is more productive than human participants, but its outputs exhibit fixation patterns closely aligned with those observed in human responses, i.e., most of its ideas fall into conventional categories and therefore lack originality. 
Compared to aggregate results from prior studies, ChatGPT-4o produces a higher proportion of fixation ideas than the weighted average reported in prior studies.
\colorbox{black}{\color{white}\scriptsize Homogenisation}
\colorbox{black}{\color{white}\scriptsize Ecosystem Volume}

\citet{doshi2024generative} conduct a randomises field study to investigate the effects of LLM assistance on creative writing. They show that the possibility of using AI-generated suggestions enhances the creativity, quality, and enjoyability of individuals’ stories, particularly for writers with initially lower creativity. 
However, the stories written with the aid of generative AI exhibit increasing stylistic convergence, reducing diversity and novelty at the system level. 
This dynamic produces a social dilemma in which individual gains come at the expense of collective originality.
\colorbox{black}{\color{white}\scriptsize Homogenisation}
\colorbox{black}{\color{white}\scriptsize Individual Diversity}

\citet{wan2025using} extend the experimental design of \citet{doshi2024generative} by introducing diversity at the AI input stage through ten generative AI personas with different traits. 
They show that this intervention preserves or enhances collective diversity in human creative outputs, while still supporting individually diverse responses. 
\colorbox{black}{\color{white}\scriptsize Individual Diversity}
\colorbox{black}{\color{white}\scriptsize Ecosystem Diversity}

\end{document}